\documentclass[reprint,superscriptaddress,amsmath,amssymb,aps,prx]{revtex4-2}
\usepackage{graphicx}
\usepackage{dcolumn}
\usepackage{bm}
\usepackage{qcircuit}
\usepackage{algpseudocode}
\usepackage{algorithm}
\usepackage{subfigure}
\usepackage[pass]{geometry}
\usepackage{float}
\usepackage{amsthm}

\newtheorem{lemma}{lemma}

\makeatletter
\renewcommand*\env@matrix[1][\arraystretch]{%
\edef\arraystretch{#1}%
\hskip -\arraycolsep
\let\@ifnextchar\new@ifnextchar
\array{*\c@MaxMatrixCols c}}
\makeatother
\begin{document}
\preprint{APS/123-QED}
\title{Unconditionally decoherence-free quantum error mitigation by density matrix vectorization}
\author{Zhong-Xia Shang}
\email{ustcszx@mail.ustc.edu.cn}
\affiliation{Hefei National Research Center for Physical Sciences at the Microscale and School of Physical Sciences,
University of Science and Technology of China, Hefei 230026, China}
\affiliation{Shanghai Research Center for Quantum Science and CAS Center for Excellence in Quantum Information and Quantum Physics,
University of Science and Technology of China, Shanghai 201315, China}
\affiliation{Hefei National Laboratory, University of Science and Technology of China, Hefei 230088, China}
\author{Zi-Han Chen}
\email{czh007@mail.ustc.edu.cn}
\affiliation{Hefei National Research Center for Physical Sciences at the Microscale and School of Physical Sciences,
University of Science and Technology of China, Hefei 230026, China}
\affiliation{Shanghai Research Center for Quantum Science and CAS Center for Excellence in Quantum Information and Quantum Physics,
University of Science and Technology of China, Shanghai 201315, China}
\affiliation{Hefei National Laboratory, University of Science and Technology of China, Hefei 230088, China}
\author{Cai-Sheng Cheng}
\email{ccs112202@mail.ustc.edu.cn}
\affiliation{Department of Chemical Physics, University of Science and Technology of China, Hefei, Anhui 230026, China}
\begin{abstract}
Fighting against noise is crucial for NISQ devices to demonstrate practical quantum applications. In this work, we give a new paradigm of quantum error mitigation based on the vectorization of density matrices. Different from the ideas of existing quantum error mitigation methods that try to distill noiseless information from noisy quantum states, our proposal directly changes the way of encoding information and maps the density matrices of noisy quantum states to noiseless pure states, which is realized by a novel and NISQ-friendly measurement protocol and a classical post-processing procedure. Our protocol requires no knowledge of the noise model, no ability to tune the noise
strength, and no ancilla qubits for complicated controlled
unitaries. Under our encoding, NISQ devices are always preparing pure quantum states which are highly desired resources for variational quantum algorithms to have good performance in many tasks. We show how this protocol can be well-fitted into variational quantum algorithms. We give several concrete ansatz constructions that are suitable for our proposal and do theoretical analysis on the sampling complexity, the expressibility, and the trainability. We also give a discussion on how this protocol is influenced by large noise and how it can be well combined with other quantum error mitigation protocols. The effectiveness of our proposal is demonstrated by various numerical experiments. 
\end{abstract}
\maketitle
\section{Introduction}
Quantum computers \cite{nielsen2010quantum} are expected to take advantage of quantum mechanical phenomena such as quantum superposition and quantum entanglement to give speedup to some classically difficult problems. Such speedup has been theoretically discovered by various quantum algorithms such as the quantum simulation \cite{feynman2018simulating}, Shor's factoring algorithm \cite{shor1999polynomial}, and Grover's searching algorithm \cite{grover1996fast}. To build a quantum computer and show such speedup in reality, the biggest challenge is to fight against noise and protect fragile quantum mechanical behaviors \cite{unruh1995maintaining}. While there have been significant theoretical developments for fault-tolerant quantum computation~\cite{shor1996fault,aharonov1997fault,dennis2002topological,bravyi2005universal,bombin2006topological,fowler2012surface,gottesman2013fault,litinski2019game} as well as recent experimental demonstrations~\cite{chen2021exponential,erhard2021entangling,ryan2021realization,krinner2022realizing,zhao2022realization,google2023suppressing,bluvstein2024logical,da2024demonstration} of quantum error correction and logical operations, in practice, in order to obtain a few logical qubits with highly suppressed logical error rate, simultaneously scaling up the number of physical qubits and maintaining a low enough physical error rate is still challenging.

Currently, we are in the Noisy Intermediate Scale Quantum (NISQ) era \cite{preskill2018quantum} where we lack the ability to build qubits and implement quantum circuits on the physical level with high qualities on a large scale that meet the thresholds for fault-tolerant quantum computing. Thus, NISQ devices are not capable of implementing most of the quantum algorithms that require a deep quantum circuit and a large number of qubits \cite{montanaro2016quantum}. To make current quantum computers useful, especially under the situation where quantum advantages have already been demonstrated using these devices \cite{arute2019quantum,zhong2020quantum}, researchers have paid much attention to designing NISQ algorithms \cite{bharti2022noisy}. One of the most popular types is the variational quantum algorithms (VQAs) \cite{cerezo2021variational} including the Variational Quantum Eigensolver \cite{peruzzo2014variational} and the Quantum Approximate Optimization Algorithm \cite{farhi2014quantum}. The workflow of VQAs is a combination of both quantum and classical computing. We typically use a parametrized quantum circuit to prepare a trial state and measure a cost function corresponding to the expectation value of an observable. The cost function is then fed into a classical optimizer to update parameters in the quantum circuit. This process will repeat until the cost function converges. The parametrized quantum circuit depth is relatively shallower than other algorithms and the optimization procedure has some degree of resilience of the noise. These features make the VQAs more friendly to NISQ quantum computers and have been designed for solving problems in various areas \cite{cerezo2021variational} including combinatorial optimization \cite{farhi2014quantum}, quantum chemistry \cite{peruzzo2014variational}, machine learning \cite{schuld2019quantum}, and simulating quantum dynamics \cite{yuan2019theory}.

While VQAs have relatively low requirements for quantum devices, the non-neglectable noise in the devices can still severely reduce their performance in approximating the solutions to the target problems. This can be seen from 3 perspectives. First, for general settings, the desired target quantum states are pure while NISQ devices can hardly prepare pure states due to the decoherence led by noise, which sets a gap between the reachable VQA results and the true results \cite{stilck2021limitations,de2023limitations}. Second, since the noise grows exponentially as the number of gates grows, NISQ devices can only build shallow quantum circuits with local connectivity to prepare quantum states with limited expressibility \cite{sim2019expressibility} and entanglement \cite{bravyi2006lieb} that may not cover the desired states, which is another gap limiting the accuracy. Third, it has been shown that noise can introduce barren plateau problems \cite{wang2021noise} to make VQAs un-trainable. 

All these issues give strong demands on developing methods beyond QECs to suppress the influence of noise. One approach for this purpose is to introduce classical strategies including neural networks \cite{zhang2022variational}, Clifford circuits \cite{shang2023schrodinger}, and tensor networks \cite{huang2023tensor} to reduce the circuit depth required on quantum devices. The other approaches \cite{temme2017error,endo2018practical,
li2017efficient,mcardle2019error,mcclean2017hybrid,huggins2021virtual,cai2021quantum,liu2024virtual}, while realized in different ways, are guided by the philosophy of distilling the noiseless information from noisy quantum states without actually recovering them, which are known as the quantum error mitigation (QEM) methods \cite{cai2023quantum}. Since the mitigation procedures are done after the occur of noise, the additional overhead of QEM is exponential \cite{cai2023quantum} rather than polynomial as in QEC, as compensation, QEM has low requirements on hardware, which makes QEM a promising tool for NISQ experiments. However, most of the existing QEM protocols have their own additional unfriendly requirements ranging from prior knowledge of the noise model \cite{temme2017error,endo2018practical} or the symmetry structures of problems \cite{mcardle2019error,mcclean2017hybrid,cai2021quantum} to the ability to tune hardware noise strength \cite{li2017efficient,temme2017error} or do complex indirect collective measurements \cite{huggins2021virtual,liu2024virtual}.

In this work, we will introduce a new QEM paradigm for VQAs based on density matrix vectorization (DMV). Unlike the philosophy of extracting noiseless information from noisy quantum states in existing QEM protocols, our protocol realizes error mitigation by changing the way of encoding information. We encode pure states into linear combinations of vectorized density matrices which directly leads to unconditionally decoherence-free pure state preparations from noisy quantum circuits. While the sampling overhead of our protocol to unconditionally eliminate decoherence is still exponential and comparable with other QEM methods such as the probabilistic error cancellation \cite{temme2017error,endo2018practical}, the way it mitigates noise might open up new possibilities to go beyond recent results on the limitations of QEM \cite{takagi2022fundamental,takagi2023universal,tsubouchi2023universal,quek2022exponentially}. By our protocol, the performance gap induced by decoherence can be directly eliminated and the gap led by low expressibility can be well mitigated without introducing additional barren plateau resources. Our protocol requires nothing but only needs 2-qubit collective unitaries before measurements to extract needed information. In the following, we will show details of our QEM protocol from its basic framework to its VQA applications. 

\section{Density matrix vectorization (DMV)}
\subsection{QEM by DMV}
In NISQ systems, since a pure quantum state can suffer from noise in the environment and thus becomes a mixed state, the wave function description is no longer universal, instead, the density matrix description is adopted to describe pure states and mixed states in a unified language. The basic idea of our protocol is inspired by the mapping:
\begin{equation}\label{1}
\rho\rightarrow |\rho\rangle
\end{equation} 
where $\rho=\sum_{ij}\rho_{ij}|i\rangle\langle j|$ and $|\rho\rangle$ is defined as $\frac{1}{C_{\rho}}\sum_{ij}\rho_{ij}|i\rangle|j\rangle$ with the normalization factor $C_\rho=||\rho||_F=\sqrt{\sum_{ij}|\rho_{ij}|^2}=\sqrt{\text{Tr}(\rho^2)}$. While DMV has been vastly used as a useful mathematical trick for simplifying many concepts in quantum information science, in this work, we see DMV in a different way. The main concept transition here is that the mapping Eq. \ref{1} means we are treating an $n$-qubit density matrix $\rho$ as a 2$n$-qubit pure state $|\rho\rangle$, which shares the same idea as in Ref. \cite{shang2024polynomial} to demonstrate a new exponential quantum speedup. By this encoding, any quantum state either pure or mixed will always be mapped to a pure state. Note that the reverse mapping $|\rho\rangle\rightarrow \rho$ has also been adopted in several works \cite{yoshioka2020variational,shang2024hermitian} for simulating open quantum systems using unitary circuits. 

\begin{figure*}[htbp]
\centering
\includegraphics[width=0.7\textwidth]{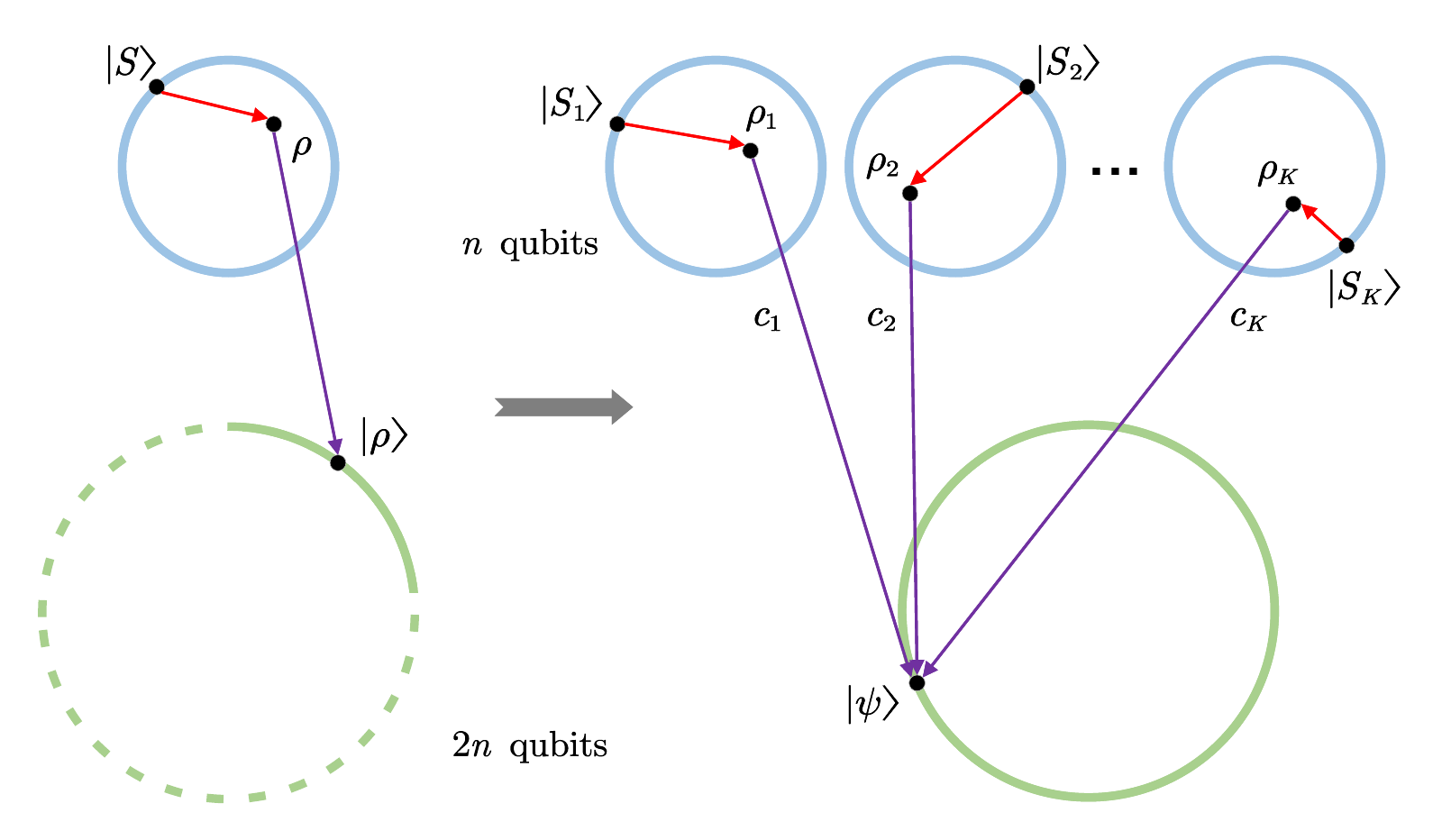}
\caption{The basic idea of quantum error mitigation by density matrix vectorization. Ideally, we want to prepare an $n$-qubit pure state $|S\rangle$ which, due to the noise in NISQ circuits, becomes $\rho$ which is generally a mixed state. Under the mapping Eq. \ref{1}, the noisy state $\rho$ is then treated as a $2n$-qubit pure state $|\rho\rangle$. Thus, in this way, any $n$-qubit state is unconditionally mapped to a $2n$-qubit pure state. Due to the Hermiticity and the positive semi-definiteness of density matrices, $|\rho\rangle$ cannot represent arbitrary $2n$-qubit pure states, thus, the mapping Eq. \ref{2} is introduced where any $2n$-qubit pure state $|\psi\rangle$ can be represented as a linear combination of many density matrices.\label{f1}}
\end{figure*}

Currently, due to the Hermiticity and the positive semi-definiteness of $\rho$, we can not encode all 2$n$-qubit pure states. To solve this problem, we can introduce a generalized mapping $f$:
\begin{eqnarray}\label{2}
&&\{\rho_1,\rho_2,...,\rho_K,c_1,c_2,...,c_K\}\\&&\rightarrow
|\psi\rangle=\frac{1}{C_\psi}\sum_{ij}(c_1\rho_{1,ij}+c_2\rho_{2,ij}+...+c_K\rho_{k,ij})|i\rangle|j\rangle \nonumber
\end{eqnarray}
where we use $K$ density matrices with $K$ complex coefficients to form a pure state $|\psi\rangle$ with the normalization factor $C_\psi=\sqrt{\sum_{ij}|c_1\rho_{1,ij}+c_2\rho_{2,ij}+...+c_k\rho_{k,ij}|^2}$. These density matrix vectorization procedures are visualized in Fig. \ref{f1}. While the value of $K$ and the value of coefficients can be chosen arbitrarily, it is sufficient to set $K=4$ with 2 real and 2 imaginary coefficients to express any pure state $|\psi\rangle$. This can be understood from the matrix form $|\psi\rangle \rightarrow \psi$ which can be decomposed first into Hermitian part and $i*$Hermitian (anti-Hermitian) part, each Hermitian part is indefinite and can be further expressed as a linear combination of two positive semi-definite density matrices. 

The mapping Eq. \ref{2} currently is only a mathematical transformation whose unnormalized version has been vastly adopted as a convenient representation of quantum elements \cite{gilchrist2009vectorization}. To give this mapping a real sense, we need to be able to extract information of $|\psi\rangle$ from $\{\rho_1,\rho_2,...,\rho_K,c_1,c_2,...,c_K\}$. Interestingly, we find that, given a Hamiltonian $H_A$, the expectation value with respect to $|\psi\rangle$ can be re-expressed as:
\begin{eqnarray}\label{3}
&&\langle\psi|H_A|\psi\rangle=\nonumber\\&&\frac{\sum_{i,j=1}^K c_i^*c_j\text{Tr}(H_B\rho_i\otimes\rho_j)}{\sum_{k,l=1}^Kc_k^*c_l\text{Tr}(\rho_k\rho_l)}
\end{eqnarray}
with 
\begin{eqnarray}\label{4}
\langle il|H_B| jk\rangle=\langle ij| H_A | kl\rangle
\end{eqnarray}
We call $H_B$ the substitute Hamiltonian of $H_A$. Eq. \ref{3} and Eq. \ref{4} means that by measuring the values of all $\text{Tr}(H_B\rho_i\otimes\rho_j)$ and $\text{Tr}(\rho_i\rho_j)$, one can obtain the value of $\langle\psi|H_A|\psi\rangle$ by classical post combination. 

\begin{table}[t]
\centering
\begin{ruledtabular}
\begin{tabular}{c|ccc} 
ID&$P$& $Q$ & Spectra of $Q$ \\\hline
1&$II$& $0.5 II+0.5XX+0.5YY+0.5ZZ$& \{1,1,1,-1\}\\
2&$XX$& $0.5 II+0.5XX-0.5YY-0.5ZZ$& \{1,1,-1,1\}\\
3&$YY$& $-0.5 II+0.5XX-0.5YY+0.5ZZ$& \{1,-1,-1,-1\}\\
4&$ZZ$& $0.5 II-0.5XX-0.5YY+0.5ZZ$& \{1,-1,1,1\}\\\hline
5&$IX$& $0.5 IX+0.5XI+0.5iYZ-0.5iZY$& \{-1,$i$,-$i$,1\}\\
6&$XI$& $0.5 IX+0.5XI-0.5iYZ+0.5iZY$& \{-1,$i$,-$i$,1\}\\
7&$YZ$& $-0.5i IX+0.5iXI+0.5YZ+0.5ZY$& \{-1,$i$,-$i$,1\}\\
8&$ZY$& $-0.5i IX+0.5iXI-0.5YZ-0.5ZY$& \{-1,$i$,-$i$,1\}\\\hline
9&$IY$& $-0.5 IY+0.5iXZ-0.5YI-0.5iZX$& \{-1,$i$,-$i$,1\}\\
10&$YI$& $0.5 IY+0.5iXZ+0.5YI-0.5iZX$& \{-1,$i$,-$i$,1\}\\
11&$XZ$& $0.5i IY+0.5XZ-0.5iYI+0.5ZX$& \{-1,$i$,-$i$,1\}\\ 
12&$ZX$& $-0.5i IY+0.5XZ+0.5iYI+0.5ZX$& \{-1,$i$,-$i$,1\}\\\hline 
13&$IZ$& $0.5 IZ+0.5iXY-0.5iYX+0.5ZI$& \{$i$,-$i$,1,-1\}\\
14&$ZI$& $0.5 IZ-0.5iXY+0.5iYX+0.5ZI$& \{$i$,-$i$,1,-1\}\\   
15&$XY$& $0.5i IZ-0.5XY-0.5YX-0.5iZI$& \{1,-1,-$i$,$i$\}\\
16&$YX$& $0.5i IZ+0.5XY+0.5YX-0.5iZI$& \{1,-1,-$i$,$i$\}\\
\end{tabular}
\end{ruledtabular}
\caption{The 16 2-qubit Pauli operators and their corresponding 2-qubit Pauli substitute operators.\label{t1}}
\end{table}

We now show an efficient measurement procedure for NISQ hardware to evaluate the expectation value Eq. \ref{3}. We can express $H_A$ under the multi-qubit Pauli basis as:
\begin{equation}\label{5}
H_A=\sum_{\alpha=1}^m g_\alpha P_\alpha
\end{equation}
where $P_\alpha$ are $2n$-qubit Pauli operators and $g_\alpha$ are real numbers. When $H_A$ is transformed into $H_B$, each $P_\alpha$ is transformed into $Q_\alpha$ as: 
\begin{equation}\label{6}
H_B=\sum_{\alpha=1}^m g_\alpha Q_\alpha
\end{equation}
In the smallest case of $n=1$, we show the basic substitute operators of all 16 two-qubit Pauli operators in Table. \ref{t1}. Note that all 16 basic substitute operators are unitary while no longer Hermitian. Since it is easy to check that the transformation rule in Eq. \ref{5} preserves the tensor structure, thus, for $n\geq 1$ cases, $P_\alpha$ as the tensor product of two-qubit Pauli operators leads to $Q_\alpha$ as the tensor product of basic substitute operators which is also unitary \cite{shang2024polynomial}. In fault-tolerant quantum computers, we can use the Hadamard test circuits \cite{aharonov2006polynomial} and swap test circuits \cite{buhrman2001quantum} to evaluate $\text{Tr}(H_B\rho_i\otimes\rho_j)$ and $\text{Tr}(\rho_i\rho_j)$. 

For NISQ hardware, it is generally difficult to build such controlled circuits with high fidelities. Therefore, we can instead adopt the idea of the operator averaging method \cite{mcclean2016theory}. For each $\text{Tr}(Q_\alpha\rho_i\otimes\rho_j)$ in $\text{Tr}(H_B\rho_i\otimes\rho_j)$, we can simply do a basis rotation circuit to the diagonal basis of $Q_\alpha$ which requires separate two-qubit gates to rotate each basic 2-qubit substitute operators to its diagonal basis and then do repeated measurements to estimate $\text{Tr}(Q_\alpha\rho_i\otimes\rho_j)$. For $\text{Tr}(\rho_i\rho_j)$, its value equals to $\text{Tr}(S\rho_i\otimes\rho_j)$ with $S$ the tensor product of $n$ two-qubit swap gates which can also be estimated in the diagonal basis of $S$. After the estimation of values of $\text{Tr}(Q_\alpha\rho_i\otimes\rho_j)$ and $\text{Tr}(S\rho_i\otimes\rho_j)$ by repeated measurements for different $\{Q_\alpha, \rho_i,\rho_j\}$, we can then do classical processing based on Eq. \ref{3} to get an estimation of $\langle\psi|H_A|\psi\rangle$. We give concrete constructions of these rotation circuits in the appendix.

\begin{figure*}[htbp]
\centering
\includegraphics[width=0.72\textwidth]{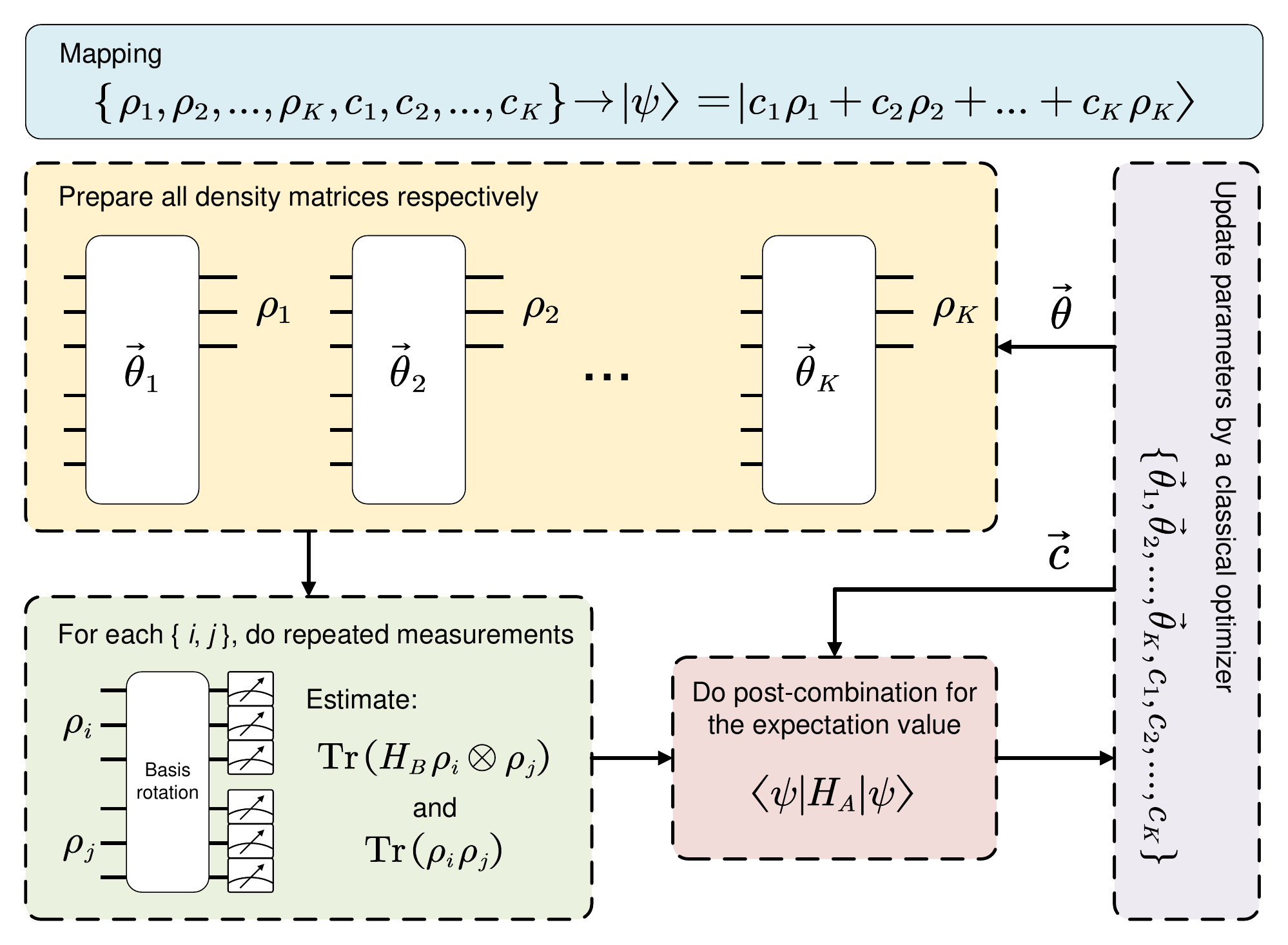}
\caption{Running VQAs with QEM by DMV. For each $n$-qubit density matrices, a $2n$-qubit parametrized quantum circuit with parameters $\vec{\theta_i}$ is assigned. These density matrices $\{\rho_1,\rho_2,...,\rho_K\}$ combined with classical coefficients $\{c_1,c_2,...,c_K\}$ form a $2n$-qubit pure state $|\psi\rangle$ according to the mapping Eq. \ref{2}. The expectation value $\langle\psi|H_A|\psi\rangle$ is evaluated by direct measurements for $\text{Tr}(H_B\rho_i\otimes\rho_j)$ and $\text{Tr}(S\rho_i\otimes\rho_j)$ with classical processing for linear combinations according to Eq. \ref{3}. The basis rotation circuits before measurements are composed of tensor products of 2-qubit rotations. The values of $\langle\psi|H_A|\psi\rangle$ as the cost function are determined by coefficients $\{\vec{\theta_1},\vec{\theta_2},...,\vec{\theta_K},c_1,c_2,...,c_K\}$ and are fed into a classical optimizer to update parameters for the next round.\label{f2}}
\end{figure*}

Being able to evaluate the value of $\langle\psi|H_A|\psi\rangle$ from $\{\rho_1,\rho_2,...,\rho_K,c_1,c_2,...,c_K\}$ by Eq. \ref{2}-\ref{6} forms the core of our proposal, which shows that we can ``prepare" unconditionally decoherence-free pure quantum states from noisy quantum circuits (Fig. \ref{f1}) and thus, can greatly mitigate the influence of noise. We should emphasize that the philosophy of our error mitigation protocol is not to extract noiseless information from noisy quantum states as in other QEM protocols but to directly encode noiseless quantum states into noisy ones. From the operational level, our protocol requires no knowledge of the noise model, no knowledge of the symmetry structures of problems, no ability to tune the noise strength, and no ancilla qubits for complicated controlled unitaries. Thus, our QEM protocol might be a promising QEM protocol for NISQ quantum applications.

\subsection{Entanglement in terms of DMV}\label{ent}
In the mapping Eq. \ref{1}, an $n$-qubit density matrix is mapped to a $2n$-qubit pure state $|\rho\rangle=\frac{1}{C_{\rho}}\sum_{ij}\rho_{ij}|i\rangle|j\rangle$ where we will call the $n$-qubit system denoted by the row index $i$ the row system and the $n$-qubit system denoted by the column index $j$ the column system. When $\rho$ is a pure state of the form $\rho=|S\rangle\langle S|$, we have $ |\rho\rangle=|S\rangle\otimes|S\rangle$ as a product state between the row system and the column system. When $\rho$ is a mixed state of the form $\rho=\sum_i p_i|s_i\rangle\langle s_i|$, the state $|\rho\rangle=\frac{1}{\sqrt{\text{Tr}(\rho^2)}}\sum_i p_i |s_i\rangle\otimes|s_i\rangle$ becomes an entangled state. Note that for the density matrix $\rho$, we can also define another commonly used pure state for it by purification \cite{preskill1998lecture} defined as $|\psi_\rho\rangle=\sum_i\sqrt{p_i}|s_i\rangle$. We proved that the Rényi entropies \cite{muller2013quantum} between these two states have the relation:
\begin{equation}\label{7}
H_\alpha(|\rho\rangle)\leq H_\alpha\left(|\psi_\rho\rangle\right)
\end{equation}
Since for $\alpha=2$, we have $ H_\alpha\left(|\psi_\rho\rangle\right)=-\log (\text{Tr}(\rho^2))$, the achievable entanglement of $|\rho\rangle$ is bound by the purity of $\rho$. When $\rho$ is the maximally mixed state, the state $|\rho\rangle$ becomes the maximally entangled state.

In the mapping Eq. \ref{2}, the entanglement of the state $|\psi\rangle$ is not only influenced by the purity of density matrices but also the number of linear combinations $K$. This can be understood as the linear combinations can enlarge the number of non-zero Schmidt coefficients and thus increase the entanglement. Indeed, we find that when $\{\rho_1,\rho_2,...,\rho_K$ are prepared randomly (roughly a unitary 1-design \cite{mele2023introduction}), the second order Rényi entropy can be enlarged by an amount $\log(K)$ at most.

As we will show later, the sampling complexity of estimating Eq. \ref{3} depends on the purity of density matrices. We require a high average of purity to make the sampling cost tolerable. Since the purity also decides the level of entanglement of pure states from DMV, this high purity restriction can thus limit the expressibility of representing highly entangled states between the row and the column system. To help solve this problem, an observation is that the purity has no restrictions on the entanglement inside the row and the column system, and for most quantum states of physical interest \cite{zeng2019quantum}, the entanglement will not be saturated between any bi-partition of the whole system. Thus, given a $2n$-qubit pure state, it is possible to find a bi-partition with low entanglement and assign the two subsystems under this bi-partition as the row and the column systems respectively to efficiently represent it. We will show this strategy in the following numerical experiments.

\section{VQAs based on DMV}
\subsection{Framework}
A direct application for QEM is VQA, which is especially suitable for our protocol where error-mitigated pure states do not correspond to the states that circuits aim to prepare when there is no noise. Combined with the classical linear combinations, it is not a trivial task to deterministically prepare a quantum state under our mapping Eq. \ref{2} since there may exist many choices of density matrices and coefficients that correspond to the same state. However, for VQAs, this weakness can be well resolved since a classical optimizer will guide us on how to adjust quantum circuits.

Now, given a Hamiltonian $H_A$, we give a framework to run VQE for its ground states using our QEM protocol. Note that this framework can also be applied to other tasks such as variational quantum simulations \cite{yuan2019theory} and variational quantum machine learning \cite{schuld2019quantum}. For each density matrix $\rho_i$ in $\{\rho_1,\rho_2,...,\rho_K\}$, we assign a parametrized quantum circuit to prepare it with parameters $\vec{\theta_i}$. If we use an $n$-qubit noisy circuit to prepare a $n$-qubit density matrix, then the purity of the density matrix can only be adjusted by uncontrollable noise which is not universal. Thus, instead, we use $2n$-qubit circuits where we call the first $n$ qubits the upper systems and the last $n$ qubits the lower systems to prepare these density matrices on the upper systems by ignoring (tracing out) the lower systems. We will show later that $2n$ is an upper bound for circuits, and in many cases, we might use a circuit of a size $n+L$ with $L\leq n$. Note that the upper and lower systems should not be confused with the row and column systems. Concrete constructions of these circuits will be introduced in the next section. 

These density matrices combined with coefficients $\{c_1,c_2,...,c_K\}$ are then used to estimate the expectation value $\langle\psi|H_A|\psi\rangle$ by the measurement and the classical combination procedures introduced before. The whole parameter set is defined as $\{\vec{\theta_1},\vec{\theta_2},...,\vec{\theta_K},c_1,c_2,...,c_K\}$ which is fed into a classical optimizer for updating with $\langle\psi|H_A|\psi\rangle$ as the cost function. When the stopping criteria are met, the virtual pure state from optimized $\{\rho_1,\rho_2,...,\rho_K,c_1,c_2,...,c_K\}$ by Eq. \ref{2} then corresponds to the approximation of the ground state of $H_A$. The whole framework is summarized in Fig. \ref{f2}. Note that since each density matrix is prepared by an at most $2n$-qubit quantum circuit and during the measurements, we only pick two density matrices for collective measurements at a time, we have no need to build $K$ $2n$-qubit circuits but only need two $2n$-qubit circuits at most that prepare different density matrices depending on different measurement tasks. 

We want to mention that the idea of classical linear combinations of quantum states has been adopted in several proposals \cite{huggins2020non,bharti2021iterative,bharti2021quantum} for variational quantum algorithms to enhance the expressibility of NISQ devices without aggravating the barren plateau problems. The key component in these protocols is the measurement strategy for values like $\langle \psi_i|O|\psi_j\rangle$ which typically require indirect hardware-challenging modified Hadamard tests \cite{huggins2020non} or complicated direct measurement strategies \cite{mitarai2019methodology}. In contrast, an interesting and nice property of our measurement procedure is that values like $Tr(H_B\rho_i\otimes\rho_j)$ that contain the unnormalized information of $\langle\rho_i|H_A|\rho_j\rangle$ can be estimated by hardware-friendly direct measurements as shown before. Thus, our proposal inherits the advantages of classical linear combinations as in those proposals and at the same time, is easier to realize on NISQ hardware.

\begin{figure*}[ht]
    \centering
    \includegraphics[width=0.9\textwidth]{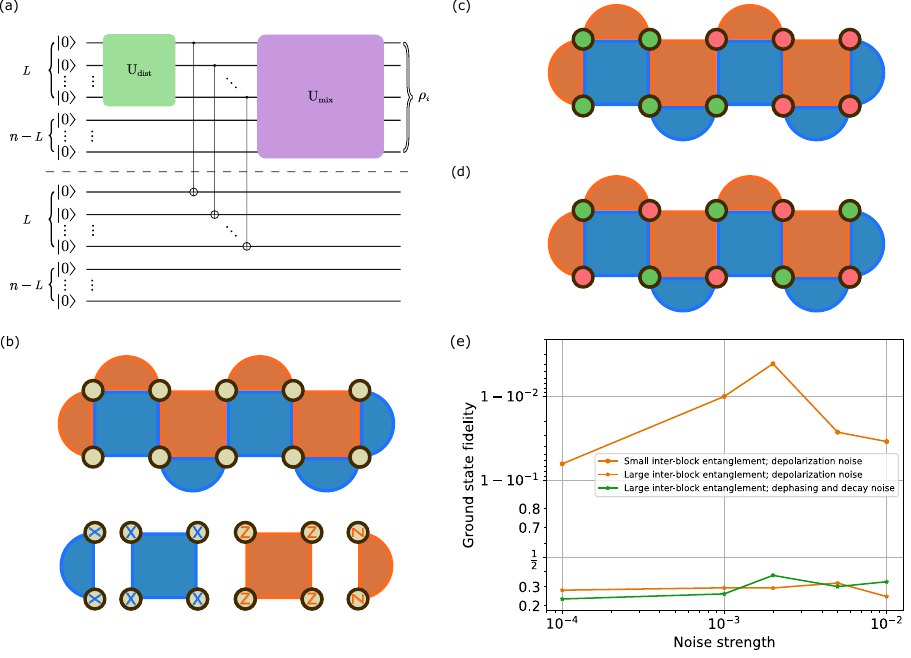}
    \caption{Schmidt ansatz and numerical experiments. (a) Structure of the Schmidt ansatz with $L$ intra-block CNOT gates. $U_{\operatorname{dist}}$ (on $L$ qubits) serves the purpose of adjusting the Schmidt coefficients and can be chosen to be a real orthogonal matrix to reduce the number of single qubit rotation gates as well as the number of parameters. $U_{\operatorname{mix}}$ (on $n$ qubits) is a parametrized unitary circuit which under certain set of parameters is expected to transform all Schmidt components into desired form. (b) 10-qubit rotated surface code with a unique ground state. (c) Partitioning of the qubits (into a green-qubit block and a red-qubit block) with small intra-block entanglement. (d) Partitioning of the qubits with large intra-block entanglement. (e) Obtained ground state fidelity under the Schmidt ansatz with $L=1$ intra-block CNOT gates for both partitions in (c) and (d) marked by dots and stars respectively. The circuit has in total $115$ gates which corresponds to a circuit fault rate of $0.115$ at $10^{-3}$ noise strength. } 
    \label{fig: maintext_Schmidt}
\end{figure*}

In the following, we will introduce several types of constructions of quantum circuit ansatz for preparing density matrices used for VQAs. We will first introduce an ansatz for general purposes which can universally prepare density matrices with minimal resources and then introduce how chemical-inspired ansatz specially designed for electronic structure problems of molecules can be well fitted into our framework. We will also give numerical examples based on these ansatzes to demonstrate the performance of VQAs using our QEM protocol under various noises. Several other strategies for constructing ansatz are discussed at the end of this section.

\subsection{Schmidt ansatz and numerical experiments}
In our proposal, we can use a $2n$-qubit quantum circuit $U$ to prepare an $n$-qubit density matrix $\rho$. When there is no noise, the output state $|\phi\rangle$ from $U$ can be expressed in a Schmidt form: $|\phi\rangle=\sum_i\lambda_i|u_i\rangle\otimes |l_i\rangle$ with $\lambda_i$ the Schmidt coefficients and $|u_i\rangle\otimes |l_i\rangle$ a product state between the upper and the lower system satisfying $\langle u_i|u_j\rangle=\langle l_i|l_j\rangle=\delta_{ij}$. The density matrix $\rho$ is then the reduced density matrix of $|\phi\rangle$ on the upper system with the form: $\rho=\sum_i |\lambda_i|^2 |u_i\rangle\langle u_i|$. From the form of $\rho$, we can find that first, the phase information of Schmidt coefficients is lost, and second, the information of $|l_i\rangle$ is lost. Thus, we can build circuits $U$ whose structure restricts the freedom of redundant information to save resources and reduce the number of useless parameters for VQAs. 

The ansatz to achieve this purpose is called the Schmidt ansatz where we aim to prepare $|\phi\rangle$ with the form: $|\phi\rangle=\sum_i\lambda_i|u_i\rangle\otimes |i\rangle$ with $\lambda_i$ real numbers and $|i\rangle$ the computational basis. The structure of the Schmidt ansatz is shown in Fig. \ref{fig: maintext_Schmidt}a. First, we implement an $L$-qubit orthogonal circuit (which is also unitary) on the first $L$ qubits in the upper system to prepare a superposition of computational basis: $\sum_{i}\lambda_i |i\rangle$ with $\lambda_i$ real numbers. Next, we do $L$ CNOT gates connecting the first $L$ qubits in the upper system and the first $L$ qubits in the lower system to generate the whole $2n$-qubit state: $\sum_{i}\lambda_i |i\rangle|0\rangle^{\otimes(n-L)}\otimes |i\rangle|0\rangle^{\otimes(n-L)}$. The reduced density matrix of this state is of the form $\sum_i |\lambda_i|^2 |i\rangle|0\rangle^{\otimes(n-L)}\langle i|\langle 0|^{\otimes(n-L)}$. This density matrix is then further fed into an $n$-qubit unitary circuit to adjust states attached to each Schmidt coefficient to get the final density matrix $\rho$. Thus, we can understand the Schmidt circuit by using the $L$-qubit orthogonal circuit to adjust $|\lambda_i|^2$ followed by using the $n$-qubit unitary circuit to adjust $|u_i\rangle$ without giving useless freedom to phases of Schmidt coefficients and $|l_i\rangle$. 

When constructing the Schmidt ansatz in real experiments, we suggest a hardware structure of the shape of a ladder (shown in the appendix) such that the CNOT gates connecting the upper and the lower systems can be built on local physical qubits. 
\begin{figure*}[ht]
    \centering
    \includegraphics[width=0.95\textwidth]{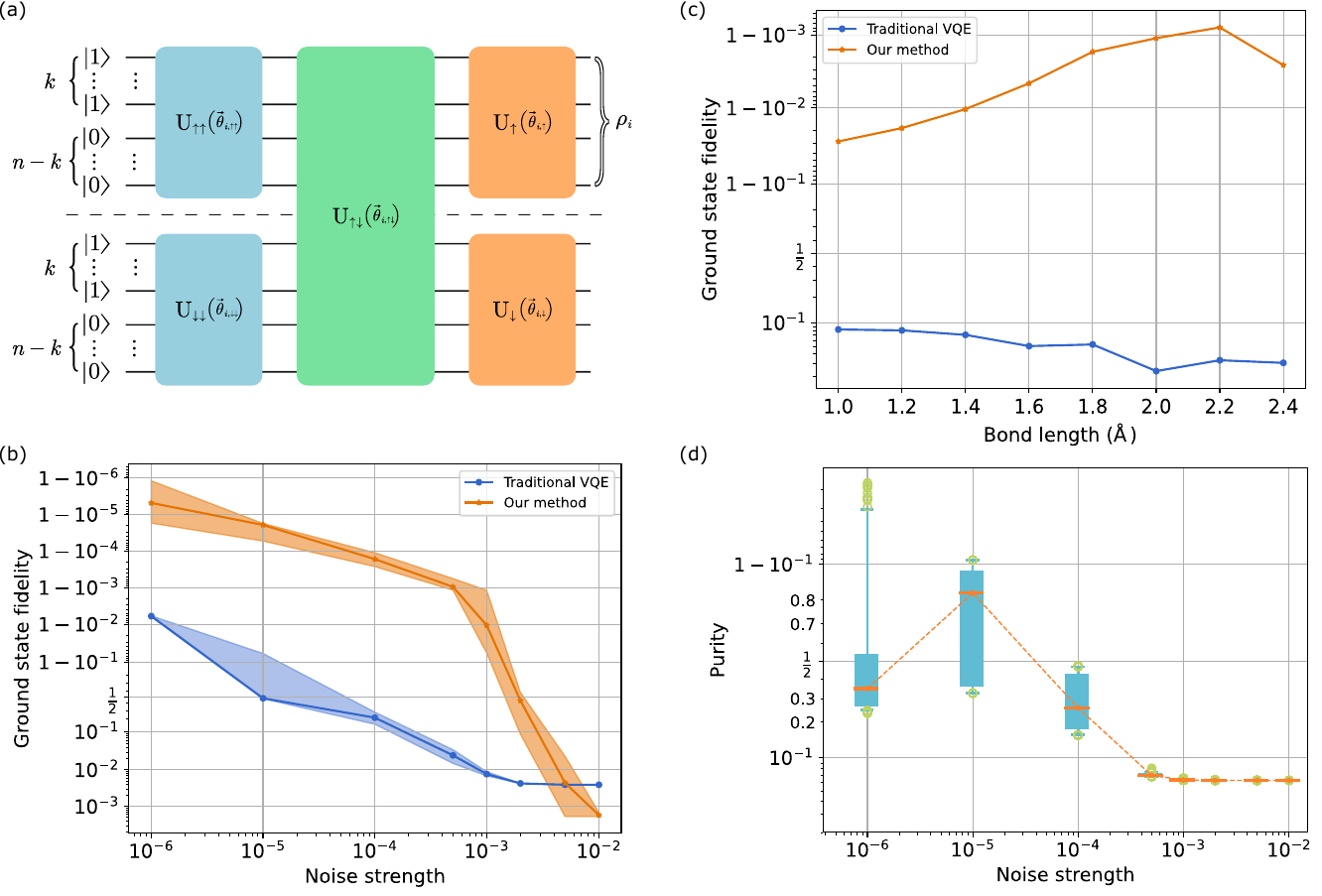}
    \caption{Solving for electronic ground states of the $\text{H}_{4}$ molecule under a spin symmetric ansatz. (a) Spin symmetric ansatz for both traditional VQE and our method. For traditional VQE, the output state is taken over all $2n$ qubits. For our protocol, the output state is the reduced density matrix of the first block of $n$ qubits. The circuit used for our numerical experiment has 5528 gates, corresponding to a circuit fault rate of $2.764$ at $0.5\times 10^{-3}$ noise strength. (b) Obtained ground state fidelity at bond length $2.0\text{\r{A}}$ under circuit level depolarization noise for various noise strengths. The unit of energy is Hartree. (c) Obtained ground state fidelity for various bond lengths under noise strength $p=0.5\times 10^{-3}$. (d) Box plots of the purity values of $\rho_{i}$ for all $i\in\{1,2,3,4\}$ throughout an optimization process under different noise strengths. At a specific noise strength, the corresponding orange line and the orange dot mark the median value of all data points, the blue box marks the range from the first quartile to the third quartile and the blue whisker (error bar) captures the range from the 1th percentile to the 99th percentile and all data points outside of the whisker are marked by green circles. }
    \label{f4}
\end{figure*}

The reason that we set a number $L$ for Schmidt coefficients generations is to restrict the purity of $\rho$ with a lower bound $2^{-L}$ which we will show has a direct influence on the sampling complexity. This also means the size of the Schmidt circuit is $n+L$ smaller than $2n$. When there is no noise, to restrict the sampling complexity into a polynomial scaling with $n$, we require $L$ to be of order $\mathcal{O}(log(n))$. As shown before, this will reduce the reachable entanglements between the row system and the column system. However, since the purity doesn't influence the entanglement inside the row system and the column system, it is possible to make the row and the column systems match a bi-partition under which the ground state of $H_A$ has low entanglement between the two systems and meets the requirement on the purity for the sampling cost. We want to mention here that, due to the low entanglement between the row system and the column system, one may expect to use circuit cutting methods \cite{peng2020simulating,harrow2024optimal} to achieve the same thing. However, those methods cannot be free of decoherence as in our method.

As an example, we use a 10-qubit rotated surface code with a single ground state (Fig.~\ref{fig: maintext_Schmidt}b). Under the partition (into a block of green qubits and another of red qubits) in Fig.~\ref{fig: maintext_Schmidt}(c), the dimension of the Schmidt decomposition of the ground state is $2$. Thus, in this case, we only need one intra-block CNOT gate in Fig.~\ref{fig: maintext_Schmidt}a. On the other hand, the dimension of the Schmidt decomposition of the ground state is $2^4$ under the partition in Fig.~\ref{fig: maintext_Schmidt}d. In our numerical experiment, we use the Schmidt ansatz with $L=1$ for both partitions (Fig.~\ref{fig: maintext_Schmidt}e). We observe that for the partition in Fig.~\ref{fig: maintext_Schmidt}c with small intra-block entanglement, the Schmidt ansatz performs well for the full range of noise strength from $10^{-4}$ to $10^{-2}$. However, for the partition in Fig.~\ref{fig: maintext_Schmidt}d with large intra-block entanglement, the performance of the Schmidt ansatz (with $L=1$) is poor for all noise strengths. Thus, generally speaking, by choosing a partition with small intra-block entanglement, we can use the Schmidt ansatz with a small enough $L$ to potentially achieve simultaneously good performance (in terms of final state fidelity) as well as low enough sample complexity.  

\subsection{Chemical-inspired ansatz and numerical experiments}

When the Schmidt ansatz is randomly chosen, we may inevitably face the barren plateau problems \cite{mcclean2018barren}. Thus, a major goal in VQAs is to develop ansatzes inspired by target problems that have high biases from random quantum circuits (that form a unitary 2-design) \cite{mcclean2018barren}. We observe that for certain problems, our method admits physically motivated ansatzes, which can be expected to achieve this goal. For instance, the electronic ground state of a paramagnetic molecule \cite{helgaker2013molecular}, in addition to particle number conservation, admits the total spin symmetry that is, the ground state is composed of the (fermionic) occupation number states with an equal number of spin-up electrons and spin-down electrons. Consider a paramagnetic molecule with $2k$ electrons and $2n$ spin orbits (the first $n$ orbits for spin-up electrons and the second $n$ orbits for spin-down electrons), we apply the Jordan-Wigner transformation \cite{mcardle2020quantum} to transform all creation and annihilation operators into Pauli string operators such that the $|1\rangle$ state ($|0\rangle$ state) on the $i$-th qubit in the first half block of all qubits indicates the presence (absence) of an electron on the $i$-th spin-up orbital and similarly for the second half block, which is reserved for spin-down orbitals. 

For traditional VQE, we can use a spin-symmetric version of the q-UCCSD ansatz \cite{mcardle2020quantum}:

\begin{equation}
\label{eq: spin symmetric ansatz}
U(\theta):=U_{\uparrow}(\vec{\theta_\uparrow})U_{\downarrow}(\vec{\theta_\downarrow}) U_{\uparrow\downarrow}(\vec{\theta_{\uparrow\downarrow}})U_{\uparrow\uparrow}(\vec{\theta_{\uparrow\uparrow}})U_{\downarrow\downarrow}(\vec{\theta_{\downarrow\downarrow}})
\end{equation}
where each term preserves the number of spin-up electrons and the number of spin-down electrons (see Fig. \ref{f4}a for the structure of the ansatz). Then, for a noiseless circuit, we can guarantee the output state to respect the spin symmetry by setting the input state to be 
\begin{equation}\label{eq: molecule input state}
|\underbrace{1,\cdots,1}_{k},\underbrace{0,\cdots,0}_{n-k},\underbrace{1\cdots,1}_{k},\underbrace{0,\cdots,0}_{n-k}\rangle
\end{equation}
with exactly $k$ electrons in each of the upper and lower systems. Interestingly, the same ansatz and the input state can also guarantee the resulting states in our protocol also satisfy the spin symmetry. We numerically test this ansatz for both traditional VQE and our protocol on the $\text{H}_4$ molecule under circuit-level depolarization noise (Fig. \ref{f4}b). While the total number of qubits in this numerical experiment is $8$, the ansatz circuit is deep and has 5528 gates (and thus 5528 faulty locations) in total. For noise probability ranging from $10^{-5}$ to $10^{-3}$, our protocol manages to achieve close to $1$ ground state fidelity and is about three orders of magnitude better than traditional VQE in terms of obtained ground state fidelity (Fig.~\ref{f4}b). Under a more experimentally relevant depolarization noise $p=0.5\times 10^{-3}$, our protocol preserves its significant advantage over traditional VQE for all bond lengths in Fig.~\ref{f4}c. Furthermore, we notice that the purity of the density matrices $\rho_{i}$ for $i\in\{1,2,3,4\}$ throughout an optimization process generally decreases (Fig~\ref{f4}d) by nearly an order of magnitude as the noise increases from the noiseless regime to the highly noisy regime ($\sim 10^{-2}$ noise strength). This can be understood as the upper system getting entangled with the environment. An interesting thing about the purity, in this case, is that in the noiseless regime, the average is much higher than the lower bound set by the number of CNOT gates connecting the lower and the upper systems, thus, a case-by-case analysis is required for the real performance of our protocol.     

It is worth noticing that the structure of the ansatz Eq. \ref{eq: spin symmetric ansatz} can be further simplified and adjusted for our protocol. The reason we keep it in its original form is to make a fair comparison between traditional VQE and ours. For example, the block $U_{\downarrow}(\theta_\downarrow)$ can be ignored since it won't change the output density matrices. At the same time, after compiling Eq. \ref{eq: spin symmetric ansatz} into digital quantum circuits, the CNOT gates connecting the upper and the lower systems can be flexibly adjusted to control the sampling complexity. What's more, we can use a locking strategy as shown in Ref. \cite{shang2024hermitian} to pair up the parameters controlling the upper and the lower blocks (for example, locking $\vec{\theta_\uparrow}$ and $\vec{\theta_\downarrow}$) and thus reduce nearly half of the number of parameters.

\subsection{Other ansatz strategy}
There are also other strategies for constructing suitable ansatz for our proposal. One is to follow the Hermitian-preserving ansatz introduced in Ref. \cite{shang2024hermitian} which aims to prepare a pure state that is Hermitian when turning it back to a matrix. This is done by pairing up the parameters in the upper and the lower systems which can also help to reduce the freedom of redundant information. However, for this type of ansatz, it is not easy to control the purity of density matrices.

The other is to utilize the idea of mixing \cite{liu2021solving} where the mixedness of density matrices is generated by tuning parameters in quantum circuits based on a classical probabilistic distribution. Note that this can be understood as using a large number of linear combinations to take the function of CNOT gates connecting the upper and the lower systems. In this way, it is possible to use an $n$-qubit circuit to generate density matrices with programmable mixedness. The problem with this approach is perhaps a trade-off between the number of parameters of optimizations and the expressibility.

\section{Properties}
\subsection{Sampling complexity}
The sampling complexity of estimating Eq. \ref{3} mainly depends on the number $K$ of classical combinations and the purity of $\{\rho_1,\rho_2,...,\rho_K\}$. When these density matrices are prepared by the ansatz introduced above, the purities are decided by the number of CNOT gates $L$ connecting the upper and the lower systems (in the Schmidt ansatz and the chemical ansatz) and the circuit fault rate $\zeta$. The number of CNOT gates will set a lower bound $2^{-L}$ for the purity in the noiseless case. The circuit fault rate $\zeta$ is defined as the sum of individual fault probabilities at all possible locations in the ansatz circuit $\zeta=\sum_i p_f$. When $\zeta\lesssim 1$ such that a Poisson distribution can be used to model the probability distribution of the number of faults that occur in the circuit, the fault-free probability is around $e^{-\zeta}$ which will set another lower bound $e^{-2\zeta}$ for the purity. 

The true lower bound is thus a competition between $2^{-L}$ and $e^{-\zeta}$, which results in the following sampling complexity:
\begin{equation}\label{8}
N\geq \max\left[e^{4\zeta},2^{2L}\right]\frac{ 3K^2}{\epsilon^2}\left( \frac{m ||H_A||_F^2}{2^{2n}}+||H_A||_2^2\right)
\end{equation}
with $\frac{||H_A||_F^2}{2^{2n}}=\sum_{\alpha=1}^m g_\alpha^2$. To get Eq. \ref{8}, we require the final $n$-qubit unitaries in the upper system in the Schmidt ansatz and the unitaries in the upper system between two CNOT gates connecting the upper and the lower systems in the chemical ansatz form at least a unitary 1-design \cite{mele2023introduction} such that the values of $\text{Tr}(\rho_i\rho_j)$ for $i\neq j$ will concentrate on $2^{-n}$ which can be ignored when $n$ in large. This also means for large $n$ cases, we have no need to measure values of $\text{Tr}(\rho_i\rho_j)$ for estimating $\langle\psi|H_A|\psi\rangle$ to save the sampling cost. Since random Pauli circuits can already form a 1-design \cite{webb2015clifford}, this requirement can be easily satisfied. Eq. \ref{8} tells the number of samples needed to estimate $\langle\psi|H_A|\psi\rangle$ within an accuracy $\epsilon$. The polynomial scaling to $K$ indicates that a large number of classical linear combinations is allowed. Note that since the purity has a lower bound $2^{-n}$, the sampling complexity is also upper bounded by:
\begin{equation}
N\leq 2^{2n}\frac{ 3K^2}{\epsilon^2}\left( \frac{m ||H_A||_F^2}{2^{2n}}+||H_A||_2^2\right)
\end{equation}

As mentioned many times, the basic idea of our QEM protocol is fundamentally different from others. However, in terms of the sampling complexity, the scaling with respect to the circuit fault rate $\zeta$ of our protocol is $e^{4\zeta}$ which interestingly coincides with the general exponential scaling behaviors of the sampling overhead in other QME protocols \cite{cai2023quantum}. Indeed, our protocol solves the decoherence by directly changing the way of encoding quantum states, however, to retrieve information under this encoding, an $e^{4\zeta}$ scaling has to be respected because of the purity restriction. Compared with other protocols, our sampling complexity shares the same scaling of $\zeta$ as using the probabilistic error cancellation method \cite{temme2017error,endo2018practical} to totally eliminate errors and as using the two-copy version of the virtual state distillation method \cite{huggins2021virtual}. When the task is to obtain information on pure states, our method is superior to the virtual state distillation since it needs a $M$-copy version with $M$ large to capture the pure state behaviors well which, however, will result in a $e^{2M\zeta}$ scaling on the sampling complexity.

\subsection{Expressibility}
As discussed earlier, when $K=1$, states $|\rho\rangle$ from $n$-qubit density matrices are not able to represent arbitrary $2n$-qubit pure states due to the Hermiticity and the positive semi-definiteness of density matrices. Thus, to enhance the expressive power, we introduce the classical linear combinations and we have shown that it is sufficient to set $K = 4$ with 2 real and 2 imaginary coefficients to express any pure $2n$-qubit states $|\psi\rangle$. As discussed in Sec. \ref{ent}, the purity of density matrices and classical linear combinations can both influence the entanglement between the row and the column systems and thus influence the expressibility. The classical linear combinations can be very useful in saving resources to ``prepare" some non-trivial states. For example, the preparation of the generalized $2n$-qubit GHZ state $\frac{1}{\sqrt{2}}(|0\rangle^{\otimes 2n}+|1\rangle^{\otimes 2n})$. While its entanglement entropy under any bi-partition is very low, its preparation requires a $\mathcal{O}(n)$-depth local unitary circuit \cite{balasubramanian2019binding}, which however, can be seen as a classical linear combination of the two easiest product states $|0\rangle^{\otimes 2n}$ and $|1\rangle^{\otimes 2n}$.

To quantitatively evaluate the expressibility of ansatz we introduced above under DMV encoding, we adopt the tools of the covering number from the statistical learning theory \cite{kolmogorov1959varepsilon} which have been used for measuring the expressibility of ansatz in standard VQAs \cite{du2022efficient}. When we do not use DMV, the upper bound for the covering number of standard VQAs is given by:
\begin{equation}
    \mathcal{N}(\mathcal{H}, \epsilon, |\cdot|) \leq \left(\frac{7 N_{gt}\|H_A\|}{\epsilon}\right)^{d^{2k}N_{gt}}
\end{equation}
where $\mathcal{H}$ is the hypothesis space for standard VQAs $\mathcal{H}=\left\{\operatorname{Tr}\left(\hat{U}(\boldsymbol{\theta})^{\dagger} H\hat{U}(\boldsymbol{\theta}) \rho\right) \mid \boldsymbol{\theta} \in \Theta\right\}$, $N_{gt}$ represents the number of trainable gates in $U(\theta)$, $d$ is the dimension of the system which is $2^{2n}$ for our cases, $\epsilon$ is a very small positive hyperparameter, and each gate $\hat{u}_i(\boldsymbol{\theta})$ acts on at most $k$ qudits.

For VQAs based on DMV as formulated before, the upper bound for the covering number is:
\begin{equation}
    \mathcal{N}(\tilde{\mathcal{H}}, \epsilon, |\cdot|) \leq 2^L \mathcal{C}\left(\frac{7 N_{GT}\|H_B\|}{\epsilon}\right)^{d^{2k}N_{GT}}
\end{equation}
where $\tilde{\mathcal{H}}$ is the hypothesis space for VQAs by DMV, $L$ is the number of CNOT gates connecting the upper and lower halves, $\mathcal{C}$ is a constant greater than 1 defined in the appendix, $N_{GT}$ is the sum of trainable and also meaningful gates in each linear combination circuit, i.e., $N_{GT}=\sum_{i,j=1}^K N_{gt(ij)}$, and the other parameters are the same as in the standard case (See Appendix).

Taking the same observable operator, since we have $\|H_B\|=\|H_A\|$. For analytical convenience, we assume that the number of trainable gates in each combination circuit is the same as in a single standard VQA circuit, i.e., $N_{GT}=KN_{gt}$. Since this term appears in the exponential part, the upper bound for the covering number increases exponentially with the number of combinations. Additionally, the constant term $2^L \mathcal{C}$ is greater than 1, further amplifying the upper bound for the covering number. It can be seen that VQAs based on DMV have higher expressibility compared to standard VQAs.

\subsection{Trainability}
Here, we give a discussion on the barren plateau issues in our proposal. The parameters used in our proposal for training VQAs are composed of $\{\vec{\theta_1},\vec{\theta_2},...,\vec{\theta_K},c_1,c_2,...,c_K\}$. The trainability for classical parameters $\{c_1,c_2,...,c_K\}$ and for quantum parameters $\{\vec{\theta_1},\vec{\theta_2},...,\vec{\theta_K}\}$ needs to be discussed separately.

For classical parameters $\{c_1,c_2,...,c_K\}$, since they appear as the classical coefficients for combining expectation values as shown in Eq. \ref{3}, they will not directly deal with the exponentially large Hilbert space
and thus have no barren plateau issues as discussed in Ref. \cite{bharti2021iterative,bharti2021quantum}. Thus, the classical linear combinations can not only enhance the expressibility but also avoid worsening the barren plateau problems, which in some sense can overcome the fundamental trade-off between the expressibility and the trainability \cite{holmes2022connecting} in variational quantum algorithms.

For quantum parameters $\{\vec{\theta_1},\vec{\theta_2},...,\vec{\theta_K}\}$, we know that in standard VQAs, when the parametrized quantum circuits are randomly enough to form a unitary 2-design, the derivatives of cost functions with respect to parameters will concentrate on exponentially small averages \cite{mcclean2018barren}. In our case, while the form of the cost function Eq. \ref{3} looks very different from those in standard VQAs, we should expect the existence of barren plateaus in highly random Schmidt ansatz. Since, we may require $L\leq n$, the barren plateau problems should be understood from the randomness of the $n$-qubit circuit in the upper systems but not the whole $2n$-qubit systems. Indeed, if an $n$-qubit parameterized circuit forms a 2-design, the average variance $\eta^2$ of derivatives will be of order $\mathcal{O}(2^{-n})$ in standard VQAs, which will be inherited in our algorithmic framework as:
\begin{equation}\label{9}
|\frac{\partial \langle\psi|H_A|\psi\rangle}{\partial \theta}|\leq 2^{L+2}K^2\sum_\alpha |g_\alpha|\eta
\end{equation}
when for example, the $n$-qubit unitary circuit in the Schmidt ansatz forms a 2-design. Eq. \ref{9} shows that when $L\ll n$, barren plateaus occur. On the other hand, when $L$ is comparable with $n$, Eq. \ref{9} should be ignored as it is a rather loose bound. Instead, we can then analyze barren plateau problems by checking whether the whole $2n$-qubit circuit forms a 2-design. 

The existence of barren plateau problems in randomized ansatz is the main motivation to let us give a construction of a chemical-inspired ansatz in the previous section. Since problem-inspired ansatz can have large derivations from unitary 2-design and thus, can potentially avoid barren plateaus with enough expressibility for the target problems at the same time.

\subsection{Noise influence under DMV}
The numerical results shown in the last section give us evidence of the effectiveness of our proposal against various types of noise. However, as we can find in Fig. \ref{f4}c-b, as the noise strength increases, our proposal inevitably loses its abilities against noise. This can be understood as the concentration effect of the output states of noisy quantum circuits \cite{stilck2021limitations,de2023limitations,wang2021noise}. For unital depolarizing noise models, the output states will concentrate on the maximally mixed state \cite{takagi2022fundamental,takagi2023universal,tsubouchi2023universal,yan2023limitations}. For non-unital models, under the assumption that the noiseless quantum circuit forms a unitary 2-design, the output states will also concentrate on some fixed states of noisy quantum circuits \cite{quek2022exponentially}. 

Under the DMV encoding, while all the output states are pure states without decoherence, the concentration effect will nevertheless limit the expressibility and trainability of our proposal for VQAs. For example, under the depolarizing noise and DMV, the output states will concentrate on the maximally entangled state $\frac{1}{\sqrt{2^n}}\sum_i|i\rangle|i\rangle$ which thus reduces the expressibility. This also explains the reason that our proposal shows weaker performance compared with traditional VQE for very large noise as shown in Fig. \ref{f4}c-b, since, unlike the maximally entangled state, the maximally mixed state can always have non-zero overlaps with arbitrary states. The concentration effect also means that the value of the cost functions will concentrate on the value corresponding to the maximally entangled state such that the derivatives with respect to the parameters $\{\vec{\theta_1},\vec{\theta_2},...,\vec{\theta_K},c_1,c_2,...,c_K\}$ will be exponentially small in general and the noise-induced barren plateaus occur \cite{wang2021noise}. Another thing we want to mention is that while by DMV, we can remove the noise-induced decoherence in the state preparation phase, the noise in the measurement phase may inevitably influence the accuracy for estimating the cost functions, thus, we recommend to combine with the measurement error mitigation methods \cite{bravyi2021mitigating, cai2023quantum}.

We want to emphasize here that since under our encoding, there is no decoherence, the expressibility and trainability we are discussing are restricted to pure states, thus the inaccuracy induced by the gap between pure states and mixed states can be completely removed.

\subsection{Combinations with other QEM protocols}
Since the way of mitigating noise in our proposal is fundamentally different from other QEM protocols, instead of making comparisons between ours and others, it is more attractive to see whether the combinations between them can give further improvements. The answer is strongly positive. Note that all good performance of our proposal in numerical results we have shown before comes directly from changing the way of encoding information (Eq. \ref{2}) which has not tried to fight against the concentration effect. In contrast, in other protocols \cite{cai2023quantum}, the basic idea is to extract desired information from concentrated output states. Thus, these are two different philosophies to mitigate noise and the combinations of them should be expected to have a better improvement compared with combinations within existing protocols that share the same philosophy. Also, as we talked about before, measurement error mitigation methods can also be introduced to help fight against measurement errors that DMV can not mitigate.

The most interesting point is that the combinations of our protocol with others are free lunches and will not introduce additional sampling complexities. Recalling Eq. \ref{8}, when there is no noise in the quantum circuit, the sampling complexity is only determined by the number of CNOT gates connecting the upper and the lower systems. Suppose an output density matrix in the noiseless and noisy cases respectively, if we directly use our QEM estimation, while the value of the cost function still corresponds to some pure state, it will have the tendency to concentrate on the value corresponding to the maximally entangled state and the noise in the circuit will inevitably increase the sampling complexity as shown in Eq. \ref{8}. Instead, we can first use some other QEM protocols to help extract the noiseless expectations shown in the numerator and denominator of the expression of the cost function in Eq. \ref{3}. This has two benefits. First, this will help fight against the concentration effect and thus enlarge the expressibility and the trainability. Second, while in this phase, we require an exponential number of samples required in other QEM protocols \cite{cai2023quantum}, in the next phase of using Eq. \ref{3} for the cost function estimation, the expectation values in Eq. \ref{3} are already from noiseless density matrices with bounded purity $2^{-L}$ and thus the exponential dependence on the circuit fault rate $\eta$ of the sampling complexity of DMV encoding in Eq. \ref{8} disappears. Thus, the combinations between our protocols with others will not only improve the overall performance but also avoid the amplification of the sampling complexity.

\section{Summary and outlook}
In summary, we propose a new QEM protocol with a philosophy different from existing ones. By using the idea of DMV, we can eliminate the decoherence of quantum states in NISQ devices from the fundamental encoding level. We give a universal and NISQ-friendly measurement technique for measuring the expectation values of operators with respect to the linear combinations of vectorized density matrices. Based on this technique, we can now run noiseless quantum experiments on noisy quantum circuits. In the example of using our protocol for VQAs, we show different ways of constructions of ansatz to extract density matrices and give various numerical results to show the effectiveness of our protocol. We give a detailed analysis of various properties of our proposal ranging from the sampling complexity, expressibility, and trainability to the noise influence in terms of DMV and its combinations with other QEM protocols. 

There are several future directions of our work. One is to find more efficient measurement strategies such as the classical shadows and those with the Heisenberg scaling to reduce the sampling complexity. Another direction is to find applications of our QEM protocols beyond the VQAs, which can also be generalized to find more possibilities of DMV beyond this work and Ref. \cite{shang2024polynomial}. We believe this work will open up new possibilities for novel quantum applications.

\textbf{Code:} We use the Qiskit package \cite{cross2018ibm} for parts of our simulations. An example code can be found online \cite{dmvqegithub}.
\begin{acknowledgements}
The authors would like to thank Chao-Yang Lu, Ming-Cheng Chen, Cheng-Cheng Yu, Yu-Xuan Yan, Zhen-Huan Liu, Zhen-Yu Cai, and Zhen-Yu Du for fruitful discussions and suggestions.
\end{acknowledgements}
\bibliographystyle{unsrt}
\bibliography{ref.bib}
\onecolumngrid
\clearpage
\appendix
\section{Preliminary: unitary design}
The Haar measure is a uniform probability measure defined on the unitary group $U(d)$ satisfying:
\begin{equation} 
\int_{Haar}f(U V)dU=\int_{Haar}f(VU)dU=\int_{Haar}f(U)dU
\end{equation}
Given the Haar measure, the $t$-th moment operator is defined as:
\begin{equation} 
\mathcal{M}_k(O)=\int_{Haar}U^{\otimes t}O U^{\dag\otimes t}dU
\end{equation}
For arbitrary $O$, the moment operator $\mathcal{M}_k(O)$ lives in the space spanned by the permutation operators $\{V_d(\pi)|\pi\in S_t\}$ with $S_t$ the $t$-th order symmetric group and thus can be expressed as:
\begin{equation} 
\mathcal{M}_k(O)=\sum_{\pi\in S_t} c_{\pi}V_d(\pi)
\end{equation}
where coefficients $c_{\pi}$ can be obtained by solving the linear equations:
\begin{equation} 
Tr(V_d(\kappa)^\dag O)=\sum_{\pi\in S_t} c_{\pi}Tr(V_d(\kappa)^\dag V_d(\pi))\quad\text{for all }\kappa \in S_t
\end{equation}
For $t=1$, the only permutation operator is $I$, and we have:
\begin{equation}\label{design1}
\mathcal{M}_1(O)=\int_{Haar}UO U^{\dag}dU=\frac{Tr(O)}{d}I
\end{equation}
For $t=2$, there are two permutation operators: $I$ and $S$ (the swap operator), and we have:
\begin{equation}\label{design2}
\mathcal{M}_2(O)=\int_{Haar}U\otimes U O U^{\dag}\otimes U^\dag dU=\frac{Tr(O)-d^{-1}Tr(SO)}{d^2-1}I+\frac{Tr(SO)-d^{-1}Tr(O)}{d^2-1}S
\end{equation}

Generating Haar random unitaries on a quantum computer generally needs an exponentially deep random quantum circuit which is unrealistic. However, if only behaviors of $t$-th order moments are needed, we can instead generate unitaries from a distribution $\nu$ that forms a unitary $t$-designs defined as:
\begin{equation} 
\int_{\nu}U^{\otimes t}O U^{\dag\otimes t}dU=\int_{Haar}U^{\otimes t}O U^{\dag\otimes t}dU
\end{equation}
for arbitrary $O$. Specifically, if $\nu$ is a 1-design, Eq. \ref{design1} is satisfied, and if $\nu$ is a 2-design, Eq. \ref{design1} and Eq. \ref{design2} are satisfied.

Assuming $U$ and $V$ are drawn from a 1-design distribution $\nu$, then, according to Eq. \ref{design1}, we have:
\begin{eqnarray}\label{gold}
\int_\nu\int_\nu Tr(U\rho_0 U^\dag V\sigma_0 V^\dag) dU dV&&=\frac{Tr(\rho_0) Tr(\sigma_0)}{d}=\frac{1}{d}
\label{ccs777}
\end{eqnarray}
Since $Tr(U\rho_0 U^\dag V\sigma_0 V^\dag)\geq 0$, we can use the Markov’s inequality: $Prob[X\geq \varepsilon]\leq \frac{E[X]}{\varepsilon}$ to show the concentration of around the average:
\begin{equation} \label{gold2}
Prob[Tr(U\rho_0 U^\dag V\sigma_0 V^\dag)\geq \varepsilon]\leq \frac{1}{d\varepsilon}
\end{equation}
When $d$ is large, Eq. \ref{gold}-\ref{gold2} indicate $Tr(U\rho_0 U^\dag V\sigma_0 V^\dag)$ will concentrate on $\frac{1}{d}$ with high probability. This is a property that will frequently be used in the following sections. 

\section{Substitute operators}
\subsection{Proof of the key equations by tensor networks}
\begin{figure}[ht]
\centering
\includegraphics[width=0.8\textwidth]{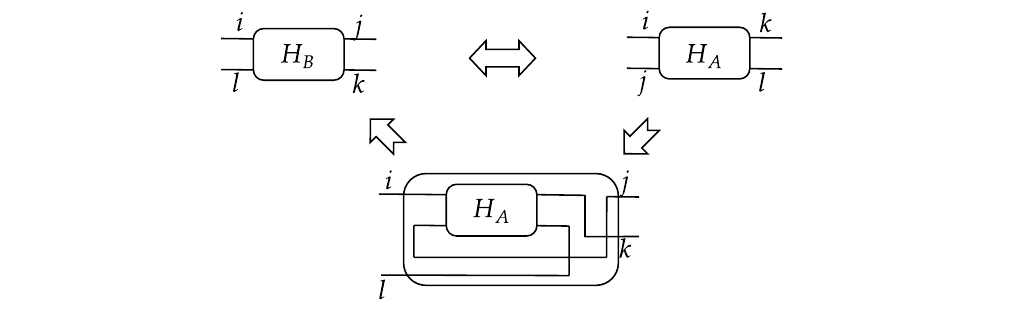}
\caption{The graphical representation of $\left\langle i l\left|H_B\right| j k\right\rangle=\left\langle i j\left|H_A\right| k l\right\rangle$ }
\label{T1}
\end{figure}

In our paper, the most crucial equation is:
\begin{equation}
\left\langle\psi\left|H_A\right| \psi\right\rangle=\frac{\sum_{i, j=1}^K c_i^* c_j \operatorname{Tr}\left(H_B \rho_i \otimes \rho_j\right)}{\sum_{k, l=1}^K c_k^* c_l \operatorname{Tr}\left(\rho_k \rho_l\right)}
\end{equation}
with
\begin{equation}
\left\langle i l\left|H_B\right| j k\right\rangle=\left\langle i j\left|H_A\right| k l\right\rangle\label{K1}
\end{equation}
In the following section, I provide a graphical proof of this equation using tensor networks. Here, let's consider the simplified case of a linear combination, i.e.,
\begin{equation}
\left\langle\rho\left|H_A\right| \rho\right\rangle=\frac{\operatorname{Tr}\left(H_B \rho \otimes \rho\right)}{\operatorname{Tr}\left(\rho^2\right)}
\end{equation}
Due to Equation \ref{K1}, we can represent the graphical equivalent of Figure \ref{T1}.

\begin{figure}[ht]
\centering
\includegraphics[width=0.8\textwidth]{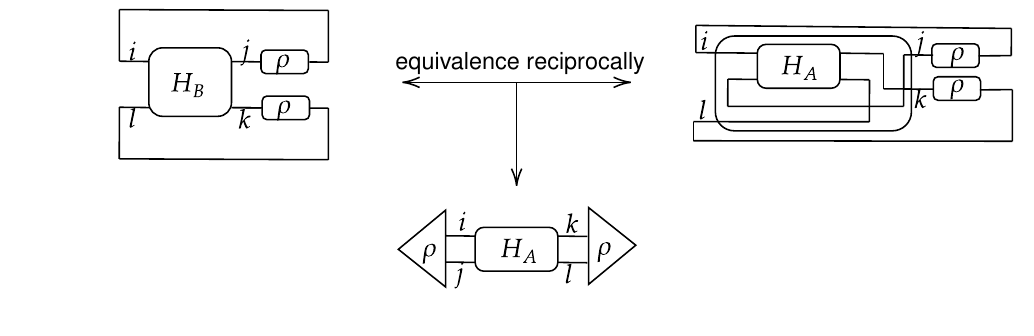}
\caption{The graphical representation of $\left\langle\rho\left|H_A\right| \rho\right\rangle=\frac{\operatorname{Tr}\left(H_B \rho \otimes \rho\right)}{\operatorname{Tr}\left(\rho^2\right)}$ }
\label{t2}
\end{figure}

Using this relationship, we can easily prove $\left\langle\rho\left|H_A\right| \rho\right\rangle=\frac{\operatorname{Tr}\left(H_B \rho \otimes \rho\right)}{\operatorname{Tr}\left(\rho^2\right)}$, as shown in Figure \ref{t2}.

\subsection{Proof of whether $H_B$ is a unitary operator by tensor networks}
\begin{figure}[ht]
\centering
\includegraphics[width=0.5\textwidth]{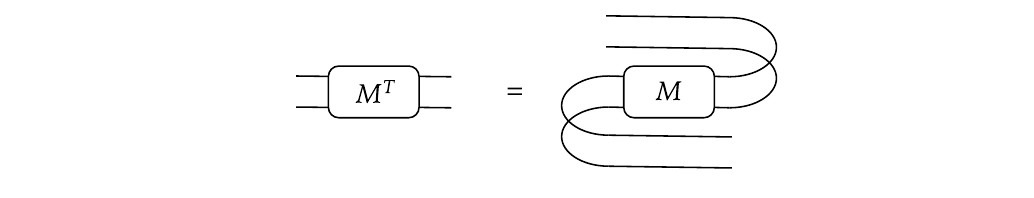}
\caption{The graphical representation of $M^T$}
\label{t3}
\end{figure}

Next, we will use tensor networks to prove that $H_B$ is unitary. Here, let's first look at this relationship: assuming $M$ is a matrix, its conjugate transpose can be graphically represented as shown in Figure \ref{t3}.

\begin{figure}[ht]
\centering
\includegraphics[width=0.65\textwidth]{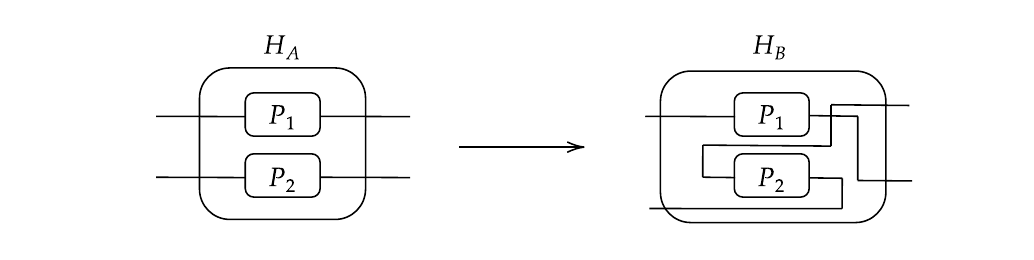}
\caption{Graphical representation of the replacement operators for two-qubit Pauli operators.}
\label{t4}
\end{figure}

For the case of the replacement operators for the two-qubit Pauli operators as given in the main text, based on the proof in Figure \ref{t2}, we can represent the replacement operators graphically as shown in Figure \ref{t4}.

\begin{figure}[ht]
\centering
\includegraphics[width=0.65\textwidth]{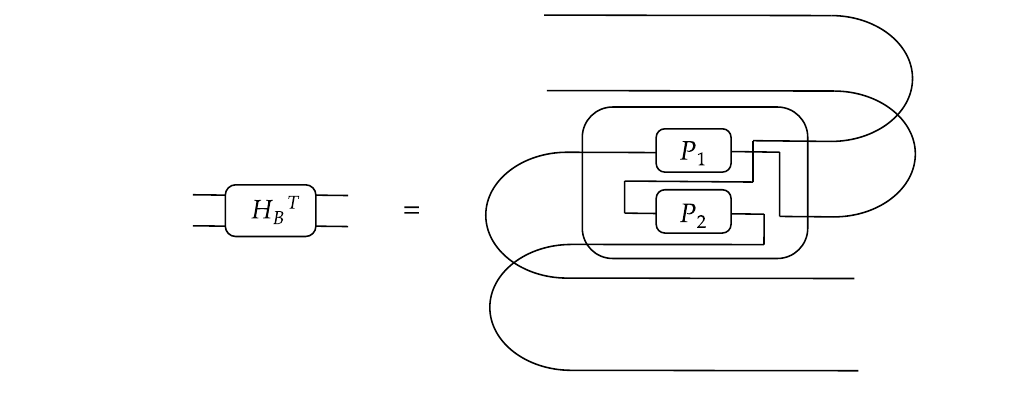}
\caption{The graphical representation of $H_B^{\dagger}$}
\label{t5}
\end{figure}

Since the criterion for a unitary operator is whether $H_B$ satisfies $H_BH_B^{\dagger}=I$, according to Figure \ref{t3}, we can provide the graphical representation of $H_B^{\dagger}$ as shown in Figure \ref{t5}.

\begin{figure}[ht]
\centering
\includegraphics[width=0.65\textwidth]{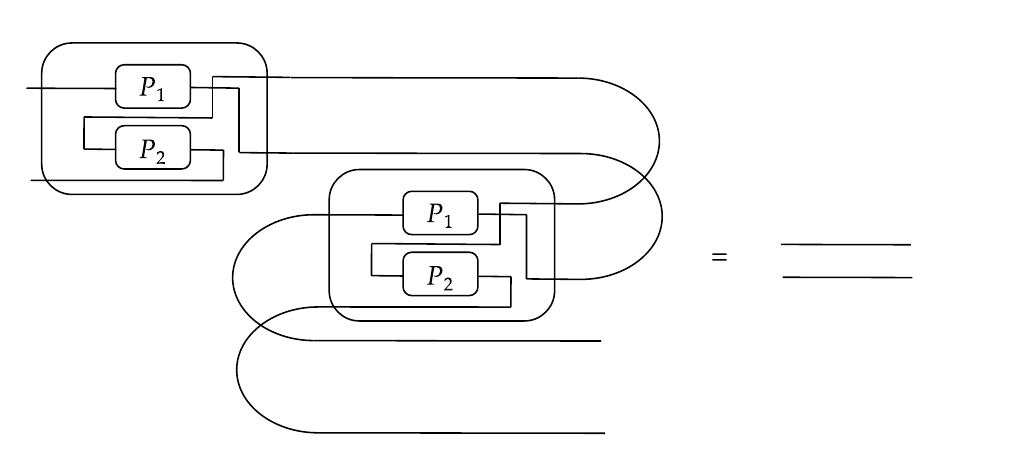}
\caption{The graphical representation of $H_BH_B^{\dagger}=I$}
\label{t6}
\end{figure}

Based on Figure \ref{t4} and Figure \ref{t5}, we can provide a graphical proof of whether $H_B$ is unitary, as shown in Figure \ref{t6}.

\section{Entanglement Rényi entropy of pure states from DMV}
Given a $2n$-qubit pure state writing in the Schmidt form: 
\begin{equation} 
|\varphi\rangle=\sum_i\lambda_i|a_i\rangle\otimes|b_i\rangle
\end{equation}
The entanglement Rényi entropy of order $\alpha$ of $|\varphi\rangle$ is then defined as:
\begin{equation} 
H_\alpha(|\varphi\rangle)=\frac{1}{1-\alpha}\log_2 (\sum_i p_i^\alpha)
\end{equation}
with $p_i=|\lambda_i|^2$. It can be proved that $H_\alpha(|\varphi\rangle)$ is non-increasing in $\alpha$ for any $|\varphi\rangle$:
\begin{equation} \label{renyy}
H_0(|\varphi\rangle)\geq H_1(|\varphi\rangle)\geq H_2(|\varphi\rangle)\geq...\geq H_\infty(|\varphi\rangle)
\end{equation}
\begin{figure}[htbp]
\begin{minipage}[t]{0.32\textwidth}
\centering
\includegraphics[width=\textwidth]{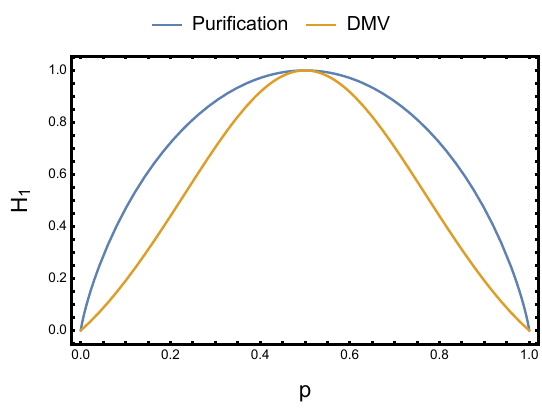}
\end{minipage}
\begin{minipage}[t]{0.32\textwidth}
\centering
\includegraphics[width=\textwidth]{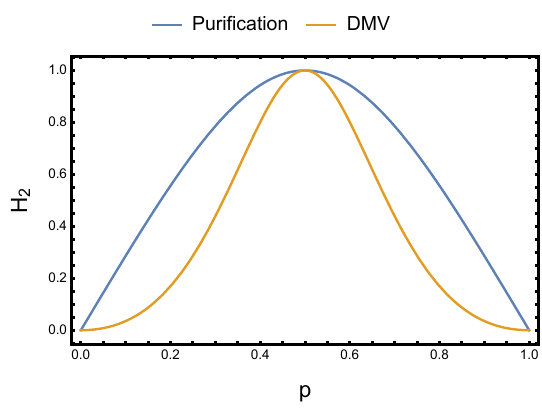}
\end{minipage}
\begin{minipage}[t]{0.32\textwidth}
\centering
\includegraphics[width=\textwidth]{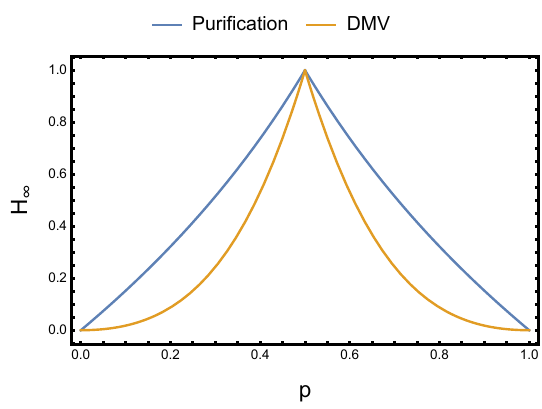}
\end{minipage}
\caption{Left: Rényi entropy of order 1 for 2-dimensional density matrices. Middle: Rényi entropy of order 2 for 2-dimensional density matrices. Right: Rényi entropy of order $\infty$ for 2-dimensional density matrices.\label{reny2}}
\end{figure}

Given an $n$-qubit density matrix $\rho=\sum_i p_i |s_i\rangle\langle s_i|$, its corresponding $2n$-qubit pure state $|\rho\rangle$ by DMV is defined as:
\begin{equation} 
|\rho\rangle=\frac{1}{\sqrt{\sum_i p_i^2}}\sum_i p_i |s_i\rangle\otimes |s_i\rangle
\end{equation}
and its corresponding $2n$-qubit pure state $|\psi_\rho\rangle$ by purification is defined as:
\begin{equation} 
|\psi_\rho\rangle=\sum_i \sqrt{p_i} |s_i\rangle\otimes |s_i\rangle
\end{equation}
Based on the relation Eq. \ref{renyy}, we can show that the $\alpha$'s order of Rényi entropy of $|\rho\rangle$ is always smaller than $|\psi_\rho\rangle$:
\begin{eqnarray}
H_\alpha(|\rho\rangle)-H_\alpha\left(|\psi_\rho\rangle\right)&&=\frac{1}{1-\alpha}\log_2 \left(\sum_i \left(\frac{p_i^2}{\sum_i p_i^2}\right)^\alpha\right)-H_\alpha\left(|\psi_\rho\rangle\right)\nonumber\\&&=\frac{1}{1-\alpha}\log_2\left(\sum_i p_i^{2\alpha}\right)-\frac{\alpha}{1-\alpha}\log_2\left(\sum_i p_i^2\right)-H_\alpha\left(|\psi_\rho\rangle\right)\nonumber\\&&=\frac{1-2\alpha}{1-\alpha}H_{2\alpha}(|\psi_\rho\rangle)+\frac{\alpha}{1-\alpha}H_2\left(|\psi_\rho\rangle\right)-H_\alpha\left(|\psi_\rho\rangle\right)\nonumber\\&&\leq \frac{1-2\alpha}{1-\alpha}H_{2\alpha}(|\psi_\rho\rangle)+\frac{\alpha}{1-\alpha}H_\alpha\left(|\psi_\rho\rangle\right)-H_\alpha\left(|\psi_\rho\rangle\right)\nonumber\\&&=\frac{1-2\alpha}{1-\alpha}\left(H_{2\alpha}(|\psi_\rho\rangle)-H_{\alpha}(|\psi_\rho\rangle)\right)\leq0
\end{eqnarray}
In Fig. \ref{reny2}, we show the comparison of Rényi entropy of order 1, 2, and $\infty$ between pure states from 2-dimensional density matrices by DMV and by purification. 

The entanglement entropy can be enlarged by classical linear combinations. Consider the mapping:
\begin{eqnarray}
\{\rho_1,\rho_2,...,\rho_K,c_1,c_2,...,c_K\}\rightarrow
|\psi\rangle=\frac{1}{C_\psi}\sum_{ij}(c_1\rho_{1,ij}+c_2\rho_{2,ij}+...+c_K\rho_{k,ij})|i\rangle|j\rangle
\end{eqnarray}
as in the main text. The reduced density matrices on the row system of $|\psi\rangle$ is:
\begin{eqnarray}
Tr_c(|\psi\rangle\langle \psi|)&&=\frac{\sum_{\alpha \beta ijk}c_\alpha c_\beta^*\rho_{\alpha,ij}\rho_{\beta,jk}|i\rangle\langle k|}{C_\psi^2}\nonumber\\&&=\frac{\sum_{\alpha \beta}c_\alpha c_\beta^*\rho_\alpha \rho_\beta}{C_\psi^2}
\end{eqnarray}
Here, we only consider the second order Rényi entropy of $|\psi\rangle$ which is:
\begin{eqnarray}
H_2(|\psi\rangle)&&=-\log\left(\frac{\sum_{abcd}c_a c_b^* c_c c_d^*Tr(\rho_a\rho_b\rho_c\rho_d)}{C_\psi^4}\right)\nonumber\\&&\approx -\log\left(\frac{\sum_{a}|c_a|^4 Tr(\rho_a^4)}{(\sum_a |c_a|^2 Tr(\rho_a^2))^2}\right)
\end{eqnarray}
where the approximation is due to Eq. \ref{gold}-\ref{gold2}. For $n$-qubit density matrices $\rho_i$ and $\rho_j$, when they are generated randomly and uncorrelatedly by circuits such as the Schmidt ansatz with the $n$-qubit unitary in the upper state forming a 1-design, we have the value of $Tr(\rho_i\rho_j)$ concentrating on $\frac{1}{2^{n}}$ which can be ignored when $n$ is large. For values like $Tr(\rho_a\rho_b\rho_c\rho_d)$, according to Eq. \ref{design1}, they will concentrate on $\frac{1}{2^{3n}}$ which can of course also be ignored. To make a comparison with the $K=1$ case, we can simply assume $Tr(\rho_i^4)=Tr(\rho^4)$
and $Tr(\rho_i^2)=Tr(\rho^2)$ for all $i$, then we have:
\begin{eqnarray}
H_2(|\psi\rangle)&&\approx-\log\left(\frac{\sum_{a}|c_a|^4 Tr(\rho^4)}{(\sum_a |c_a|^2 Tr(\rho^2))^2}\right)\nonumber\\&&=-\log\left(\frac{Tr(\rho^4)}{Tr(\rho^2)^2}\right)-\log\left(\frac{\sum_{a}|c_a|^4}{(\sum_a |c_a|^2)^2}\right)\nonumber\\&&\leq H_2(|\rho\rangle)-\log\left(\frac{\sum_{a}|c_a|^4}{(\sum_a |c_a|^4)K}\right)\nonumber\\&&=H_2(|\rho\rangle)+\log(K)
\end{eqnarray}
Thus, the classical linear combination can give an enhancement no more than $\log(K)$ for entanglement entropy $H_2$.

\section{Ansatz}
\subsection{Schmidt ansatz}
Based on the fact that only the reduced density matrices $\rho_i$ supported on the first n-qubit block are used in our proposal, we developed a new ansatz, Schmidt ansatz, in the main text to achieve a better control on the purity of $\rho_{i}$ as well as to reduce the number of parameters. Here, we demonstrate that as long as the dimension $\chi$ of the Schmidt decomposition of the desired ground state $|\Psi_{\operatorname{gs}}\rangle$ is upper bounded by $2^{L}$ (assuming that the system is partitioned into two blocks that contain the first n qubits and the second n qubits respectively), then it suffices to use $L$ CNOT gates between the two blocks in the Schmidt ansatz.

Under the Schmidt decomposition, $|\Psi_{\operatorname{gs}}\rangle$ can be represented as
\begin{equation}\label{append_eq: SVD decomposition of desired ground state}
    |\Psi_{\operatorname{gs}}\rangle = \sum_{i=1}^{\chi} \lambda_{i}|\phi_{i}\rangle |\psi_{i}\rangle,
\end{equation}
where $\lambda_{i}>0$ for $\forall i\in\{1,\cdots,\chi\}$ and $\{|\phi_{i}\rangle| i\in\{1,\cdots, \chi\}\}$ and $\{|\psi_{i}\rangle|i\in\{1,\cdots, \chi\}\}$ are both sets of orthonormal n-qubit states. We observe that Eq.~\ref{append_eq: SVD decomposition of desired ground state} can be rewritten as
\begin{equation}\label{append_eq: decompose into four combinations}
    \begin{split}
    |\Psi_{\operatorname{gs}}\rangle = & \frac{1}{4}\sum_{i=1}^{\chi} \lambda_{i} \left(|\phi_i\rangle+|\psi_{i}^{*}\rangle\right) \left(|\phi_i^{*}\rangle+|\psi_{i}\rangle\right) \\ 
    & - \frac{1}{4}\sum_{i=1}^{\chi} \lambda_{i} \left(|\phi_i\rangle-|\psi_{i}^{*}\rangle\right) \left(|\phi_i^{*}\rangle-|\psi_{i}\rangle\right) \\ 
    & + \frac{i}{4} \sum_{i=1}^{\chi} \lambda_{i} \left(-i|\phi_i\rangle + |\psi_{i}^{*}\rangle\right)\left(i|\phi_{i}^{*}\rangle + |\psi_{i}\rangle\right) \\
    & - \frac{i}{4} \sum_{i=1}^{\chi} \lambda_{i} \left(-i|\phi_i\rangle - |\psi_{i}^{*}\rangle\right)\left(i|\phi_{i}^{*}\rangle - |\psi_{i}\rangle\right),
    \end{split}
\end{equation}
such that we can find four density matrices $\rho_{1},\cdots, \rho_{4}$ with $|\Psi_{\operatorname{gs}}\rangle = c_1|\rho_{1}\rangle - c_{2}|\rho_{2}\rangle + ic_{3}|\rho_{3}\rangle-ic_{4}|\rho_{4}\rangle$ where $c_{i}\geq 0$ and $|\rho_{i}\rangle$ is proportional to the $i$-th row in the RHS of Eq.~\ref{append_eq: decompose into four combinations} for $\forall i\in\{1,2,3,4\}$. It is straight forward to see that $\operatorname{rank}\rho_{i}\leq \chi$ for $\forall i\in\{1,2,3,4\}$. Then as long as $\chi\leq 2^{L}$, each $\rho_i$ can be prepared via the Schmidt ansatz with $L$ CNOT gates between the upper block and the lower block. 

\subsection{Chemical-inspired ansatz}\label{App_subsec: Chemical-inspired ansatz} 

The electric ground state of a paramagnetic molecule has an equal number of spin-up electrons and spin-down electrons. Consider a paramagnetic molecule with $2k$ electrons and $2n$ spin orbits (the first $n$ orbits for spin-up electrons and the second $n$ orbits for spin-down electrons),  we use $a_{i,\uparrow}^{\dagger}\ (a_{i,\uparrow})$ to denote the creation (annihilation) operator for the $i$-th spin-up orbit and similarly, $a_{i,\downarrow}^{\dagger}\ (a_{i,\downarrow})$ the creation (annihilation) operator for the $i$-th spin-down orbit. To account for its spin symmetry, we use the following ansatz circuit based on the q-UCCSD ansatz \cite{mcardle2020quantum}:
\begin{equation}
\label{append_eq: spin symmetric ansatz}
    U(\theta):=U_{\uparrow}(\theta)U_{\downarrow}(\theta) U_{\uparrow\downarrow}(\theta)U_{\uparrow\uparrow}(\theta)U_{\downarrow\downarrow}(\theta),
\end{equation}
with
\begin{eqnarray}
    \label{append_eq: each term in the spin symmetric ansatz}
        U_{\uparrow}(\theta):=&&\prod_{i>j}\exp\left(\theta_{ij}^{\uparrow}(a_{i,\uparrow}^{\dagger}a_{j,\uparrow}-a_{j,\uparrow}^{\dagger}a_{i,\uparrow})\right), \nonumber \\
        U_{\downarrow}(\theta):=&& \prod_{i>j}\exp\left(\theta_{ij}^{\downarrow}(a_{i,\downarrow}^{\dagger}a_{j,\downarrow}-a_{j,\downarrow}^{\dagger}a_{i,\downarrow})\right), \nonumber \\
        U_{\uparrow\downarrow}(\theta):=&& \prod_{i>l,j>k}\exp\left(\theta_{ijkl}^{\uparrow\downarrow}(a_{i,\uparrow}^{\dagger}a_{j,\downarrow}^{\dagger}a_{k,\downarrow}a_{l,\uparrow}-a_{l,\uparrow}^{\dagger}a_{k,\downarrow}^{\dagger}a_{j,\downarrow}^{\dagger}a_{i,\uparrow}^{\dagger})\right),\nonumber \\
        U_{\uparrow\uparrow}(\theta):=&& \prod_{i>j>k>l}\exp\left(\theta_{ijkl}^{\uparrow\uparrow}(a_{i,\uparrow}^{\dagger}a_{j,\uparrow}^{\dagger}a_{k,\uparrow}a_{l,\uparrow}-a_{l,\uparrow}^{\dagger}a_{k,\uparrow}^{\dagger}a_{j,\uparrow}a_{i,\uparrow})\right), \text{ and }\nonumber \\
        U_{\downarrow\downarrow}(\theta):=&& \prod_{i>j>k>l} \exp\left(\theta_{ijkl}^{\downarrow\downarrow}(a_{i,\downarrow}^{\dagger}a_{j,\downarrow}^{\dagger}a_{k,\downarrow}a_{l,\downarrow}-a_{l,\downarrow}^{\dagger}a_{k,\downarrow}^{\dagger}a_{j,\downarrow}a_{i,\downarrow})\right), 
\end{eqnarray}
where each term above conserves the number of spin-up electrons and the number of spin-down electrons. We can use a qubit system to simulate this fermionic system by using the JW transformation, 
\begin{eqnarray}
        a_{i,\uparrow}^{\dagger}=&& \sigma_{i,\uparrow}^{-}\prod_{r=1}^{i-1}Z_{r,\uparrow} = \frac{1}{2}(X_{i,\uparrow}-iY_{i,\uparrow})\prod_{r=1}^{i-1}Z_{r,\uparrow}, \nonumber\\ 
    a_{i,\uparrow} =&& \sigma_{i,\uparrow}^{+}\prod_{r=1}^{i-1}Z_{r,\uparrow} = \frac{1}{2}(X_{i,\uparrow}+iY_{i,\uparrow})\prod_{r=1}^{i-1}Z_{r,\uparrow},\nonumber \\
    a_{i,\downarrow}^{\dagger}=&& \sigma_{i,\downarrow}^{-}\prod_{r=1}^{i-1}Z_{r,\downarrow}\prod_{m=1}^{n}Z_{m,\uparrow} = \frac{1}{2}(X_{i,\downarrow}-iY_{i,\downarrow})\prod_{r=1}^{i-1}Z_{r,\downarrow}\prod_{m=1}^{n}Z_{m,\uparrow}, \text{ and }\nonumber\\
    a_{i,\downarrow}=&&\sigma_{i,\downarrow}^{+}\prod_{r=1}^{i-1}Z_{r,\downarrow}\prod_{m=1}^{n}Z_{m,\uparrow} = \frac{1}{2}(X_{i,\downarrow}+iY_{i,\downarrow})\prod_{r=1}^{i-1}Z_{r,\downarrow}\prod_{m=1}^{n}Z_{m,\uparrow},
\end{eqnarray}
where we use $2n$ qubits and a Pauli operator $P$ on the $i$-th qubit in the first (second) block of $n$ qubits is denoted by $P_{i,\uparrow}\ (P_{i,\downarrow})$. In this way, the $|1\rangle$ state on the $i$-th qubit in the first (second) block indicates the presence of an electron on the $i$-th spin-up (spin-down) orbital. Thus for traditional VQE (on a qubit system), to restrict output states to respect the spin symmetry in the noiseless setting, we simply need to start with the initial state $|1,1,\cdots,1,0,\cdots,0,1,1\cdots,1,0,\cdots,0\rangle$ with $k$ $1$s in the first and second block respectively and run the corresponding parametrized circuit of Eq.~\ref{append_eq: spin symmetric ansatz}. (See Eq.~\ref{append_eq: unitary after JW I} and Eq.~\ref{append_eq: unitary after JW II} for explicit forms of the unitaries used in the parametrized circuit.) 

As for our algorithm, we obtain the output state $|\rho(\theta)\rangle$ as a combination of four states $|\rho(\theta)\rangle=\sum_{i=1}^{4}c_{i}|\rho_{i}(\theta)\rangle$ where each $\rho_{i}(\theta)$ is the reduced density matrix on the first half of all output qubits. Denote the output state corresponding to $\rho_{i}(\theta)$ as $|\psi_i(\theta)\rangle$. Suppose the Schmidt decomposition of $|\psi_{i}(\theta)\rangle$ reads $|\psi_{i}(\theta)\rangle=\sum_{\ell}\lambda_{\ell}|\psi_{\ell,\uparrow}\rangle|\psi_{\ell,\downarrow}\rangle$ with $\lambda_{\ell}\in \mathbb{R}^{+}$, $\{|\psi_{\ell,\uparrow}\rangle\}$ supported on the first $n$ qubits and $\{|\psi_{\ell,\uparrow}\rangle\}$ supported on the rest of the qubits, we obtain $|\rho_{i}(\theta)\rangle\propto \sum_{i}\lambda_{i}^{2}|\psi_{\ell,\uparrow}(\theta)\rangle |\psi_{\ell,\uparrow}(\theta)\rangle$. Then, as long as each $|\psi_{\ell,\uparrow}\rangle$ is only composed of states with $k$ spin-up electrons, each $|\rho_{i}(\theta)\rangle$ satisfy the spin symmetry and the final state $|\rho_{i}(\theta)\rangle$ also satisfy the spin symmetry. Thus, we can see that we may simply use the same strategy as the traditional VQE, namely the same parametrized circuit (Eq.~\ref{append_eq: spin symmetric ansatz}) and the same initialization to guarantee, in the noiseless circuit setting, the final state $|\rho(\theta)\rangle$ possess the spin symmetry.

In the following, we describe explicitly the parametrized circuit ansatz that realizes $U(\theta)$ with (parametrized) single-qubit gates and CNOT gates. After the JW transformation, the five unitary operators on the RHS of Eq. \ref{append_eq: each term in the spin symmetric ansatz} become
\begin{eqnarray}\label{append_eq: unitary after JW I}
               U_{\uparrow}(\theta) =&& \prod_{i>j}\exp\left(\frac{i \theta_{ij}^{\uparrow}}{2}\prod_{r=j+1}^{i-1}Z_{r}(X_{i}Y_{j})\right) \cdot \exp\left(-\frac{i \theta_{ij}^{\uparrow}}{2}\prod_{r=j+1}^{i-1}Z_{r}(Y_{i}X_{j})\right),    \nonumber \\
    U_{\uparrow\uparrow}(\theta)  =&& \prod_{i>j>k>l}\exp\left(\frac{i\theta_{ijkl}^{\uparrow\uparrow}}{8}\prod_{r=j+1}^{i-1}Z_{r,\uparrow}\prod_{r'=l+1}^{k-1}Z_{r',\uparrow}\  (X_{i,\uparrow}Y_{j,\uparrow}X_{k,\uparrow}X_{l,\uparrow})\right) \nonumber \\
    && \times \exp\left(\frac{i\theta_{ijkl}^{\uparrow\uparrow}}{8}\prod_{r=j+1}^{i-1}Z_{r,\uparrow}\prod_{r'=l+1}^{k-1}Z_{r',\uparrow}\  (Y_{i,\uparrow}X_{j,\uparrow}X_{k,\uparrow}X_{l,\uparrow})\right)\nonumber \\
        && \times \exp\left(\frac{i\theta_{ijkl}^{\uparrow\uparrow}}{8}\prod_{r=j+1}^{i-1}Z_{r,\uparrow}\prod_{r'=l+1}^{k-1}Z_{r',\uparrow}\  (Y_{i,\uparrow}Y_{j,\uparrow}X_{k,\uparrow}Y_{l,\uparrow})\right)\nonumber \\
            && \times \exp\left(\frac{i\theta_{ijkl}^{\uparrow\uparrow}}{8}\prod_{r=j+1}^{i-1}Z_{r,\uparrow}\prod_{r'=l+1}^{k-1}Z_{r',\uparrow}\  (Y_{i,\uparrow}Y_{j,\uparrow}Y_{k,\uparrow}X_{l,\uparrow})\right)\nonumber \\
    && \times \exp\left(-\frac{i\theta_{ijkl}^{\uparrow\uparrow}}{8}\prod_{r=j+1}^{i-1}Z_{r,\uparrow}\prod_{r'=l+1}^{k-1}Z_{r',\uparrow}\  (X_{i,\uparrow}X_{j,\uparrow}X_{k,\uparrow}Y_{l,\uparrow})\right) \nonumber\\
    && \times \exp\left(-\frac{i\theta_{ijkl}^{\uparrow\uparrow}}{8}\prod_{r=j+1}^{i-1}Z_{r,\uparrow}\prod_{r'=l+1}^{k-1}Z_{r',\uparrow}\  (X_{i,\uparrow}X_{j,\uparrow}Y_{k,\uparrow}X_{l,\uparrow})\right)\nonumber \\
    && \times \exp\left(-\frac{i\theta_{ijkl}^{\uparrow\uparrow}}{8}\prod_{r=j+1}^{i-1}Z_{r,\uparrow}\prod_{r'=l+1}^{k-1}Z_{r',\uparrow}\  (X_{i,\uparrow}Y_{j,\uparrow}Y_{k,\uparrow}Y_{l,\uparrow})\right)\nonumber \\
    && \times \exp\left(-\frac{i\theta_{ijkl}^{\uparrow\uparrow}}{8}\prod_{r=j+1}^{i-1}Z_{r,\uparrow}\prod_{r'=l+1}^{k-1}Z_{r',\uparrow}\  (Y_{i,\uparrow}X_{j,\uparrow}Y_{k,\uparrow}Y_{l,\uparrow})\right), 
    \end{eqnarray}
    \begin{eqnarray}\label{append_eq: unitary after JW II}
            U_{\uparrow\downarrow}(\theta) =&& \prod_{i>j>k>l}\exp\left(\frac{i\theta_{ijkl}^{\uparrow\downarrow}}{8}\prod_{r=l+1}^{i-1}Z_{r,\uparrow}\prod_{r'=k+1}^{j-1}Z_{r',\downarrow}\  (X_{i,\uparrow}Y_{l,\uparrow}X_{j,\downarrow}X_{k,\downarrow})\right)\nonumber \\
    && \times \exp\left(\frac{i\theta_{ijkl}^{\uparrow\downarrow}}{8}\prod_{r=l+1}^{i-1}Z_{r,\uparrow}\prod_{r'=k+1}^{j-1}Z_{r',\downarrow}\  (X_{i,\uparrow}X_{l,\uparrow}X_{j,\downarrow}Y_{k,\downarrow})\right) \nonumber\\
    && \times \exp\left(\frac{i\theta_{ijkl}^{\uparrow\downarrow}}{8}\prod_{r=l+1}^{i-1}Z_{r,\uparrow}\prod_{r'=k+1}^{j-1}Z_{r',\downarrow}\  (X_{i,\uparrow}Y_{l,\uparrow}Y_{j,\downarrow}Y_{k,\downarrow})\right)\nonumber \\
    && \times \exp\left(\frac{i\theta_{ijkl}^{\uparrow\downarrow}}{8}\prod_{r=l+1}^{i-1}Z_{r,\uparrow}\prod_{r'=k+1}^{j-1}Z_{r',\downarrow}\  (Y_{i,\uparrow}Y_{l,\uparrow}X_{j,\downarrow}Y_{k,\downarrow})\right)\nonumber \\
    && \times \exp\left(-\frac{i\theta_{ijkl}^{\uparrow\downarrow}}{8}\prod_{r=l+1}^{i-1}Z_{r,\uparrow}\prod_{r'=k+1}^{j-1}Z_{r',\downarrow}\  (Y_{i,\uparrow}X_{l,\uparrow}X_{j,\downarrow}X_{k,\downarrow})\right) \nonumber\\
    && \times \exp\left(-\frac{i\theta_{ijkl}^{\uparrow\downarrow}}{8}\prod_{r=l+1}^{i-1}Z_{r,\uparrow}\prod_{r'=k+1}^{j-1}Z_{r'\downarrow}\  (X_{i,\uparrow}X_{l,\uparrow}Y_{j,\downarrow}Y_{k,\downarrow})\right)\nonumber \\
    && \times \exp\left(-\frac{i\theta_{ijkl}^{\uparrow\downarrow}}{8}\prod_{r=l+1}^{i-1}Z_{r,\uparrow}\prod_{r'=k+1}^{j-1}Z_{r',\downarrow}\  (Y_{i,\uparrow}X_{l,\uparrow}Y_{j,\downarrow}Y_{k,\downarrow})\right) \nonumber\\
    && \times \exp\left(-\frac{i\theta_{ijkl}^{\uparrow\downarrow}}{8}\prod_{r=l+1}^{i-1}Z_{r,\downarrow}\prod_{r'=k+1}^{j-1}Z_{r',\downarrow}\  (Y_{i,\uparrow}Y_{l,\uparrow}Y_{j,\downarrow}X_{k,\downarrow})\right), 
    \end{eqnarray}
and $U_{\downarrow}(\theta)$ and $U_{\downarrow\downarrow}(\theta)$ are similar to $U_{\uparrow}(\theta)$ and $U_{\uparrow\uparrow}(\theta)$ respectively and are not displayed here. We observe that $U(\theta)$ comprises Pauli-string rotation operators of the form $\exp\left(i\frac{\theta}{2} \mathbf{P}\right)$ where $\mathbf{P}:=\prod_{r\in \operatorname{supp}}P_{r}$ is a product of single-qubit Pauli operators. For single qubit Pauli rotation gates, we can transform $R_{x}(\theta):=\exp\left(-i\frac{\theta}{2}X\right)$ to $R_{z}(\theta):=\exp\left(-i\frac{\theta}{2}Z\right)$ and $R_{y}(\theta):=\exp\left(-i\frac{\theta}{2}Y\right)$ to $R_{z}(\theta)$ using the following two equalities 
\begin{equation}
    H R_{x}(\theta) H = R_{z}(\theta)  \text{ and } R_{x}\left(\pi/2\right)R_{y}(\theta)R_{x}\left(-\pi/2\right)=R_{z}(\theta),
\end{equation}
which are the result of these two facts: $HXH=Z$ and $R_{x}\left(\pi/2\right)YR_{x}\left(-\pi/2\right)=Z$.
By generalizing the above argument, we can use single qubit rotations to transform $\exp(i\frac{\theta}{2}\mathbf{P})$ to $\exp(i\frac{\theta}{2}\prod_{r\in\operatorname{supp}}Z_{r})$, which then can be realized by sandwiching a single $R_{z}$ gate between two staircases of CNOT gates. For instance, $\exp(i\frac{\theta}{2} Z_{1}\otimes \cdots \otimes Z_{4})$ can be realized by the circuit in Eq.~\ref{eq: circuit for ZZZZ rotation}. 
\begin{equation}
\label{eq: circuit for ZZZZ rotation}
    \Qcircuit @C=1em @R=.7em {
&\ctrl{1} & \qw & \qw & \qw & \qw & \qw & \ctrl{1} & \qw\\
& \targ{} & \ctrl{1} & \qw & \qw & \qw & \ctrl{1} & \targ{} & \qw\\
& \qw & \targ{} & \ctrl{1} & \qw & \ctrl{1} & \targ{} & \qw & \qw\\
& \qw & \qw & \targ{} & \gate{\operatorname{R_z}(-\theta)} & \targ{} & \qw & \qw & \qw
}
\end{equation}

\section{Sampling Complexity}
\subsection{Ratio estimatior}
Given 2 random variables $X$ and $Y$ with mean values $\mu_x$ and $\mu_y$, the ratio $\frac{X}{Y}$ can be expanded around the point $(\mu_x$,$\mu_y)$ by the Taylor series. The approximation of the expectation value $E[\frac{X}{Y}]$ up to the second order Taylor expansion is:
\begin{eqnarray}\label{estim1}
E\left[\frac{X}{Y}\right]&&\approx E\left[\frac{\mu_x}{\mu_y}+\frac{1}{\mu_y}(X-\mu_x)-\frac{\mu_x}{\mu_y^2}(Y-\mu_y)+0-\frac{1}{\mu_y^2}(X-\mu_x)(Y-\mu_y)+\frac{\mu_x}{\mu_y^3}(Y-\mu_y)^2\right]\nonumber\\&&=\frac{\mu_x}{\mu_y}-\frac{1}{\mu_y^2}Cov[X,Y]+\frac{\mu_x}{\mu_y^3}Var[Y]
\end{eqnarray}
For the variance $Var\left[\frac{X}{Y}\right]=E\left[\left(\frac{X}{Y}-\frac{\mu_x}{\mu_y}\right)^2\right]$, we can similarly obtain its approximation up to the first order Taylor expansion:
\begin{eqnarray}\label{estim2}
Var\left[\frac{X}{Y}\right]&&\approx E\left[\left(\frac{1}{\mu_y}(X-\mu_x)-\frac{\mu_x}{\mu_y^2}(Y-\mu_y))\right)^2\right]\nonumber\\&&=E\left[\frac{1}{\mu_y^2}(X-\mu_x)^2-2\frac{\mu_x}{\mu_y^3}(X-\mu_x)(Y-\mu_y)+\frac{\mu_x^2}{\mu_y^4}(Y-\mu_y)^2\right]\nonumber\\&&=\frac{1}{\mu_y^2}Var[X]-2\frac{\mu_x}{\mu_y^3}Cov[X,Y]+\frac{\mu_x^2}{\mu_y^4}Var[Y]
\end{eqnarray}
The approximations in Eq. \ref{estim1} and Eq. \ref{estim2} are valid as long as the variances $Var[X]$ and $Var[Y]$ and the covariance $Cov[X,Y]$ are small enough. Eq. \ref{estim1} and Eq. \ref{estim2} indicate the following estimator:
\begin{itemize}
\item Estimator for $\frac{\mu_x}{\mu_y}$: Assuming $X$ and $Y$ are two independent random variables, the value of the ratio of their means $\frac{\mu_x}{\mu_y}$ can be estimated by an asymptotically unbiased estimator $\frac{\overline{X}}{\overline{Y}}$ where $\overline{X}$ and $\overline{Y}$ are the averages of $X$ and $Y$ from sampling. The expectation and the variance of this estimator are:
\begin{eqnarray}
E\left[\frac{\overline{X}}{\overline{Y}}\right]&&=\frac{\mu_x}{\mu_y}+Bias\left[\frac{\overline{X}}{\overline{Y}}\right]\approx\frac{\mu_x}{\mu_y}+\frac{\mu_x}{\mu_y^3}Var[\overline{Y}]\label{m1}\\
Var\left[\frac{\overline{X}}{\overline{Y}}\right]&&\approx \frac{1}{\mu_y^2}Var[\overline{X}]+\frac{\mu_x^2}{\mu_y^4}Var[\overline{Y}]
\end{eqnarray}
The variance and the bias give the mean squared error (MSE):
\begin{eqnarray}\label{mse}
MSE\left[\frac{\overline{X}}{\overline{Y}}\right]&&=Bias\left[\frac{\overline{X}}{\overline{Y}}\right]^2+Var\left[\frac{\overline{X}}{\overline{Y}}\right]\nonumber\\&&\approx Var\left[\frac{\overline{X}}{\overline{Y}}\right]\approx \frac{1}{\mu_y^2}Var[\overline{X}]+\frac{\mu_x^2}{\mu_y^4}Var[\overline{Y}]
\end{eqnarray}
where the first approximation is valid when $Var[\overline{Y}]^2\ll Var[\overline{Y}]$.

\end{itemize}

\subsection{Overall analysis}
We now analyze the sampling complexity of estimating:
\begin{eqnarray}
\langle\psi|H_A|\psi\rangle=\frac{\sum_{i,j=1}^K c_i^*c_jTr(H_B\rho_i\otimes\rho_j)}{C_\psi^2}=\frac{\sum_{i,j=1}^K c_i^*c_jTr(H_B\rho_i\otimes\rho_j)}{\sum_{kl}c_k^*c_lTr(S\rho_k\otimes\rho_l)}
\end{eqnarray}
where $H_A=\sum_{\alpha=1}^m g_\alpha P_\alpha$ and $H_B=\sum_{\alpha=1}^m g_\alpha Q_\alpha$ where $P_\alpha$ and $Q_\alpha$ are $2n$-qubit Pauli operators and their substitute operators respectively. Since $\langle\psi|H_A|\psi\rangle$ is real, for each $g_\alpha c_i^*c_jTr(Q_\alpha\rho_i\otimes\rho_j)$, we can separate it into $g_\alpha Re[c_i^*c_j]Tr(Re[Q_\alpha]\rho_i\otimes\rho_j)$ and $-g_\alpha Im[c_i^*c_j]Tr(Im[Q_\alpha]\rho_i\otimes\rho_j)$. Since the eigenvalues of all Pauli substitute operators belong to $\{1,-1,i,-i\}$, during the measurements, each $Re[Q_\alpha]$ (and $Im[Q_\alpha]$) can be seen a random variable: $X$ with an expectation $E[X]=p_0*0+p_1*1+p_{-1}*(-1)$ and a bounded variance $Var[X]=p_1(1-p_1)+p_{-1}(1-p_{-1})+2p_1p_{-1}\leq 1$. For each of such a part, we can assign $N_n/(2K^2m)$ measurements to estimate its value which means there are $N_n$ measurements in total for estimating the numerator. (We add the zero imaginary parts when $i=j$ for simplicity.) For the denominator, since $Tr(S\rho_k\otimes\rho_l)$ is real, we only need to estimate $Re[c_k^*c_l]Tr(S\rho_k\otimes\rho_l)$. During the measurements, $Tr(S\rho_k\otimes\rho_l)$ can be seen as a random variable: $Z_d$ with a expectation $E[Z_d]=p_1*1+p_{-1}*(-1)$ and a bounded variance $Var[Z_d]=4p_1(1-p_1)\leq 1$. For each of such a part, we can assign $N_d/(K^2)$ measurements to estimate its value which means there are $N_d$ measurements in total for estimating the denominator. Following these settings, the MSE of estimating $\langle\psi|H_A|\psi\rangle$ using Eq. \ref{mse} is:
\begin{eqnarray}\label{sample1}
MSE[\langle\psi|A|\psi\rangle]&&\approx \frac{1}{C_{\psi}^4}\sum_{ij\alpha}\left(Re[c_i^*c_j]^2g_\alpha^2Var[\overline{Tr(Re[Q_\alpha]\rho_i\otimes\rho_j)}]+Im[c_i^*c_j]^2g_\alpha^2Var[\overline{Tr(Im[Q_\alpha]\rho_i\otimes\rho_j)}]\right)\nonumber\\&&+\frac{\langle\psi|H_A|\psi\rangle^2}{C_{\psi}^4}\sum_{ij}Re[c_i^* c_j]^2Var[\overline{Tr(S\rho_i\otimes\rho_j)}]\nonumber\\&&\leq \frac{2K^2 m}{C_{\psi}^4 N_n}\sum_{ij\alpha}g_\alpha^2|c_i|^2|c_j|^2+\frac{\langle\psi|H_A|\psi\rangle^2K^2}{C_{\psi}^4 N_d}\sum_{ij}|c_i|^2|c_j|^2\nonumber\\&&=\frac{\sum_{ij}|c_i|^2|c_j|^2}{C_{\psi}^4}\left(\frac{2K^2 m}{N_n}\sum_{\alpha}g_\alpha^2 +\frac{K^2}{N_d}\langle\psi|H_A|\psi\rangle^2\right)\nonumber\\&&\leq \frac{\sum_{ij}|c_i|^2|c_j|^2}{C_{\psi}^4}\left(\frac{2K^2 m ||H_A||_F^2}{2^{2n}N_n} +\frac{K^2||H_A||_2^2}{N_d}\right)=\chi\left(\frac{2K^2 m ||H_A||_F^2}{2^{2n}N_n} +\frac{K^2||H_A||_2^2}{N_d}\right)
\end{eqnarray}
where $||H_A||_2$ and $||H_A||_F$ are the 2-norm and the Frobenius-norm of $H_A$. To further estimate the value of $\chi=\frac{\sum_{ij}|c_i|^2|c_j|^2}{C_{\psi}^4}$, we need to separate the noiseless and noisy cases. In the following, we will take the case of the Schmidt ansatz as an example, which can be easily generalized to other types of ansatz.

\subsection{Noiseless case}
 Starting from $|0\rangle^{\otimes n}\otimes |0\rangle^{\otimes n}$, the random Schmidt ansatz will give an output state:
\begin{equation} 
U_S|0\rangle^{\otimes n}\otimes |0\rangle^{\otimes n}=\sum_{i=0}^{2^L-1}\lambda_i (U(n)|i\rangle)\otimes |i\rangle
\end{equation}
with $\lambda_i$ generated by a $L$-qubit random orthogonal circuit and $U$ a random $n$-qubit unitary circuit. According to our encoding, only the reduced density matrix $\rho=\sum_{i=0}^{2^L-1}\lambda_i^2U|i\rangle\langle i|U^\dag$ is needed. In the expression of $\chi$, we need to evaluate the values like $Tr(\rho \sigma)$ with $\sigma=\sum_{i=0}^{2^L-1}\gamma_i^2V|i\rangle\langle i|U^\dag$. If $U$ and $V$ are drawn from a 1-design $\nu$, according to Eq. \ref{gold}-\ref{gold2}, $Tr(\rho_i\rho_j)$ will concentrate on an exponentially small average:
\begin{equation} 
Prob[Tr(\rho\sigma)\geq \varepsilon]\leq \frac{1}{2^n\varepsilon}
\end{equation}
which means the probability of $Tr(\rho\sigma)$ greater than a polynomially small $\varepsilon$ is exponentially small. 

On the other hand, the form of the Schmidt circuit gives a lower bound on the purity: $Tr(\rho^2)\geq\frac{1}{2^L}$ since each CNOT gate connecting the upper and lower systems can be decomposed into two local tensors with a bond dimension equal to 2: $CNOT=|0\rangle\langle 0|\otimes I+|1\rangle\langle 1|\otimes X$. Note that these discussions hold for the other two ansatzes as long as there are $L$ 2-qubit entangled gates connecting the upper and lower systems in total and the local circuits between two such gates form a 1-design, which can be easily satisfied since random Pauli circuits can already form a 1-design.

Using $Tr(\rho^2)\geq\frac{1}{2^L}$ and $Tr(\rho_i \rho_j)\approx 0$ when $n$ is large, we have:
\begin{eqnarray}\label{sample2}
\chi&&=\frac{\sum_{ij}|c_i|^2|c_j|^2}{(\sum_{kl}c_k^*c_lTr(S\rho_k\otimes\rho_l))^2}\approx \frac{\sum_{ij}|c_i|^2|c_j|^2}{(\sum_{k}|c_k|^2 Tr(\rho_k^2))^2}\nonumber\\&&\leq 2^{2L}\frac{\sum_{ij}|c_i|^2|c_j|^2}{(\sum_{k}|c_k|^2)^2}=2^{2L}
\end{eqnarray}
By putting this bound into Eq. \ref{sample1}, we get:
\begin{eqnarray}
MSE[\langle\psi|A|\psi\rangle]\leq 2^{2L}K^2\left(\frac{2 m ||H_A||_F^2}{2^{2n}N_n} +\frac{||H_A||_2^2}{N_d}\right)
\end{eqnarray}
Suppose $N_n=\frac{2N}{3}$ and $N_d=\frac{N}{3}$, to achieve an accuracy $\epsilon$, the required number of measurements $N$ should be:
\begin{eqnarray}
N\geq \frac{2^{2L}3K^2}{\epsilon^2}\left( \frac{m ||H_A||_F^2}{2^{2n}}+||H_A||_2^2\right)
\end{eqnarray}

\subsection{Noisy case}
For the noisy case, we can think the original density matrix $\rho=\sum_{i=0}^{2^L-1}\lambda_i^2U|i\rangle\langle i|U^\dag$ suffers from a noise channel:
\begin{equation} 
\mathcal{N}(\rho)=\sum_\alpha\sum_{i=0}^{2^L-1}\lambda_i^2 K_\alpha U|i\rangle\langle i|U^\dag K_\alpha^\dag
\end{equation} 
with $\sum_\alpha K_\alpha^\dag K_\alpha=I$. Under this noise channel, the average of $Tr(\mathcal{N}(\rho)\mathcal{N}(\sigma))$ can be estimated assuming $U$ and $V$ form a 1-design:
\begin{eqnarray}\label{sample4}
\int_\nu\int_\nu Tr(\mathcal{N}(\rho)\mathcal{N}(\sigma)) dU dV&&=\sum_{ij\alpha\beta}\lambda_i^2\gamma_j^2Tr\left(\int_\nu K_\alpha U|i\rangle\langle i|U^\dag  K_\alpha^\dag d U \int_\nu K_\beta V|j\rangle\langle j |V^\dag K_\beta^\dag dV\right)\nonumber\\&&=\sum_{ij\alpha\beta}\lambda_i^2\gamma_j^2Tr\left(\int_\nu  U|i\rangle\langle i|U^\dag   d U  K_\alpha^\dag K_\beta \int_\nu V|j\rangle\langle j |V^\dag dV K_\beta^\dag K_\alpha \right)\nonumber\\&&=\sum_{ij\alpha\beta}\lambda_i^2\gamma_j^2\frac{1}{2^{2n}}Tr(K_\alpha K_\alpha^\dag K_\beta  K_\beta^\dag )
\end{eqnarray}
Now suppose the noise channel is unital i.e. $\mathcal{N}(I)=I$, which means $\sum_\alpha K_\alpha K_\alpha^\dag=I$ then we have: $E(Tr(\mathcal{N}(\rho)\mathcal{N}(\sigma)))=\frac{1}{2^n}$ which coincides with the noiseless case. For non-unital noise, we need to require the channel to be close to the identity map to get the same conclusion. Note that if the channel is implemented before the unitary $U$, according to Eq. \ref{gold}-\ref{gold2}, we can find that $\frac{1}{2^n}$ is the average no matter whether the channel is unital or not:
\begin{eqnarray}
\sum_{ij}\lambda_i^2\gamma_j^2Tr\left(\int_\nu  U \sum_\alpha K_\alpha|i\rangle\langle i|K_\alpha^\dag U^\dag   d U \int_\nu  V\sum_\beta K_\beta|j\rangle\langle j |K_\beta^\dag V^\dag  dV\right)\nonumber\\=\sum_{ij}\lambda_i^2\gamma_j^2\frac{1}{2^{2n}}Tr(I)=\frac{1}{2^n}\sum_{ij}\lambda_i^2\gamma_j^2=\frac{1}{2^n}
\end{eqnarray}
This means the unital requirement is only for the noise channel at the end of the circuit. These discussions also work for the alternate gate-noise models.

When the quantum circuit is noisy, the purity of the reduced density matrices in the upper system prepared by the circuit can be smaller than $\frac{1}{2^L}$ since the system can have entanglements with the environment. In this case, we can model noise as discrete, probabilistic faults that can happen at the various locations in the circuit including gates, idling steps, and measurements. Suppose at a location $a$, the probability corresponding to no faults is $(1-p_a)$, then the fault-free probability of the whole circuit is defined as $P_0=\Pi_a (1-p_a)$ and the purity will be greater than $P_0^2$. If $\zeta=\sum_a p_a\sim 1$ and $p_a$ at different locations are the same, then the probability that
$l$ faults occur in the circuit is given by the Poisson distribution $P_l = e^{-\zeta}\frac{\zeta^l}{l!}$ due to the Le Cam's theorem with:
\begin{equation} 
P_0=e^{-\zeta}
\end{equation}
Thus, we have:
\begin{equation} 
\chi\leq \max\left[e^{4\zeta},2^{2L}\right]
\end{equation}
and:
\begin{eqnarray}
MSE[\langle\psi|A|\psi\rangle]&&\leq \max\left[e^{4\zeta},2^{2L}\right]K^2\left(\frac{2 m ||H_A||_F^2}{2^{2n}N_n} +\frac{||H_A||_2^2}{N_d}\right)\\
N&&\geq \max\left[e^{4\zeta},2^{2L}\right]\frac{ 3K^2}{\epsilon^2}\left( \frac{m ||H_A||_F^2}{2^{2n}}+||H_A||_2^2\right)
\end{eqnarray}

\section{Expressibility by covering number}
\subsection{covering numbers and packing number}

\begin{figure}[t]
	\centering
	\includegraphics[width=0.8\textwidth]{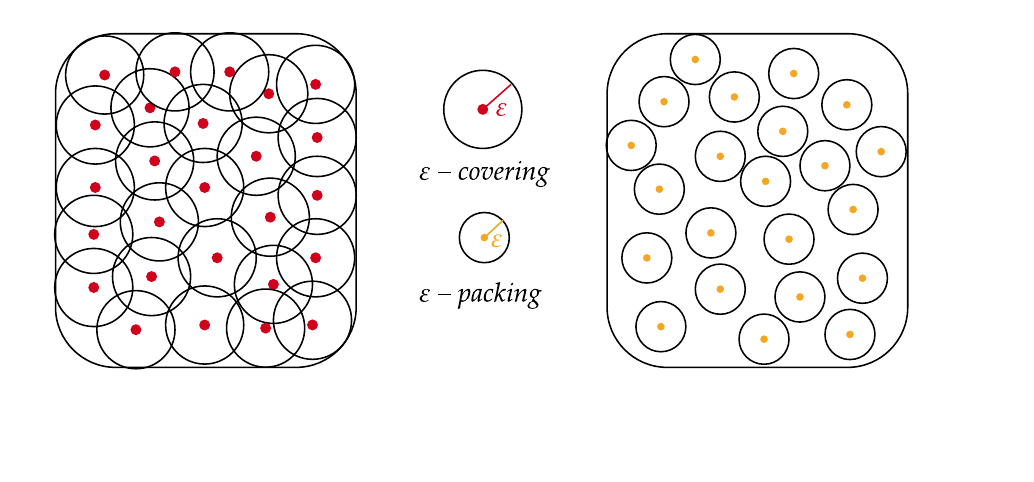}
	\caption{In the left figure, an $\varepsilon$-cover $\mathcal{Q}$ of the metric space $(\mathcal{M}, d)$ consists of points in $\mathcal{M}$ such that every point $z \in \mathcal{M}$ is no more than $\varepsilon$ away from some point $x \in \mathcal{Q}$. The size of the smallest $\varepsilon$-cover, denoted $\mathcal{N}(\mathcal{M}, d, \varepsilon)$, is termed the $\varepsilon$-covering number of $(\mathcal{M}, d)$. In the right figure, an $\varepsilon$-packing $\mathcal{Q}$ is a subset of $\mathcal{M}$ where the distance between any pair of distinct points $x$ and $y$ in $\mathcal{Q}$ exceeds $\varepsilon$.}
	\label{ccsf1}
\end{figure}

Figure \ref{ccsf1} provides a geometric illustration of covering and packing numbers. A subset $\mathcal{Q} \subset \mathcal{M}$ in the set $\mathcal{M}$ is defined as an $\varepsilon$-covering set if for every point $z \in \mathcal{M}$, there exists an $x \in \mathcal{Q}$ such that $d(x, z) \leq \varepsilon$. Conversely, if the distance between any two distinct points $x$ and $y$ in $\mathcal{Q}$ is greater than $\varepsilon$, then this subset $\mathcal{Q} \subset \mathcal{M}$ is termed $\varepsilon$-packing set.

Consider the following problem in the metric space $(\mathcal{M}, d)$. We cannot guarantee the simultaneous existence of an $\varepsilon$-covering set $\mathcal{Q}$ with cardinality $N$ and a $2\varepsilon$-packing set $\mathcal{Q}'$ with cardinality $\bar{N} > N$. If such sets exist, then within $\mathcal{Q}'$, there must be at least two elements $y_1$ and $y_2$ that lie within a $\varepsilon$-ball centered at some $x \in \mathcal{Q}$, leading to a distance $d(y_1, y_2) \leq 2\varepsilon$. This contradiction implies that covering and packing numbers satisfy the following inequality \cite{kolmogorov1959varepsilon}:
\begin{equation}
	\overline{\mathcal{N}}(\mathcal{M}, d, 2 \varepsilon) \leq \mathcal{N}(\mathcal{M}, d, \varepsilon) \leq \overline{\mathcal{N}}(\mathcal{M}, d, \varepsilon) .
	\label{ccs1}
\end{equation}

And it is important to highlight that $\varepsilon$ is a predefined hyperparameter, specifically a small constant within the interval $(0,1)$, independent of any other factors. \cite{kolmogorov1959varepsilon} Consequently, we can select $\varepsilon$ to be arbitrarily small to meet our needs in the following proof.

We will use two lemmas from \cite{barthel2018fundamental}, which we include here for the reader's convenience.

 \begin{lemma}[lemma 5 of \cite{barthel2018fundamental}]
  Let $f: \mathcal{M}_1 \rightarrow \mathcal{M}_2 $ be a bi-Lipschitz mapping between the metric spaces $\left(\mathcal{M}_1, d_1\right)$ and $\left(\mathcal{M}_2, d_2\right)$, satisfying $ f\left(\mathcal{M}_1\right)=\mathcal{M}_2 $ and the following conditions:
\begin{eqnarray}
&& d_2(f(x), f(y)) \leq k_1 d_1(x, y) \quad \forall x, y \in \mathcal{M}_1 \text{ and } \nonumber \\
&& d_2(f(x), f(y)) \geq k_2 d_1(x, y) \quad \forall x, y \in \mathcal{M}_1 \text{ when } d_1(x, y) \leq r
\end{eqnarray}
Then, their covering numbers can be derived from each other and satisfy the following relationship:
\begin{equation}
\mathcal{N}\left(\mathcal{M}_1, d_1, \frac{2\varepsilon}{k_2}\right) \leq \mathcal{N}\left(\mathcal{M}_2, d_2, \varepsilon\right) \leq \mathcal{N}\left(\mathcal{M}_1, d_1, \frac{\varepsilon}{k_1}\right)
\end{equation}
And the first inequality requires $\varepsilon \leq \frac{k_2 r}{2}$.
 \end{lemma}

\begin{lemma}[lemma 1 of \cite{barthel2018fundamental}]
	 For $0 < \tilde{\varepsilon} \leq 1/10$, the $\tilde{\varepsilon}$-covering number of the unitary group $U(n)$ with respect to the operator norm distance satisfies the following inequality:
	\begin{equation}
	\left(\frac{3}{4 \bar{\varepsilon}}\right)^{n^2} \leq \mathcal{N}(U(n),\|\cdot\|, \bar{\varepsilon}) \leq\left(\frac{7}{\bar{\varepsilon}}\right)^{n^2} .
\end{equation}
\end{lemma}

Lemma 2 provided an expression for the upper bound of the covering number of the unitary group. Now, by leveraging Lemma 1, which gives the proportion $k_1$  relating the metric of density matrix vectorization (DMV) to the metric of the unitary group, we derive the upper bound for the covering number of DMV in the following subsection.

\subsection{The expressiveness of VQAs by DMV}

This part will present the expressiveness of VQE under DMV. Let's start by reviewing some knowledge of matrix theory and some inequalities we will use:
	
	Discrete form of Holder's inequality: if $p, q > 1$, and $\frac{1}{p} + \frac{1}{q} = 1$, we have
	\begin{equation}
		\left|\sum_{k=1}^n a_k b_k\right| \leq\left(\sum_{k=1}^n\left|a_k\right|^p\right)^{\frac{1}{p}}\left(\sum_{k=1}^n\left|b_k\right|^q\right)^{\frac{1}{q}}
	\end{equation}
	When $p=q=2$, we obtain the Cauchy-Schwarz Inequality
	\begin{equation}
		\sum |x_i y_i \mid \leq\left(\sum |x_i|^2\right)^{1 / 2}\left(\sum |y_j|^2\right)^{1 / 2}\label{ccs21}
	\end{equation}
	
	In the content of this section, the norm is defined as follows: Suppose $A$ is an $n \times n$ matrix. The operator norm $A$ is defined as
	$$
	\|A\|=\sup _{\|\boldsymbol{x}\|_2=1, \boldsymbol{x} \in \mathbb{C}^n}\|A \boldsymbol{x}\| .
	$$
	Alternatively, $\|A\|=\sqrt{\lambda_1\left(A A^{\dagger}\right)}$, where $\lambda_i\left(A A^{\dagger}\right)$ is the $i$-th largest eigenvalue of the matrix $A A^{\dagger}$.
 
When using unconditionally decoherence-free quantum error mitigation, our output can be written as
\begin{equation}
	\left\langle\psi\left|H_A\right| \psi\right\rangle=\frac{\sum_{i, j=1}^K c_i^* c_j \operatorname{Tr}\left(H_B \rho_i \otimes \rho_j\right)}{C_\psi^2}=\frac{\sum_{i, j=1}^K c_i^* c_j \operatorname{Tr}\left(H_B \rho_i \otimes \rho_j\right)}{\sum_{k l} c_k^* c_l \operatorname{Tr}\left(S \rho_k \otimes \rho_l\right)}
\end{equation}
Please note that here,
\begin{eqnarray}
		 &&\operatorname{Tr}\left(H_B \rho_i \otimes \rho_j\right) \nonumber\\&&=\operatorname{Tr}\left(H_B^{1,2} \otimes I^{3,4} \rho_I^{1,3} \otimes \rho_J^{2,4}\right) \nonumber\\&&= \operatorname{Tr}\left(H_B^{1,2} \otimes I^{3,4}  U(\theta_I) \rho_{I 0}^{1,3} U(\theta_I)^{\dagger} \otimes U(\theta_J) \rho_{J 0}^{2,4} U(\theta_J)^{\dagger}\right)
\end{eqnarray}
In all the equations below, we follow the equation relations given above, namely $H_B=H_B^{1,2} \otimes I^{3,4}$ and $\rho_i \otimes \rho_j=U\left(\theta_I\right) \rho_{I 0}^{1,3} U\left(\theta_I\right)^{\dagger} \otimes U\left(\theta_J\right) \rho_{J 0}^{2,4} U\left(\theta_J\right)^{\dagger}$.
So, following the form of Lemma 1, we have
\begin{eqnarray}
		&&d_2\left(\frac{\sum_{i, j=1}^K c_i^* c_j \operatorname{Tr}\left(H_B \rho_i \otimes \rho_j\right)}{\sum_{k l} c_k^* c_l \operatorname{Tr}\left(S \rho_k \otimes \rho_l\right)}, \frac{\sum_{i, j=1}^K c_i^* c_j \operatorname{Tr}\left(H_B \rho_{i\varepsilon} \otimes \rho_{j\varepsilon}\right)}{\sum_{k l} c_k^* c_l \operatorname{Tr}\left(S \rho_{k\varepsilon} \otimes \rho_{l\varepsilon}\right)}\right) \nonumber\\&&=  \left|\frac{\sum_{i, j=1}^K c_i^* c_j \operatorname{Tr}\left(H_B \rho_i \otimes \rho_j\right)}{\sum_{k l} c_k^* c_l \operatorname{Tr}\left(S \rho_k \otimes \rho_l\right)}-\frac{\sum_{i, j=1}^K c_i^* c_j \operatorname{Tr}\left(H_B \rho_{i\varepsilon} \otimes \rho_{j\varepsilon}\right)}{\sum_{k l} c_k^* c_l \operatorname{Tr}\left(S \rho_{k\varepsilon} \otimes \rho_{l\varepsilon}\right)}\right|
\end{eqnarray}

In Appendix A, we provide equation\ref{ccs777}, So if $U$ and $V$ are drawn from a 1-design $\nu, \operatorname{Tr}\left(\rho_i \rho_j\right)$ will be exponentially small. In this case, we can simplify the numerator of the expression
\begin{eqnarray}
		&&d_2\left(\frac{\sum_{i, j=1}^K c_i^* c_j \operatorname{Tr}\left(H_B \rho_i \otimes \rho_j\right)}{\sum_{k l} c_k^* c_l \operatorname{Tr}\left(S \rho_k \otimes \rho_l\right)}, \frac{\sum_{i, j=1}^K c_i^* c_j \operatorname{Tr}\left(H_B \rho_{i\varepsilon} \otimes \rho_{j\varepsilon}\right)}{\sum_{k l} c_k^* c_l \operatorname{Tr}\left(S \rho_{k\varepsilon} \otimes \rho_{l\varepsilon}\right)}\right) \nonumber\\&&\approx  \left|\frac{\sum_{i, j=1}^K c_i^* c_j \operatorname{Tr}\left(H_B \rho_i \otimes \rho_j\right)}{\sum_k\left|c_k\right|^2 \operatorname{Tr}\left(\rho_k^2\right)}-\frac{\sum_{i, j=1}^K c_i^* c_j \operatorname{Tr}\left(H_B \rho_{i\varepsilon} \otimes \rho_{j\varepsilon}\right)}{\sum_k\left|c_k\right|^2 \operatorname{Tr}\left(\rho_{k\varepsilon}^2\right)}\right|
\end{eqnarray}
In Appendix A, we know that $\varepsilon$ is a predefined hyperparameter, i.e., a very small constant, satisfying $\varepsilon \in (0,1)$, and is independent of any factor. To satisfy Lemma 2, it is sufficient that $0 < \varepsilon \leq 1/10$. Therefore, for the sake of convenience in derivation, let us consider $\varepsilon$ approaching 0, so that
\begin{equation}
	\operatorname{Tr}\left(\rho_k^2\right)=\operatorname{Tr}\left(\rho_{k \varepsilon}^2\right)
\end{equation}

In this case, our equation can be further simplified to:
\begin{eqnarray}
		 && d_2\left(\frac{\sum_{i, j=1}^K c_i^* c_j \operatorname{Tr}\left(H_B \rho_i \otimes \rho_j\right)}{\sum_{k l} c_k^* c_l \operatorname{Tr}\left(S \rho_k \otimes \rho_l\right)}, \frac{\sum_{i, j=1}^K c_i^* c_j \operatorname{Tr}\left(H_B \rho_{i\varepsilon} \otimes \rho_{j\varepsilon}\right)}{\sum_{k l} c_k^* c_l \operatorname{Tr}\left(S \rho_{k\varepsilon} \otimes \rho_{l\varepsilon}\right)}\right) \nonumber\\&&\approx \left|\frac{\sum_{i, j=1}^K c_i^* c_j \operatorname{Tr}\left(H_B \rho_i \otimes \rho_j\right)}{\sum_k\left|c_k\right|^2 \operatorname{Tr}\left(\rho_k^2\right)}-\frac{\sum_{i, j=1}^K c_i^* c_j \operatorname{Tr}\left(H_B \rho_{i\varepsilon} \otimes \rho_{j\varepsilon}\right)}{\sum_k\left|c_k\right|^2 \operatorname{Tr}\left(\rho_{k}^2\right)}\right| \nonumber\\&&=  \left|\frac{\sum_{i, j=1}^K c_i^* c_j\left( \operatorname{Tr}\left(H_B \rho_i \otimes \rho_j\right)-\operatorname{Tr}\left(H_B \rho_{i \varepsilon} \otimes \rho_{j \varepsilon}\right)\right)}{\sum_k\left|c_k\right|^2 \operatorname{Tr}\left(\rho_k^2\right)}\right| \nonumber\\&&\leq  \left|\frac{\left(\sum_{i, j=1}^K\left|c_i^* c_j\right|^2\right)^{1 / 2}\left(\sum_{i, j=1}^K\left|\operatorname{Tr}\left(H_B \rho_i \otimes \rho_j\right)-\operatorname{Tr}\left(H_B \rho_{i \varepsilon} \otimes \rho_{j \varepsilon}\right)\right|^2\right)^{1 / 2}}{\sum_k\left|c_k\right|^2 \operatorname{Tr}\left(\rho_k^2\right)}\right| \nonumber\\&&\leq   2^{ L}\left(\sum_{i, j=1}^K\left|\operatorname{Tr}\left(H_B \rho_i \otimes \rho_j\right)-\operatorname{Tr}\left(H_B \rho_{i \varepsilon} \otimes \rho_{j \varepsilon}\right)\right|^2\right)^{1 / 2} \nonumber\\&&=  2^{ L}\left(\sum_{i, j=1}^K\left|\operatorname{Tr}\left(\hat{U}(\theta_{ij})^{\dagger} H_B \hat{U}(\theta_{ij}) \rho_{i0} \otimes \rho_{j0}\right)-\operatorname{Tr}\left(\hat{U}_{\varepsilon}(\theta_{ij})^{\dagger} H_B \hat{U}_{\varepsilon}(\theta_{ij}) \rho_{i0 } \otimes \rho_{j0 }\right)\right|^2\right)^{1 / 2} \nonumber\\&&\leq  2^{ L}\left(\sum_{i, j=1}^K\left|\left|\hat{U}(\theta_{ij})^{\dagger} H_B \hat{U}(\theta_{ij})-\hat{U}_{\varepsilon}(\theta_{ij})^{\dagger} H_B \hat{U}_{\varepsilon}(\theta_{ij}) \right|\right|^2\operatorname{Tr}\left( \rho_{i0 } \otimes \rho_{j0 }\right)\right)^{1 / 2} \nonumber\\&&	=   2^{ L}\left(\sum_{i, j=1}^K\left|\left|\hat{U}(\theta_{ij})^{\dagger} H_B \hat{U}(\theta_{ij})-\hat{U}_{\varepsilon}(\theta_{ij})^{\dagger} H_B \hat{U}_{\varepsilon}(\theta_{ij}) \right|\right|^2\right)^{1 / 2} \nonumber\\&&\leq   2^{ L}\left(\left(\sum_{i, j=1}^K\left|\left|\hat{U}(\theta_{ij})^{\dagger} H_B \hat{U}(\theta_{ij})-\hat{U}_{\varepsilon}(\theta_{ij})^{\dagger} H_B \hat{U}_{\varepsilon}(\theta_{ij}) \right|\right|\right)^2\right)^{1 / 2} \nonumber\\&&	=   2^{ L}\sum_{i, j=1}^K\left|\left|\hat{U}(\theta_{ij})^{\dagger} H_B \hat{U}(\theta_{ij})-\hat{U}_{\varepsilon}(\theta_{ij})^{\dagger} H_B \hat{U}_{\varepsilon}(\theta_{ij}) \right|\right| \nonumber\\&&=   2^{ L}\mathcal{C}\left|\left|\sum_{i, j=1}^K\hat{U}(\theta_{ij})^{\dagger} H_B \hat{U}(\theta_{ij})-\sum_{i, j=1}^K\hat{U}_{\varepsilon}(\theta_{ij})^{\dagger} H_B \hat{U}_{\varepsilon}(\theta_{ij}) \right|\right| \nonumber\\&&=  2^{ L}\mathcal{C} d_1\left(\sum_{i, j=1}^K\hat{U}(\theta_{ij})^{\dagger} H_B \hat{U}(\theta_{ij}),\sum_{i, j=1}^K\hat{U}_{\varepsilon}(\theta_{ij})^{\dagger} H_B \hat{U}_{\varepsilon}(\theta_{ij}) \right)
\end{eqnarray}
The first inequality stems from the Cauchy-Schwarz Inequality \ref{ccs21}, and the second inequality arises from what we mentioned in Appendix D:the form of the Schmidt circuit gives a lower bound on the purity: $\operatorname{Tr}\left(\rho^2\right) \geq \frac{1}{2^L}$ since each CNOT gate connecting the upper and lower systems can be decomposed into two local tensors with a bond dimension equal to 2: $C N O T=|0\rangle\langle 0|\otimes I+| 1\rangle\langle 1| \otimes X$.So we have:
\begin{equation}
		\frac{\sum_{i j}\left|c_i\right|^2\left|c_j\right|^2}{\left(\sum_k\left|c_k\right|^2 \operatorname{Tr}\left(\rho_k^2\right)\right)^2} 
		\leq 2^{2 L} \frac{\sum_{i j}\left|c_i\right|^2\left|c_j\right|^2}{\left(\sum_k\left|c_k\right|^2\right)^2}=2^{2 L}
\end{equation}
Taking the square root yields our second inequality.The third inequality arises from the triangle inequality, while the fourth equality is due to $\left\|\hat{U}\left(\theta_{ij}\right)^{\dagger} H_B \hat{U}\left(\theta_{ij}\right)-\hat{U}_{\epsilon}\left(\theta_{ij}\right)^{\dagger} H_B \hat{U}_{\epsilon}\left(\theta_{ij}\right)\right\|\geq0$.The last equality arises from the triangle inequality of matrix norms:$\quad\|A+B\| \leq\|A\|+\|B\|$Hence, we have: 
\begin{equation}
	\sum_{i, j=1}^K\left\|\hat{U}\left(\theta_{i j}\right)^{\dagger} H_B \hat{U}\left(\theta_{i j}\right)-\hat{U}_e\left(\theta_{i j}\right)^{\dagger} H_B \hat{U}_{\varepsilon}\left(\theta_{i j}\right)\right\|\geq
	\left\|\sum_{i, j=1}^K \hat{U}\left(\theta_{i j}\right)^{\dagger} H_B \hat{U}\left(\theta_{i j}\right)-\sum_{i, j=1}^K \hat{U}_e\left(\theta_{i j}\right)^{\dagger} H_B \hat{U}_e\left(\theta_{i j}\right)\right\|
\end{equation}
Here, we introduce a constant such that:
\begin{equation}
	\sum_{i, j=1}^K\left\|\hat{U}\left(\theta_{i j}\right)^{\dagger} H_B \hat{U}\left(\theta_{i j}\right)-\hat{U}_e\left(\theta_{i j}\right)^{\dagger} H_B \hat{U}_{\varepsilon}\left(\theta_{i j}\right)\right\|=\mathcal{C}
	\left\|\sum_{i, j=1}^K \hat{U}\left(\theta_{i j}\right)^{\dagger} H_B \hat{U}\left(\theta_{i j}\right)-\sum_{i, j=1}^K \hat{U}_e\left(\theta_{i j}\right)^{\dagger} H_B \hat{U}_e\left(\theta_{i j}\right)\right\|
\end{equation}
Here, $\mathcal{C} \geq 1$. Thus, we obtain the last equality.

From the previous inequalities, we obtain $k_1 = 2^L \mathcal{C}$, which is a global property. Now, let's examine the covering number of $\sum_{i,j=1}^K \hat{U}(\theta_{ij})^\dagger H_B \hat{U}(\theta_{ij})$, where the trainable unitary operator $\hat{U}(\boldsymbol{\theta}) = \prod_{h=1}^{N_g} \hat{u}_h(\boldsymbol{\theta}_h)$ contains $N_{gt(ij)}$ trainable gates and $N_{fix(ij)}=N_{g(ij)} - N_{gt(ij)}$ fixed gates. (For the case of the Schmidt ansatz as given in the main text, please note that when there are additional quantum gates to the right of the CNOT in the lower half, we consider these gates redundant as they do not enhance our expressiveness. Therefore, we should remove these gates, and let the number of such gates be denoted as $N_{n(ij)}$. In this case, the number of trainable gates that affect our expressive power is $N_{gt(ij)}=N_{g(ij)} - N_{fix(ij)} - N_{n(ij)}$.) To achieve this goal, we consider a fixed $\epsilon$-covering set $\mathcal{S}$ of the set $\mathcal{N}(U(d^k), \epsilon, \|\cdot\|)$, and define the set as:
\begin{equation}
	\tilde{\mathcal{S}}:=\left\{\sum_{i, j=1}^K \prod_{h \in\left\{N_{g t}\right\}} \hat{u}_h\left(\boldsymbol{\theta}_h\right) \prod_{k \in\left\{N_{\mathrm{g}}-N_{g t}\right\}} \hat{u}_k \mid \hat{u}_h\left(\boldsymbol{\theta}_h\right) \in \mathcal{S}\right\},
\end{equation}
where $\hat{u}_h(\boldsymbol{\theta}_h)$ and $\hat{u}_k$ respectively specify the trainable and fixed quantum gates used in the ansatz. Please note that for any circuit $\hat{U}(\boldsymbol{\theta}) = \prod_{i=1}^{N_g} \hat{u}_h(\boldsymbol{\theta}_h)$, we can always find a $\hat{U}_\epsilon(\boldsymbol{\theta}) \in \tilde{\mathcal{S}}$, where each trainable gate $\hat{u}_i(\boldsymbol{\theta}_i)$ is replaced by the closest element in the covering set $\mathcal{S}$, and the difference  satisfies:

\begin{eqnarray}
		 &&\left\|\sum_{i, j=1}^K \hat{U}\left(\theta_{i j}\right)^{\dagger} H_B \hat{U}\left(\theta_{i j}\right)-\sum_{i, j=1}^K \hat{U}_e\left(\theta_{i j}\right)^{\dagger} H_B \hat{U}_e\left(\theta_{i j}\right)\right\| \nonumber\\&&
		\leq \sum_{i, j=1}^K\left\|\hat{U}\left(\theta_{i j}\right)^{\dagger} H_B \hat{U}\left(\theta_{i j}\right)-\hat{U}_{\varepsilon}\left(\theta_{i j}\right)^{\dagger} H_B \hat{U}_{\varepsilon}\left(\theta_{i j}\right)\right\| \nonumber\\&&
		\leq \sum_{i, j=1}^K\left\|  \hat{U}\left(\theta_{i j}\right)-  \hat{U}_{\varepsilon}\left(\theta_{i j}\right)\right\|\left\|H_B\right\| \nonumber\\&&
		\leq \sum_{i, j=1}^KN_{g t(ij)}\|H_B\| \epsilon 
		=N_{GT}\|H_B\| \epsilon 
\end{eqnarray}
where $N_{GT} = \sum_{i,j=1}^K N_{gt(ij)}$. The first two inequalities utilize the triangle inequality, and the third inequality stems from $\left\|\hat{U}-\hat{U}_\epsilon\right\| \leq N_{gt} \epsilon$. Therefore, by definition, we know that $\tilde{\mathcal{S}}$ is an $N_{GT}\|H_B\| \epsilon$-covering set for $\mathcal{H}_{\text{circ}}$. 

Recalling the upper bound given in Lemma 1, $|\mathcal{S}| \leq \left(\frac{7}{\epsilon}\right)^{d^{2k}}$. Since there are $|\mathcal{S}|^{N_{GT}}$ possible topological combinations of gates in $\tilde{\mathcal{S}}$, we have $|\tilde{\mathcal{S}}| \leq \left(\frac{7}{\epsilon}\right)^{d^{2k} N_{GT}}$, and the covering number of $\mathcal{H}_{\text{circ}}$ satisfies:
\begin{equation}
	\mathcal{N}\left(\mathcal{H}_{\text {circ }}, N_{GT}\|H_B\| \epsilon,\|\cdot\|\right) \leq\left(\frac{7}{\epsilon}\right)^{d^{2 k} N_{GT}}
\end{equation}
An equivalent expression of the above inequalities is:
\begin{equation}
	\mathcal{N}\left(\mathcal{H}_{\text {circ }}, \epsilon,\|\cdot\|\right) \leq\left(\frac{7 N_{GT}\|H_B\|}{\epsilon}\right)^{d^{2 k} N_{GT}}
\end{equation}
Therefore, by Lemma 1, we can conclude that the upper bound on the covering number under our QEM algorithm is:
\begin{equation}
	\mathcal{N}(\tilde{\mathcal{H}}, \epsilon,|\cdot|) \leq2^L \mathcal{C}\left(\frac{7 N_{GT}\|H_B\|}{\epsilon}\right)^{d^{2 k} N_{GT}}
\end{equation}
In contrast, the cover number of traditional  VQAs is \cite{du2022efficient}
\begin{equation}
    \mathcal{N}(\mathcal{H}, \epsilon, |\cdot|) \leq \left(\frac{7 N_{gt}\|H_A\|}{\epsilon}\right)^{d^{2k}N_{gt}}
\end{equation}

\section{Trainability}
The cost function used in this work is parameterized by $\{\vec{\theta},\vec{c}\}$ with $\vec{\theta}$ the parameters that decide $\{\rho_1,\rho_2,...,\rho_k\}$. The derivatives with respect to $\vec{c}$ are trivial since they will not directly deal with the exponentially large Hilbert space and thus have no barren plateau issues. For $\vec{\theta}$, since we have:
\begin{equation} 
|\frac{\partial \langle\psi|H_A|\psi\rangle}{\partial \theta}|\leq\sum_{i,j=1}^K|\frac{\partial C_{ij}}{\partial \theta}|
\end{equation}
with $C_{ij}=c_i^*c_j\frac{Tr(H_B\rho_i\otimes\rho_j)}{\sum_{kl}c_k^*c_lTr(S\rho_k\otimes\rho_l)}$, and for each $C_{ij}$, we have:
\begin{eqnarray}\label{train2}
\left|\frac{\partial C_{ij}}{\partial \theta}\right|&&\leq \left|\frac{c_i^*c_j}{\sum_{kl}c_k^*c_lTr(S\rho_k\otimes\rho_l)}\right|\left|\frac{\partial Tr(H_B\rho_i\otimes\rho_j)}{\partial \theta}\right|\nonumber\\&&+\left|\frac{C_{ij}}{\sum_{kl}c_k^*c_lTr(S\rho_k\otimes\rho_l)}\right|\left|\frac{\partial \sum_{kl}c_k^*c_lTr(S\rho_k\otimes\rho_l)}{\partial \theta}\right|\nonumber\\&&\approx \left|\frac{c_i^*c_j}{\sum_{k}|c_k|^2Tr(S\rho_k\otimes\rho_k)}\right|\left|\frac{\partial Tr(H_B\rho_i\otimes\rho_j)}{\partial \theta}\right|\nonumber\\&&+\left|\frac{C_{ij}}{\sum_{k}|c_k|^2Tr(S\rho_k\otimes\rho_k)}\right|\left|\frac{\partial \sum_{kl}c_k^*c_l Tr(S\rho_k\otimes\rho_l)}{\partial \theta}\right|\nonumber\\&&\leq 2^L \left|\frac{c_i^* c_j}{|c_i|^2+|c_j|^2}\right|\left|\frac{\partial Tr(H_B\rho_i\otimes\rho_j)}{\partial \theta}\right|+2^L|C_{ij}|\sum_{kl}\left|\frac{c_k^* c_l}{|c_k|^2+|c_l|^2}\right|\left|\frac{\partial Tr(S\rho_k\otimes\rho_l)}{\partial \theta}\right|\nonumber\\&&\leq 2^{L+1} \left|\frac{\partial Tr(H_B\rho_i\otimes\rho_j)}{\partial \theta}\right|+2^{L+1}|C_{ij}|\sum_{kl}\left|\frac{\partial Tr(S\rho_k\otimes\rho_l)}{\partial \theta}\right|
\nonumber\\&&\leq 2^{L+1} \left|\frac{\partial Tr(H_B\rho_i\otimes\rho_j)}{\partial \theta}\right|+2^{2L+1}\left|\frac{c_i^* c_j}{|c_i|^2+|c_j|^2}\right|\left|Tr(H_B\rho_i\otimes\rho_j)\right|\sum_{kl}\left|\frac{\partial Tr(S\rho_k\otimes\rho_l)}{\partial \theta}\right|\nonumber\\&&\leq 2^{L+1} \left|\frac{\partial Tr(H_B\rho_i\otimes\rho_j)}{\partial \theta}\right|+2^{2L+2}\left|Tr(H_B\rho_i\otimes\rho_j)\right|\sum_{kl}\left|\frac{\partial Tr(S\rho_k\otimes\rho_l)}{\partial \theta}\right|\nonumber\\&&\approx 2^{L+1} \left|\frac{\partial Tr(H_B\rho_i\otimes\rho_j)}{\partial \theta}\right|+2^{2L+2-n}\left|\langle\rho_i|H_A|\rho_j\rangle\right|\sum_{kl}\left|\frac{\partial Tr(S\rho_k\otimes\rho_l)}{\partial \theta}\right|\nonumber\\&&\leq 2^{L+1} \left|\frac{\partial Tr(H_B\rho_i\otimes\rho_j)}{\partial \theta}\right|+2^{2L+2-n}||H_A||_2\sum_{kl}\left|\frac{\partial Tr(S\rho_k\otimes\rho_l)}{\partial \theta}\right|
\end{eqnarray}
where $|\rho\rangle$ is defined as $|\rho\rangle=\frac{1}{\sqrt{\sum_{ij}|\rho_{ij}|^2}}\sum_{ij}\rho_{ij}|i\rangle|j\rangle$. During the derivation, we assume the  noiseless case where there are $L$ CNOT gates connecting the upper and the lower system and use $Tr(\rho_i\rho_j)\approx\frac{1}{2^n}$. For noisy cases, one can use $e^{-2\zeta}$ to replace $2^L$. In our settings, each $\theta$ only controls a single density matrix, thus, suppose $\theta$ is a parameter for $\rho$, to further estimate the gradients, we need to consider two types of derivatives: $\left|\frac{\partial Tr(O\rho\otimes\sigma)}{\partial \theta}\right|$ and $\left|\frac{\partial Tr(O\rho\otimes\rho)}{\partial \theta}\right|$. If we have $O=\sum_\alpha o_\alpha P_\alpha$ with $P_\alpha=P_{\alpha,up}\otimes P_{\alpha,low}$ $2n$-qubit Pauli operators, then we have:
\begin{eqnarray} 
\left|\frac{\partial Tr(O\rho\otimes\sigma)}{\partial \theta}\right|&&\leq\sum_\alpha |o_\alpha|\left|\frac{\partial Tr(P_{\alpha,up}\rho)}{\partial \theta}\right|\left|Tr(P_{\alpha,low}\sigma)\right|\nonumber\\&&\leq \sum_\alpha |o_\alpha|\left|\frac{\partial Tr(P_{\alpha,up}\rho)}{\partial \theta}\right| 
\end{eqnarray}
\begin{eqnarray} 
\left|\frac{\partial Tr(O\rho\otimes\rho)}{\partial \theta}\right|&&\leq\sum_\alpha |o_\alpha|\left(\left|\frac{\partial Tr(P_{\alpha,up}\rho)}{\partial \theta}\right|\left|Tr(P_{\alpha,low}\rho)\right|+\left|\frac{\partial Tr(P_{\alpha,low}\rho)}{\partial \theta}\right|\left|Tr(P_{\alpha,up}\rho)\right|\right)\nonumber\\&&\leq\sum_\alpha |o_\alpha|\left(\left|\frac{\partial Tr(P_{\alpha,up}\rho)}{\partial \theta}\right|+\left|\frac{\partial Tr(P_{\alpha,low}\rho)}{\partial \theta}\right|\right)
\end{eqnarray}
where during the derivation, we use the fact $\left|Tr(P_{\alpha,low}\sigma)\right|\leq 1$ for arbitrary density matrix $\sigma$. For $\rho$, we can model it as:
\begin{equation}\label{train1}
\rho=\sum_{\alpha\beta}K_\alpha U(\theta) L_\beta |0\rangle\langle 0|^{\otimes n}L_\beta^\dag U(\theta)^\dag K_\alpha^\dag=\sum_{\alpha}K_\alpha U(\theta) \rho_0 U(\theta)^\dag K_\alpha^\dag
\end{equation}
with $\sum_\alpha K_\alpha^\dag K_\alpha=\sum_\beta L_\beta^\dag L_\beta=I$ and $\rho_L=\sum_\beta L_\beta |0\rangle\langle 0|^{\otimes n}L_\beta^\dag$. This form can be understood as $U(\theta)$ sandwiched in between two CNOT gates that connect the upper and lower systems. The influence of these CNOT gates and the rest of the circuits can be formulated as quantum channels $\{K_\alpha\}$ and $\{L_\beta\}$. $\frac{\partial Tr(P_{\alpha,up(low)}\rho)}{\partial \theta}$ can have barren plateau issues from various resources. For example, if $U(\theta)$ forms a 2-design, then $E\left(\left(\frac{\partial Tr(P_{\alpha,up(low)}\rho)}{\partial \theta}\right)^2\right)\leq \eta^2$ with $\eta^2$ of order $\mathcal{O}(2^{-n})$. By the Markov’s inequality, we have: 
\begin{equation}
Prob[\left|\frac{\partial Tr(P_{\alpha,up(low)}\rho)}{\partial \theta}\right|\geq\varepsilon]=Prob[\left(\frac{\partial Tr(P_{\alpha,up(low)}\rho)}{\partial \theta}\right)^2\geq\varepsilon^2]\leq \frac{\eta}{\varepsilon^2}
\end{equation}
Thus, $\frac{\partial Tr(P_{\alpha,up(low)}\rho)}{\partial \theta}$ will be concentrated to an exponentially small value, which leads to:
\begin{equation} 
\left|\frac{\partial Tr(O\rho\otimes\sigma)}{\partial \theta}\right|\lesssim  \sum_\alpha |o_\alpha|\eta=\frac{\eta||O||_{1,1}}{2^n}
\end{equation}
\begin{equation} 
\left|\frac{\partial Tr(O\rho\otimes\rho)}{\partial \theta}\right|\lesssim\sum_\alpha 2|o_\alpha|\eta=\frac{2\eta ||O||_{1,1}}{2^n}
\end{equation}
Where $||\cdot||_{1,1}$ denotes the $L_{1,1}$ norm. For $2n$-qubit SWAP operator $S$, we have $||S||_{1,1}=2^n$ since the 2-qubit SWAP operator has the decomposition: $\frac{1}{2}(II+XX+YY+ZZ)$. For $H_B=\sum_{\alpha=1}^m g_\alpha Q_\alpha$, note that each $Q_\alpha$ is a tensor product of $n$ basic 2-qubit Pauli substitute operators that form four groups: 
\begin{eqnarray}
\text{Group 1: }&&0.5 II+0.5XX+0.5YY+0.5ZZ\nonumber\\&&0.5 II+0.5XX-0.5YY-0.5ZZ\nonumber\\&&-0.5 II+0.5XX-0.5YY+0.5ZZ\nonumber\\&&0.5 II-0.5XX-0.5YY+0.5ZZ\nonumber\\\text{Group 2: }&&0.5 IX+0.5XI+0.5iYZ-0.5iZY\nonumber\\&&0.5 IX+0.5XI-0.5iYZ+0.5iZY\nonumber\\&&-0.5i IX+0.5iXI+0.5YZ+0.5ZY\nonumber\\&&-0.5i IX+0.5iXI-0.5YZ-0.5ZY\nonumber\\\text{Group 3: }&&-0.5 IY+0.5iXZ-0.5YI-0.5iZX\nonumber\\&&0.5 IY+0.5iXZ+0.5YI-0.5iZX\nonumber\\&&0.5i IY+0.5XZ-0.5iYI+0.5ZX\nonumber\\&&-0.5i IY+0.5XZ+0.5iYI+0.5ZX\nonumber\\\text{Group 4: }&&0.5 IZ+0.5iXY-0.5iYX+0.5ZI\nonumber\\&&0.5 IZ-0.5iXY+0.5iYX+0.5ZI\nonumber\\&&0.5i IZ-0.5XY-0.5YX-0.5iZI\nonumber\\&&0.5i IZ+0.5XY+0.5YX-0.5iZI
\end{eqnarray}
For two such 2-qubit operators $B_1$ and $B_2$ in different groups, since they share no common Pauli operators, we have $Tr(B_1^\dag B_2)=0$. When they are in the same group, the coefficients before the Pauli operators assure the same result $Tr(B_1^\dag B_2)=0$. Since $Tr(Q^\dag_\alpha Q_\alpha)=4^n$, we have:
\begin{equation}\label{train3}
||H_B||_{1,1}=2^n\sum_\alpha |g_\alpha|=||H_A||_{1,1}
\end{equation}
Based on these discussions, Eq. \ref{train2} can be further reduced to:
\begin{eqnarray}
\left|\frac{\partial C_{ij}}{\partial \theta}\right|&&\leq2^{L+1} \left|\frac{\partial Tr(H_B\rho_i\otimes\rho_j)}{\partial \theta}\right|+2^{2L+2-n}||H_A||_2\sum_{kl}\left|\frac{\partial Tr(S\rho_k\otimes\rho_l)}{\partial \theta}\right|\nonumber\\&&\leq \frac{2^{L+2}\eta||H_A||_{1,1}}{2^n}+\sum_{kl}\frac{2^{2L+3}||H_A||_2 \eta||S||_{1,1}}{2^{2n}}\nonumber\\&&=\left(\frac{2^{L+2}||H_A||_{1,1}}{2^n}+ \frac{2^{2L+3}K^2||H_A||_2}{2^{n}}\right)\eta
\end{eqnarray}
By noticing $||H_A||_2\leq\sum_\alpha |g_\alpha|$ and Eq. \ref{train3}, we have:
\begin{equation}
\left|\frac{\partial C_{ij}}{\partial \theta}\right|\leq \left(2^{L+2}+\frac{2^{2L+3}K^2}{2^{n}}\right)\sum_\alpha |g_\alpha|\eta\approx 2^{L+2}\sum_\alpha |g_\alpha|\eta
\end{equation}
where in the last approximation, we assume $2L\ll n$ to limit the sampling complexity. Finally, the whole derivative is bounded by:
\begin{equation}\label{train5}
|\frac{\partial \langle\psi|H_A|\psi\rangle}{\partial \theta}|\leq\sum_{i,j=1}^K|\frac{\partial C_{ij}}{\partial \theta}|\leq 2^{L+2}K^2\sum_\alpha |g_\alpha|\eta
\end{equation}
Note that, generally, we have $\sum_\alpha |g_\alpha|$ of order $\mathcal{O}(1)$ and $\eta^2$ of order $\mathcal{O}(2^{-n})$, thus, when we require $2L\ll n$ and $U(\theta)$ forms a 2-design, the derivative will be exponentially small and thus barren plateau problems occur. Since Eq. \ref{train5} is a rather loose bound, it can't be used to estimate cases where $L$ is comparable with $n$ and cases where the existence of noise leads to exponentially small purities. For $L$ is comparable with $n$, it is possible that the whole $2n$-qubit unitary circuit forms a 2-design, which will then also lead to barren plateau problems. For other resources of barren plateaus, we refer to Ref. \cite{wang2021noise,cerezo2021cost,ragone2023unified}.

\section{Rotation unitaries for measurement}

\begin{table*}[ht]
	\centering
	\begin{tabular}{|c|c|p{0.5\textwidth}<{\centering}|c|}\hline
		ID&$P$ (Hermite)& $Q$ (Unitary)& Eigenvalues of Q \\\hline
		1&$II$& $0.5 II+0.5XX+0.5YY+0.5ZZ$& diag(1,1,1,-1)\\\hline
		2&$XX$& $0.5 II+0.5XX-0.5YY-0.5ZZ$& diag(1,1,-1,1)\\\hline 
		3&$YY$& $-0.5 II+0.5XX-0.5YY+0.5ZZ$& diag(1,-1,-1,-1)\\\hline
		4&$ZZ$& $0.5 II-0.5XX-0.5YY+0.5ZZ$& diag(1,-1,1,1)\\\hline 
		
		\multicolumn{4}{|c|}{Q matrix}\\\hline
		\multicolumn{4}{|c|}{
			\begin{tabular}{@{} p{0.25\textwidth}|p{0.25\textwidth}|p{0.25\textwidth}|p{0.25\textwidth} @{}}
				1 & 2& 3& 4
		\end{tabular}} 
		\\
		\multicolumn{4}{|c|}{
			\begin{tabular}{@{} p{0.25\textwidth}<{\centering}|p{0.25\textwidth}<{\centering}|p{0.25\textwidth}<{\centering}|p{0.25\textwidth}<{\centering} @{}}
				$\begin{pmatrix}
					1 & 0 & 0 & 0\\
					0 & 0 & 1 & 0\\
					0 & 1 & 0 & 0\\
					0 & 0 & 0 & 1
				\end{pmatrix}$&
				$\begin{pmatrix}
					0 & 0 & 0 & 1\\
					0 & 1 & 0 & 0\\
					0 & 0 & 1 & 0\\
					1 & 0 & 0 & 0
				\end{pmatrix}$&
				$\begin{pmatrix}
					0 & 0 & 0 & 1\\
					0 & -1 & 0 & 0\\
					0 & 0 & -1 & 0\\
					1 & 0 & 0 & 0
				\end{pmatrix}$&
				$\begin{pmatrix}
					1 & 0 & 0 & 0\\
					0 & 0 & -1 & 0\\
					0 & -1 & 0 & 0\\
					0 & 0 & 0 & 1
				\end{pmatrix}$\\
				& & & 
		\end{tabular}} 
		\\\hline
		
		\multicolumn{4}{|c|}{Q circuit}\\\hline
		\multicolumn{4}{|c|}{
			\begin{tabular}{@{} p{0.25\textwidth}|p{0.25\textwidth}|p{0.25\textwidth}|p{0.25\textwidth} @{}}
				1 & 2& 3& 4
		\end{tabular}} 
		\\
		\multicolumn{4}{|c|}{
			\begin{tabular}{@{} p{0.25\textwidth}<{\centering}|p{0.25\textwidth}<{\centering}|p{0.25\textwidth}<{\centering}|p{0.25\textwidth}<{\centering} @{}}
				\
				\Qcircuit @C=3.5em @R=3.5em {
					\lstick{} & \qswap  & \qw \\
					\lstick{} & \qswap \qwx & \qw 
				}&
				\
				\Qcircuit @C=2em @R=2em {
					\lstick{} & \gate{X} & \qswap  & \qw \\
					\lstick{} & \gate{X} & \qswap \qwx & \qw 
				}&
				\
				\Qcircuit @C=2em @R=2em {
					\lstick{} & \gate{i Y} & \qswap  & \qw \\
					\lstick{} & \gate{i Y} & \qswap \qwx & \qw 
				}&
				\
				\Qcircuit @C=2em @R=2em {
					\lstick{} & \gate{Z} & \qswap  & \qw \\
					\lstick{} & \gate{Z} & \qswap \qwx & \qw 
				}\\
				& & &
		\end{tabular}} 
		\\\hline
		
		\multicolumn{4}{|c|}{Diagonalizing matrix}\\\hline
		\multicolumn{4}{|c|}{
			\begin{tabular}{@{} p{0.25\textwidth}|p{0.25\textwidth}|p{0.25\textwidth}|p{0.25\textwidth} @{}}
				1 & 2& 3& 4
		\end{tabular}} 
		\\
		\multicolumn{4}{|c|}{
			\begin{tabular}{@{} p{0.25\textwidth}<{\centering}|p{0.25\textwidth}<{\centering}|p{0.25\textwidth}<{\centering}|p{0.25\textwidth}<{\centering} @{}}
				$\begin{pmatrix}
					0 & 0 & 1 & 0 \\
					\frac{1}{\sqrt{2}} & \frac{1}{\sqrt{2}} & 0 & 0 \\
					\frac{1}{\sqrt{2}} & -\frac{1}{\sqrt{2}} & 0 & 0 \\
					0 & 0 & 0 & 1
				\end{pmatrix}$&
				$\begin{pmatrix}
					\frac{1}{\sqrt{2}} & \frac{1}{\sqrt{2}} & 0 & 0 \\
					0 & 0 & 1 & 0 \\
					0 & 0 & 0 & 1 \\
					\frac{1}{\sqrt{2}} & -\frac{1}{\sqrt{2}} & 0 & 0
				\end{pmatrix}$&
				$\begin{pmatrix}
					\frac{1}{\sqrt{2}} & \frac{1}{\sqrt{2}} & 0 & 0 \\
					0 & 0 & 1 & 0 \\
					0 & 0 & 0 & 1 \\
					\frac{1}{\sqrt{2}} & -\frac{1}{\sqrt{2}} & 0 & 0
				\end{pmatrix}$&
				$\begin{pmatrix}
					0 & 0 & 1 & 0 \\
					\frac{1}{\sqrt{2}} & \frac{1}{\sqrt{2}} & 0 & 0 \\
					-\frac{1}{\sqrt{2}} & \frac{1}{\sqrt{2}} & 0 & 0 \\
					0 & 0 & 0 & 1
				\end{pmatrix}$\\
				& & & 
		\end{tabular}} 
		\\\hline
		\multicolumn{4}{|c|}{Diagonalizing matrix circuit}\\\hline

		\multicolumn{4}{|c|}{
			\begin{tabular}{@{} p{1.02\textwidth} @{}}
				1 
		\end{tabular}} 
		\\
		\multicolumn{4}{|c|}{
			\begin{tabular}{@{} p{0.6\textwidth}<{\centering} @{}}
				\\
				\Qcircuit @C=0.2em @R=0.6em {
					\lstick{} & \gate{RZ(\pi)}  & \gate{RY(-\frac{3\pi}{4})} & \gate{RX(\frac{\pi}{2})}  &  \ctrl{1}  & \gate{RX(-\frac{\pi}{2})} & \ctrl{1} & \gate{RX(-\frac{\pi}{2})} & \gate{RZ(-\pi)} & \gate{RY(-\frac{\pi}{2})} & \gate{RZ(\frac{\pi}{2})} & \gate{U1(\frac{3\pi}{2})} & \gate{RZ(-\frac{3\pi}{2})} & \qw \\
					\lstick{} & \gate{RZ(-\frac{\pi}{2})}  & \gate{RY(-\frac{\pi}{2})}  &  \gate{RZ(-\frac{\pi}{2})} & \targ & \gate{RY(-\frac{\pi}{4})} & \targ & \gate{RZ(-\pi)} & \gate{RY(-\pi)} & \gate{RZ(-\pi)} & \qw & \qw & \qw & \qw & \qw\\
				} 
				
				\\
				
			\end{tabular}
			
		} 
		\\\hline
		
		\multicolumn{4}{|c|}{
			\begin{tabular}{@{} p{1.02\textwidth} @{}}
				2 
		\end{tabular}} 
		\\
		\multicolumn{4}{|c|}{
			\begin{tabular}{@{} p{0.6\textwidth}<{\centering} @{}}
				\\
				\Qcircuit @C=0.2em @R=0.6em {
					\lstick{} & \gate{RZ(\pi)}  & \gate{RY(-\frac{3\pi}{4})} & \gate{RZ(-\pi)}  & \gate{RX(\frac{\pi}{2})} & \ctrl{1}  & \gate{RX(-\frac{\pi}{2})} & \ctrl{1} & \gate{RX(-\frac{\pi}{2})} & \gate{RZ(\pi)} & \gate{RY(-\frac{\pi}{2})} & \gate{RZ(\frac{\pi}{2})} & \gate{U1(\frac{3\pi}{2})} & \gate{RZ(-\frac{3\pi}{2})} & \qw \\
					\lstick{} & \gate{RZ(\frac{\pi}{2})}  & \gate{RY(-\frac{\pi}{2})}  &  \gate{RZ(\frac{\pi}{2})} & \qw & \targ & \gate{RY(-\frac{\pi}{4})} & \targ & \gate{RY(-\pi)} & \qw & \qw & \qw & \qw & \qw & \qw & \qw\\
				} 
				
				\\
				
			\end{tabular}
			
		} 
		\\\hline
		
		\multicolumn{4}{|c|}{
			\begin{tabular}{@{} p{1.02\textwidth} @{}}
				3 
		\end{tabular}} 
		\\
		\multicolumn{4}{|c|}{
			\begin{tabular}{@{} p{0.6\textwidth}<{\centering} @{}}
				\\
				\Qcircuit @C=0.2em @R=0.6em {
					\lstick{} & \gate{RZ(\pi)}  & \gate{RY(-\frac{3\pi}{4})} & \gate{RZ(-\pi)}  & \gate{RX(\frac{\pi}{2})} & \ctrl{1}  & \gate{RX(-\frac{\pi}{2})} & \ctrl{1} & \gate{RX(-\frac{\pi}{2})} & \gate{RZ(\pi)} & \gate{RY(-\frac{\pi}{2})} & \gate{RZ(\frac{\pi}{2})} & \gate{U1(\frac{3\pi}{2})} & \gate{RZ(-\frac{3\pi}{2})} & \qw \\
					\lstick{} & \gate{RZ(\frac{\pi}{2})}  & \gate{RY(-\frac{\pi}{2})}  &  \gate{RZ(\frac{\pi}{2})} & \qw & \targ & \gate{RY(-\frac{\pi}{4})} & \targ & \gate{RY(-\pi)} & \qw & \qw & \qw & \qw & \qw & \qw & \qw\\
				}
				
				\\
				
			\end{tabular}
			
		} 
		\\\hline

		\multicolumn{4}{|c|}{
			\begin{tabular}{@{} p{1.02\textwidth} @{}}
				4 
		\end{tabular}} 
		\\
		\multicolumn{4}{|c|}{
			\begin{tabular}{@{} p{0.6\textwidth}<{\centering} @{}}
				\\
				\Qcircuit @C=0.2em @R=0.6em {
					\lstick{} & \gate{RZ(-\pi)}  & \gate{RY(-\frac{3\pi}{4})} & \gate{RX(\frac{\pi}{2})}  &  \ctrl{1}  & \gate{RX(-\frac{\pi}{2})} & \ctrl{1} & \gate{RX(-\frac{\pi}{2})} & \gate{RZ(\pi)} & \gate{RY(-\frac{\pi}{2})}  & \qw \\
					\lstick{} & \gate{RZ(\pi)}  & \gate{RY(-\frac{\pi}{2})}  &  \gate{RZ(-\frac{\pi}{2})} & \targ & \gate{RY(-\frac{\pi}{4})} & \targ & \gate{RY(-\pi)} & \gate{RZ(\frac{\pi}{2})} & \qw & \qw & \qw\\
				} 
				
				\\
				
			\end{tabular}
			
		} 
		\\\hline
		
	\end{tabular}
	\caption{\textbf{Set 1}}
\end{table*}

\clearpage

\begin{table*}[b]
	\centering
	\begin{tabular}{|c|c|p{0.5\textwidth}<{\centering}|c|}\hline
		ID&$P$ (Hermite)& $Q$ (Unitary)& Eigenvalues of Q \\\hline
		5&$IX$& $0.5 IX+0.5XI+0.5iYZ-0.5iZY$& \{-1,$i$,-$i$,1\}\\\hline 
		6&$XI$& $0.5 IX+0.5XI-0.5iYZ+0.5iZY$& \{-1,$i$,-$i$,1\}\\\hline 
		7&$YZ$& $-0.5i IX+0.5iXI+0.5YZ+0.5ZY$& \{-1,$i$,-$i$,1\}\\\hline
		8&$ZY$& $-0.5i IX+0.5iXI-0.5YZ-0.5ZY$&\{-1,$i$,-$i$,1\}\\\hline
		
		\multicolumn{4}{|c|}{Q matrix}\\\hline
		\multicolumn{4}{|c|}{
			\begin{tabular}{@{} p{0.25\textwidth}|p{0.25\textwidth}|p{0.25\textwidth}|p{0.25\textwidth} @{}}
				5 & 6& 7& 8
		\end{tabular}} 
		\\
		\multicolumn{4}{|c|}{
			\begin{tabular}{@{} p{0.25\textwidth}<{\centering}|p{0.25\textwidth}<{\centering}|p{0.25\textwidth}<{\centering}|p{0.25\textwidth}<{\centering} @{}}
				$\begin{pmatrix}
					0 & 0 & 1 & 0\\
					1 & 0 & 0 & 0\\
					0 & 0 & 0 & 1\\
					0 & 1 & 0 & 0
				\end{pmatrix}$&
				$\begin{pmatrix}
					0 & 1 & 0 & 0\\
					0 & 0 & 0 & 1\\
					1 & 0 & 0 & 0\\
					0 & 0 & 1 & 0
				\end{pmatrix}$&
				$\begin{pmatrix}
					0 & -i & 0 & 0\\
					0 & 0 & 0 & i\\
					i & 0 & 0 & 0\\
					0 & 0 & -i & 0
				\end{pmatrix}$&
				$\begin{pmatrix}
					0 & 0 & i & 0\\
					-i & 0 & 0 & 0\\
					0 & 0 & 0 & -i\\
					0 & i & 0 & 0
				\end{pmatrix}$\\
				& & & 
		\end{tabular}} 
		\\\hline
		
		\multicolumn{4}{|c|}{Q circuit}\\\hline
		\multicolumn{4}{|c|}{
			\begin{tabular}{@{} p{0.25\textwidth}|p{0.25\textwidth}|p{0.25\textwidth}|p{0.25\textwidth} @{}}
				5 & 6& 7& 8
		\end{tabular}} 
		\\
		\multicolumn{4}{|c|}{
			\begin{tabular}{@{} p{0.25\textwidth}<{\centering}|p{0.25\textwidth}<{\centering}|p{0.25\textwidth}<{\centering}|p{0.25\textwidth}<{\centering} @{}}
				\
				\Qcircuit @C=2.2em @R=2.2em {
					\lstick{} & \gate{X} & \qswap  & \qw \\
					\lstick{} & \qw & \qswap \qwx & \qw 
				}&
				\
				\Qcircuit @C=2.2em @R=2.2em {
					\lstick{} & \qw & \qswap  & \qw \\
					\lstick{} & \gate{X} & \qswap \qwx & \qw 
				}&
				\
				\Qcircuit @C=2em @R=2em {
					\lstick{} & \gate{Z} & \qswap  & \qw \\
					\lstick{} & \gate{Y} & \qswap \qwx & \qw 
				}&
				\
				\Qcircuit @C=2em @R=2em {
					\lstick{} & \gate{i Y} & \qswap  & \qw \\
					\lstick{} & \gate{i Z} & \qswap \qwx & \qw 
				}\\
				& & & 
		\end{tabular}} 
		\\\hline
		
		\multicolumn{4}{|c|}{Diagonalizing matrix}\\\hline
		\multicolumn{4}{|c|}{
			\begin{tabular}{@{} p{0.25\textwidth}|p{0.25\textwidth}|p{0.25\textwidth}|p{0.25\textwidth} @{}}
				5 & 6& 7& 8
		\end{tabular}} 
		\\
		\multicolumn{4}{|c|}{
			\begin{tabular}{@{} p{0.25\textwidth}<{\centering}|p{0.25\textwidth}<{\centering}|p{0.25\textwidth}<{\centering}|p{0.25\textwidth}<{\centering} @{}}
				$\begin{pmatrix}
					0.5 & -0.5 i & 0.5 & 0.5 \\
					-0.5 & -0.5 & 0.5 i & 0.5 \\
					-0.5  & 0.5 & -0.5i & 0.5 \\
					0.5 & 0.5 i & -0.5 & 0.5
				\end{pmatrix}$&
				$\begin{pmatrix}
					0.5 & -0.5 i & 0.5 & 0.5 \\
					-0.5  & 0.5 & -0.5i & 0.5 \\
					-0.5 & -0.5 & 0.5 i & 0.5 \\
					0.5 & 0.5 i & -0.5 & 0.5
				\end{pmatrix}$&
				$\begin{pmatrix}
					0.5 & -0.5  & 0.5 & 0.5 \\
					-0.5 i & 0.5 & 0.5 & 0.5 i \\
					-0.5 i & -0.5 & -0.5 & 0.5 i \\
					0.5 & 0.5  & -0.5 & 0.5
				\end{pmatrix}$&
				$\begin{pmatrix}
					0.5 & 0.5  & 0.5 & 0.5 \\
					0.5 i & -0.5 & 0.5 & -0.5 i \\
					0.5 i & 0.5 & -0.5 & -0.5 i \\
					0.5 & -0.5  & -0.5 & 0.5
				\end{pmatrix}$\\
				& & & 
		\end{tabular}} 
		\\\hline
		\multicolumn{4}{|c|}{Diagonalizing matrix circuit}\\\hline

		\multicolumn{4}{|c|}{
			\begin{tabular}{@{} p{1.02\textwidth} @{}}
				5 
		\end{tabular}} 
		\\
		\multicolumn{4}{|c|}{
			\begin{tabular}{@{} p{0.6\textwidth}<{\centering} @{}}
				\\
				\Qcircuit @C=0.2em @R=0.6em {
					\lstick{} & \gate{RZ(\frac{\pi}{4})}  & \gate{RY(-\frac{\pi}{2})} & \gate{RZ(-\pi)}  &  \ctrl{1}  & \qw & \qw & \ctrl{1} & \gate{RX(-\frac{\pi}{2})} & \gate{RZ(\frac{\pi}{2})} & \gate{RY(-\frac{\pi}{2})} & \gate{RZ(\frac{\pi}{2})} & \gate{U1(-2\pi)} & \gate{RZ(2\pi)} & \qw \\
					\lstick{} & \gate{RZ(\frac{3\pi}{4})}  & \gate{RY(-\frac{\pi}{2})}  &  \gate{RZ(-\pi)} & \targ & \gate{RZ(\frac{\pi}{4})} & \gate{RX(\frac{\pi}{2})} & \targ & \gate{RZ(\frac{\pi}{2})} & \gate{RY(-\frac{\pi}{2})} & \gate{RZ(-\frac{3\pi}{2})} & \qw & \qw & \qw & \qw & \qw\\
				} 
				
				\\
				
			\end{tabular}
			
		} 
		\\\hline
		
		\multicolumn{4}{|c|}{
			\begin{tabular}{@{} p{1.02\textwidth} @{}}
				6 
		\end{tabular}} 
		\\
		\multicolumn{4}{|c|}{
			\begin{tabular}{@{} p{0.6\textwidth}<{\centering} @{}}
				\\
				\Qcircuit @C=0.2em @R=0.6em {
					\lstick{} & \gate{RZ(\frac{7\pi}{8})}  & \gate{RZ(\frac{7\pi}{8})}   & \gate{RX(\frac{\pi}{2})} & \ctrl{1}  & \gate{RX(-\frac{\pi}{4})} & \ctrl{1} & \gate{RX(-\frac{\pi}{2})} & \gate{RY(-\frac{\pi}{2})} & \gate{RZ(-2\pi)} & \qw \\
					\lstick{} & \gate{RZ(-\frac{3\pi}{8})}  & \gate{RZ(-\frac{3\pi}{8})}  &   \qw & \targ & \gate{RY(-\frac{\pi}{4})} & \targ & \gate{RZ(\pi)} & \gate{RY(-\frac{\pi}{2})}  & \qw & \qw\\
				} 
				
				\\
				
			\end{tabular}
			
		} 
		\\\hline
		
		\multicolumn{4}{|c|}{
			\begin{tabular}{@{} p{1.02\textwidth} @{}}
				7 
		\end{tabular}} 
		\\
		\multicolumn{4}{|c|}{
			\begin{tabular}{@{} p{0.6\textwidth}<{\centering} @{}}
				\\
				\Qcircuit @C=0.2em @R=0.6em {
					\lstick{} & \gate{RZ(\frac{\pi}{2})}  & \gate{RY(-\pi)} & \gate{RZ(-\frac{\pi}{2})}   & \ctrl{1}  & \qw & \qw  & \targ & \gate{RX(\frac{\pi}{4})} & \ctrl{1}   & \gate{RX(-\frac{\pi}{2})} & \gate{RZ(-\frac{3\pi}{2})} & \gate{RY(-\frac{\pi}{2})}  & \gate{U1(-\frac{\pi}{4})} & \gate{RZ(\frac{\pi}{4})} & \qw \\
					\lstick{} & \qw  & \qw   & \qw & \targ & \gate{RZ(\frac{3\pi}{4})} & \gate{RY(\frac{\pi}{2})} & \ctrl{-1} & \gate{RY(\frac{\pi}{4})} & \targ & \gate{RZ(\frac{\pi}{2})} & \gate{RY(-\frac{\pi}{2})} & \gate{RZ(-\pi)} &  \qw & \qw & \qw & \qw\\
				}
				
				\\
				
			\end{tabular}
			
		} 
		\\\hline

		\multicolumn{4}{|c|}{
			\begin{tabular}{@{} p{1.02\textwidth} @{}}
				8 
		\end{tabular}} 
		\\
		\multicolumn{4}{|c|}{
			\begin{tabular}{@{} p{0.6\textwidth}<{\centering} @{}}
				\\
				\Qcircuit @C=0.2em @R=0.6em {
					\lstick{} &  \gate{RY(-\pi)}    & \ctrl{1}  & \qw & \qw  & \targ & \gate{RX(\frac{\pi}{4})} & \ctrl{1}   & \gate{RX(-\frac{\pi}{2})} & \gate{RZ(-\pi)} & \gate{RY(-\frac{\pi}{2})}  & \gate{U1(-\frac{5\pi}{4})} & \gate{RZ(\frac{5\pi}{4})} & \qw \\
					\lstick{} & \qw   & \targ & \gate{RZ(\frac{3\pi}{4})} & \gate{RY(\frac{\pi}{2})} & \ctrl{-1} & \gate{RY(\frac{\pi}{4})} & \targ & \gate{RY(-\frac{\pi}{2})} & \gate{RZ(-\pi)} & \qw &  \qw & \qw & \qw & \qw\\
				}
				
				\\
				
			\end{tabular}
			
		} 
		\\\hline
	\end{tabular}
	\caption{\textbf{Set 2}}
\end{table*}

\clearpage

\begin{table*}[t]   
	\centering 
	\begin{tabular}{|c|c|p{0.5\textwidth}<{\centering}|c|}\hline
		ID&$P$ (Hermite)& $Q$ (Unitary)& Eigenvalues of Q \\\hline
		9&$IY$& $-0.5 IY+0.5iXZ-0.5YI-0.5iZX$& \{-1,$i$,-$i$,1\}\\\hline 
		10&$YI$& $0.5 IY+0.5iXZ+0.5YI-0.5iZX$& \{-1,$i$,-$i$,1\}\\\hline 
		11&$XZ$& $0.5i IY+0.5XZ-0.5iYI+0.5ZX$& \{-1,$i$,-$i$,1\}\\\hline 
		12&$ZX$& $-0.5i IY+0.5XZ+0.5iYI+0.5ZX$& \{-1,$i$,-$i$,1\}\\\hline  
		
		\multicolumn{4}{|c|}{Q matrix}\\\hline
		\multicolumn{4}{|c|}{
			\begin{tabular}{@{} p{0.25\textwidth}|p{0.25\textwidth}|p{0.25\textwidth}|p{0.25\textwidth} @{}}
				9 & 10& 11& 12
		\end{tabular}} 
		\\
		\multicolumn{4}{|c|}{
			\begin{tabular}{@{} p{0.25\textwidth}<{\centering}|p{0.25\textwidth}<{\centering}|p{0.25\textwidth}<{\centering}|p{0.25\textwidth}<{\centering} @{}}
				$\begin{pmatrix}
					0 & 0 & i & 0\\
					-i & 0 & 0 & 0\\
					0 & 0 & 0 & i\\
					0 & -i & 0 & 0
				\end{pmatrix}$&
				$\begin{pmatrix}
					0 & -i & 0 & 0\\
					0 & 0 & 0 & -i\\
					i & 0 & 0 & 0\\
					0 & 0 & i & 0
				\end{pmatrix}$&
				$\begin{pmatrix}
					0 & 1 & 0 & 0\\
					0 & 0 & 0 & -1\\
					1 & 0 & 0 & 0\\
					0 & 0 & -1 & 0
				\end{pmatrix}$&
				$\begin{pmatrix}
					0 & 0 & 1 & 0\\
					1 & 0 & 0 & 0\\
					0 & 0 & 0 & -1\\
					0 & -1 & 0 & 0
				\end{pmatrix}$\\
				& & & 
		\end{tabular}} 
		\\\hline
		
		\multicolumn{4}{|c|}{Q circuit}\\\hline
		\multicolumn{4}{|c|}{
			\begin{tabular}{@{} p{0.25\textwidth}|p{0.25\textwidth}|p{0.25\textwidth}|p{0.25\textwidth} @{}}
				9 & 10& 11& 12
		\end{tabular}} 
		\\
		\multicolumn{4}{|c|}{
			\begin{tabular}{@{} p{0.25\textwidth}<{\centering}|p{0.25\textwidth}<{\centering}|p{0.25\textwidth}<{\centering}|p{0.25\textwidth}<{\centering} @{}}
				\
				\Qcircuit @C=2.2em @R=2.2em {
					\lstick{} & \gate{-Y} & \qswap  & \qw \\
					\lstick{} & \qw & \qswap \qwx & \qw 
				}&
				\
				\Qcircuit @C=2.2em @R=2.2em {
					\lstick{} & \qw & \qswap  & \qw \\
					\lstick{} & \gate{Y} & \qswap \qwx & \qw 
				}&
				\
				\Qcircuit @C=2em @R=2em {
					\lstick{} & \gate{Z} & \qswap  & \qw \\
					\lstick{} & \gate{X} & \qswap \qwx & \qw 
				}&
				\
				\Qcircuit @C=2em @R=2em {
					\lstick{} & \gate{X} & \qswap  & \qw \\
					\lstick{} & \gate{Z} & \qswap \qwx & \qw 
				}\\
				& & & 
		\end{tabular}} 
		\\\hline
		
		\multicolumn{4}{|c|}{Diagonalizing matrix}\\\hline
		\multicolumn{4}{|c|}{
			\begin{tabular}{@{} p{0.25\textwidth}|p{0.25\textwidth}|p{0.25\textwidth}|p{0.25\textwidth} @{}}
				9 & 10& 11& 12
		\end{tabular}} 
		\\
		\multicolumn{4}{|c|}{
			\begin{tabular}{@{} p{0.25\textwidth}<{\centering}|p{0.25\textwidth}<{\centering}|p{0.25\textwidth}<{\centering}|p{0.25\textwidth}<{\centering} @{}}
				$\begin{pmatrix}
					0.5 & 0.5  & 0.5 & -0.5 \\
					0.5 i & -0.5 & 0.5 & 0.5 i \\
					0.5 i & 0.5 & -0.5 & 0.5 i \\
					-0.5 & 0.5  & 0.5 & 0.5
				\end{pmatrix}$&
				$\begin{pmatrix}
					0.5 & -0.5  & 0.5 & -0.5 \\
					-0.5 i & 0.5 & 0.5 & -0.5 i \\
					-0.5 i & -0.5 & -0.5 & -0.5 i \\
					-0.5 & -0.5  & 0.5 & 0.5
				\end{pmatrix}$&
				$\begin{pmatrix}
					0.5 & -0.5 i & 0.5 & -0.5 \\
					-0.5  & 0.5 & -0.5i & -0.5 \\
					-0.5 & -0.5  & 0.5i & -0.5 \\
					-0.5 & -0.5 i & 0.5 & 0.5
				\end{pmatrix}$&
				$\begin{pmatrix}
					0.5 & -0.5 i & 0.5 & -0.5 \\
					-0.5 & -0.5 & 0.5i & -0.5  \\
					-0.5  & 0.5 & -0.5i & -0.5 \\
					-0.5 & -0.5 i & 0.5 & 0.5
				\end{pmatrix}$\\
				& & & 
		\end{tabular}} 
		\\\hline
		\multicolumn{4}{|c|}{Diagonalizing matrix circuit}\\\hline

		\multicolumn{4}{|c|}{
			\begin{tabular}{@{} p{1.02\textwidth} @{}}
				9 
		\end{tabular}} 
		\\
		\multicolumn{4}{|c|}{
			\begin{tabular}{@{} p{0.6\textwidth}<{\centering} @{}}
				\\
				\Qcircuit @C=0.2em @R=0.6em {
					\lstick{} & \gate{RZ(\pi)}  & \gate{RY(-\frac{\pi}{2})} & \gate{RZ(\frac{\pi}{2})}   & \ctrl{1}  & \qw & \qw  & \targ & \gate{RX(\frac{\pi}{4})} & \ctrl{1}   & \gate{RX(-\frac{\pi}{2})} & \gate{RY(-\frac{\pi}{2})} & \gate{RZ(\pi)}  & \gate{U1(-\frac{5\pi}{4})} & \gate{RZ(\frac{5\pi}{4})} & \qw \\
					\lstick{} & \gate{RY(-\frac{\pi}{2})}  & \gate{RZ(-\frac{\pi}{2})}   & \qw & \targ & \gate{RZ(\frac{3\pi}{4})} & \gate{RY(\frac{\pi}{2})} & \ctrl{-1} & \gate{RY(\frac{\pi}{4})} & \targ & \gate{RY(-\frac{\pi}{2})} & \qw & \qw &  \qw & \qw & \qw & \qw\\
				}
				
				\\
				
			\end{tabular}
			
		} 
		\\\hline
		
		\multicolumn{4}{|c|}{
			\begin{tabular}{@{} p{1.02\textwidth} @{}}
				10 
		\end{tabular}} 
		\\
		\multicolumn{4}{|c|}{
			\begin{tabular}{@{} p{0.6\textwidth}<{\centering} @{}}
				\\
				\Qcircuit @C=0.2em @R=0.6em {
					\lstick{} & \gate{RZ(\pi)}  & \gate{RY(-\frac{\pi}{2})} & \gate{RZ(\frac{\pi}{2})}   & \ctrl{1}  & \qw & \qw  & \targ & \gate{RX(\frac{\pi}{4})} & \ctrl{1}   & \gate{RX(-\frac{\pi}{2})} & \gate{RZ(-\pi)} & \gate{RZ(-\pi)}  & \gate{U1(-\frac{5\pi}{4})} & \gate{RZ(\frac{5\pi}{4})} & \qw \\
					\lstick{} & \gate{RY(-\frac{\pi}{2})}  & \gate{RZ(-\frac{\pi}{2})}   & \qw & \targ & \gate{RZ(\frac{3\pi}{4})} & \gate{RY(\frac{\pi}{2})} & \ctrl{-1} & \gate{RY(\frac{\pi}{4})} & \targ & \gate{RZ(-\frac{\pi}{2})} & \gate{RZ(-\frac{\pi}{2})} & \qw &  \qw & \qw & \qw & \qw\\
				} 
				
				\\
				
			\end{tabular}
			
		} 
		\\\hline
		
		\multicolumn{4}{|c|}{
			\begin{tabular}{@{} p{1.02\textwidth} @{}}
				11 
		\end{tabular}} 
		\\
		\multicolumn{4}{|c|}{
			\begin{tabular}{@{} p{0.6\textwidth}<{\centering} @{}}
				\\
				\Qcircuit @C=0.2em @R=0.6em {
					\lstick{} & \gate{RY(-\frac{\pi}{2})}  & \qw  & \qw & \qw   & \ctrl{1}  & \qw & \qw  & \targ & \gate{RX(0)} & \ctrl{1}   & \gate{RX(-\frac{\pi}{2})} & \gate{RZ(-\frac{\pi}{2})} & \gate{RY(-\frac{3\pi}{4})} & \gate{RZ(\pi)} & \gate{U1(-\frac{3\pi}{2})} & \gate{RZ(\frac{3\pi}{2})} & \qw \\
					\lstick{} & \gate{RZ(\pi)} & \gate{RY(-\frac{\pi}{2})}  & \gate{RZ(-\pi)}   & \qw & \targ & \gate{RZ(\frac{\pi}{4})} & \gate{RX(\frac{\pi}{2})} & \ctrl{-1} & \gate{RY(\frac{\pi}{4})} & \targ & \gate{RZ(\frac{3\pi}{2})} & \gate{RY(-\frac{3\pi}{4})} &\qw & \qw & \qw & \qw & \qw & \qw\\
				}
				
				\\
				
			\end{tabular}
			
		} 
		\\\hline

		\multicolumn{4}{|c|}{
			\begin{tabular}{@{} p{1.02\textwidth} @{}}
				12 
		\end{tabular}} 
		\\
		\multicolumn{4}{|c|}{
			\begin{tabular}{@{} p{0.6\textwidth}<{\centering} @{}}
				\\
				\Qcircuit @C=0.2em @R=0.6em {
					\lstick{} & \gate{RZ(-0.982)}  & \gate{RY(-\pi)} & \gate{RZ(0.982)}  & \gate{RX(\frac{\pi}{2})} & \ctrl{1}  & \gate{RX(-\frac{\pi}{4})} & \ctrl{1} & \gate{RX(-\frac{\pi}{2})} & \gate{RZ(-\frac{7\pi}{8})} & \gate{RY(-\frac{\pi}{2})} & \gate{RZ(\frac{\pi}{2})} & \gate{U1(-\frac{3\pi}{2})} & \gate{RZ(\frac{3\pi}{2})} & \qw \\
					\lstick{} & \gate{RZ(-0.589)}  & \gate{RZ(-0.589)}  &  \qw & \qw & \targ & \gate{RY(-\frac{\pi}{4})} & \targ & \gate{RZ(\frac{\pi}{8})} & \gate{RY(-\frac{\pi}{2})} & \gate{RZ(-\frac{\pi}{2})} & \qw & \qw & \qw & \qw & \qw\\
				} 
				
				\\
				
			\end{tabular}
			
		} 
		\\\hline
	\end{tabular}
	\caption{\textbf{Set 3}}
\end{table*}
\clearpage

\begin{table*}[b]
	\centering
	\begin{tabular}{|c|c|p{0.5\textwidth}<{\centering}|c|}\hline
		ID&$P$ (Hermite)& $Q$ (Unitary)& Eigenvalues of Q \\\hline
		13&$IZ$& $0.5 IZ+0.5iXY-0.5iYX+0.5ZI$& \{$i$,-$i$,1,-1\}\\\hline 
		14&$ZI$& $0.5 IZ-0.5iXY+0.5iYX+0.5ZI$& \{$i$,-$i$,1,-1\}\\\hline   
		15&$XY$& $0.5i IZ-0.5XY-0.5YX-0.5iZI$& \{1,-1,-$i$,$i$\}\\\hline
		16&$YX$& $0.5i IZ+0.5XY+0.5YX-0.5iZI$& \{1,-1,-$i$,$i$\}\\\hline
		
		\multicolumn{4}{|c|}{Q matrix}\\\hline
		\multicolumn{4}{|c|}{
			\begin{tabular}{@{} p{0.25\textwidth}|p{0.25\textwidth}|p{0.25\textwidth}|p{0.25\textwidth} @{}}
				13 & 14& 15& 16
		\end{tabular}} 
		\\
		\multicolumn{4}{|c|}{
			\begin{tabular}{@{} p{0.25\textwidth}<{\centering}|p{0.25\textwidth}<{\centering}|p{0.25\textwidth}<{\centering}|p{0.25\textwidth}<{\centering} @{}}
				$\begin{pmatrix}
					1 & 0 & 0 & 0\\
					0 & 0 & -1 & 0\\
					0 & 1 & 0 & 0\\
					0 & 0 & 0 & -1
				\end{pmatrix}$&
				$\begin{pmatrix}
					1 & 0 & 0 & 0\\
					0 & 0 & 1 & 0\\
					0 & -1 & 0 & 0\\
					0 & 0 & 0 & -1
				\end{pmatrix}$&
				$\begin{pmatrix}
					0 & 0 & 0 & i\\
					0 & -i & 0 & 0\\
					0 & 0 & i & 0\\
					-i & 0 & 0 & 0
				\end{pmatrix}$&
				$\begin{pmatrix}
					0 & 0 & 0 & -i\\
					0 & -i & 0 & 0\\
					0 & 0 & i & 0\\
					i & 0 & 0 & 0
				\end{pmatrix}$\\
				& & & 
		\end{tabular}} 
		\\\hline
		
		\multicolumn{4}{|c|}{Q circuit}\\\hline
		\multicolumn{4}{|c|}{
			\begin{tabular}{@{} p{0.25\textwidth}|p{0.25\textwidth}|p{0.25\textwidth}|p{0.25\textwidth} @{}}
				13 & 14& 15& 16
		\end{tabular}} 
		\\
		\multicolumn{4}{|c|}{
			\begin{tabular}{@{} p{0.25\textwidth}<{\centering}|p{0.25\textwidth}<{\centering}|p{0.25\textwidth}<{\centering}|p{0.25\textwidth}<{\centering} @{}}
				\
				\Qcircuit @C=2.2em @R=2.2em {
					\lstick{} & \gate{Z} & \qswap  & \qw \\
					\lstick{} & \qw & \qswap \qwx & \qw 
				}&
				\
				\Qcircuit @C=2.2em @R=2.2em {
					\lstick{} & \qw & \qswap  & \qw \\
					\lstick{} & \gate{Z} & \qswap \qwx & \qw 
				}&
				\
				\Qcircuit @C=2em @R=2em {
					\lstick{} & \gate{iY} & \qswap  & \qw \\
					\lstick{} & \gate{iX} & \qswap \qwx & \qw 
				}&
				\
				\Qcircuit @C=2em @R=2em {
					\lstick{} & \gate{X} & \qswap  & \qw \\
					\lstick{} & \gate{Y} & \qswap \qwx & \qw 
				}\\
				& & & 
		\end{tabular}} 
		\\\hline
		
		\multicolumn{4}{|c|}{Diagonalizing matrix}\\\hline
		\multicolumn{4}{|c|}{
			\begin{tabular}{@{} p{0.25\textwidth}|p{0.25\textwidth}|p{0.25\textwidth}|p{0.25\textwidth} @{}}
				13 & 14& 15& 16
		\end{tabular}} 
		\\
		\multicolumn{4}{|c|}{
			\begin{tabular}{@{} p{0.25\textwidth}<{\centering}|p{0.25\textwidth}<{\centering}|p{0.25\textwidth}<{\centering}|p{0.25\textwidth}<{\centering} @{}}
				$\begin{pmatrix}
					0 & 0 & 1 & 0 \\
					\frac{1}{\sqrt{2}}i & \frac{1}{\sqrt{2}} & 0 & 0 \\
					\frac{1}{\sqrt{2}} & \frac{1}{\sqrt{2}}i & 0 & 0 \\
					0 & 0 & 0 & 1
				\end{pmatrix}$&
				$\begin{pmatrix}
					0 & 0 & 1 & 0 \\
					-\frac{1}{\sqrt{2}}i & \frac{1}{\sqrt{2}} & 0 & 0 \\
					\frac{1}{\sqrt{2}} & -\frac{1}{\sqrt{2}}i & 0 & 0 \\
					0 & 0 & 0 & 1
				\end{pmatrix}$&
				$\begin{pmatrix}
					\frac{1}{\sqrt{2}}i & \frac{1}{\sqrt{2}} & 0 & 0 \\
					0 & 0 & 1 & 0 \\
					0 & 0 & 0 & 1 \\
					\frac{1}{\sqrt{2}} & \frac{1}{\sqrt{2}}i & 0 & 0
				\end{pmatrix}$&
				$\begin{pmatrix}
					-\frac{1}{\sqrt{2}}i  & \frac{1}{\sqrt{2}} & 0 & 0 \\
					0 & 0 & 1 & 0 \\
					0 & 0 & 0 & 1 \\
					\frac{1}{\sqrt{2}} & -\frac{1}{\sqrt{2}}i & 0 & 0
				\end{pmatrix}$\\
				& & & 
		\end{tabular}} 
		\\\hline
		\multicolumn{4}{|c|}{Diagonalizing matrix circuit}\\\hline

		\multicolumn{4}{|c|}{
			\begin{tabular}{@{} p{1.02\textwidth} @{}}
				13 
		\end{tabular}} 
		\\
		\multicolumn{4}{|c|}{
			\begin{tabular}{@{} p{0.6\textwidth}<{\centering} @{}}
				\\
				\Qcircuit @C=0.2em @R=0.6em {
					\lstick{} & \gate{RZ(-\frac{\pi}{2})}  & \gate{RY(-\frac{\pi}{4})} & \gate{RZ(\pi)}  & \gate{RX(\frac{\pi}{2})} & \ctrl{1}  & \gate{RX(-\frac{\pi}{2})} & \ctrl{1} & \gate{RX(-\frac{\pi}{2})} &  \gate{RY(-\frac{\pi}{2})} &  \gate{U1(\frac{3\pi}{2})} & \gate{RZ(-\frac{3\pi}{2})} & \qw \\
					\lstick{} & \gate{RZ(\frac{\pi}{2})}  & \gate{RY(-\frac{\pi}{2})}  &  \gate{RZ(-\frac{\pi}{2})} & \qw & \targ & \gate{RY(-\frac{\pi}{4})} & \targ & \gate{RZ(2.35)} & \gate{RZ(3.93)} &   \qw & \qw & \qw & \qw\\
				} 
				
				\\
				
			\end{tabular}
			
		} 
		\\\hline
		
		\multicolumn{4}{|c|}{
			\begin{tabular}{@{} p{1.02\textwidth} @{}}
				14 
		\end{tabular}} 
		\\
		\multicolumn{4}{|c|}{
			\begin{tabular}{@{} p{0.6\textwidth}<{\centering} @{}}
				\\
				\Qcircuit @C=0.2em @R=0.6em {
					\lstick{} & \gate{RZ(-\frac{\pi}{2})}  & \gate{RY(-\frac{3\pi}{4})} & \gate{RZ(-\pi)}  & \gate{RX(\frac{\pi}{2})} & \ctrl{1}  & \gate{RX(-\frac{\pi}{2})} & \ctrl{1} & \gate{RX(-\frac{\pi}{2})} &  \gate{RY(-\frac{\pi}{2})} &  \gate{U1(\frac{3\pi}{2})} & \gate{RZ(-\frac{3\pi}{2})} & \qw \\
					\lstick{} & \gate{RZ(\frac{3\pi}{2})}  & \gate{RY(-\frac{\pi}{2})}  &  \gate{RZ(\frac{\pi}{2})} & \qw & \targ & \gate{RY(-\frac{\pi}{4})} & \targ & \gate{RY(-\pi)} & \gate{RZ(\pi)} &   \qw & \qw & \qw & \qw\\
				}
				
				\\
				
			\end{tabular}
			
		} 
		\\\hline
		
		\multicolumn{4}{|c|}{
			\begin{tabular}{@{} p{1.02\textwidth} @{}}
				15 
		\end{tabular}} 
		\\
		\multicolumn{4}{|c|}{
			\begin{tabular}{@{} p{0.6\textwidth}<{\centering} @{}}
				\\
				\Qcircuit @C=0.2em @R=0.6em {
					\lstick{} & \gate{RZ(\frac{3\pi}{2})}  & \gate{RY(-\frac{\pi}{4})} &  \gate{RX(\frac{\pi}{2})} & \ctrl{1}  & \gate{RX(-\frac{\pi}{2})} & \ctrl{1} & \gate{RX(-\frac{\pi}{2})} &  \gate{RY(-\frac{\pi}{2})} & \gate{RZ(\pi)} & \gate{U1(\frac{3\pi}{2})} & \gate{RZ(-\frac{3\pi}{2})} & \qw \\
					\lstick{} & \gate{RZ(-\frac{\pi}{2})}  & \gate{RY(-\frac{\pi}{2})}  &  \gate{RZ(\frac{\pi}{2})} &  \targ & \gate{RY(-\frac{\pi}{4})} & \targ & \gate{RY(\pi)} & \gate{RZ(\pi)} & \qw &  \qw & \qw & \qw & \qw\\
				}
				
				\\
				
			\end{tabular}
			
		} 
		\\\hline

		\multicolumn{4}{|c|}{
			\begin{tabular}{@{} p{1.02\textwidth} @{}}
				16 
		\end{tabular}} 
		\\
		\multicolumn{4}{|c|}{
			\begin{tabular}{@{} p{0.6\textwidth}<{\centering} @{}}
				\\
				\Qcircuit @C=0.2em @R=0.6em {
					\lstick{} & \gate{RZ(-\frac{\pi}{2})}  & \gate{RY(-\frac{3\pi}{4})} & \gate{RZ(\pi)}  & \gate{RX(\frac{\pi}{2})} & \ctrl{1}  & \gate{RX(-\frac{\pi}{2})} & \ctrl{1} & \gate{RX(-\frac{\pi}{2})} & \gate{RZ(-\pi)} & \gate{RY(-\frac{\pi}{2})} & \gate{RZ(-\pi)} & \gate{U1(-\frac{5\pi}{2})} & \gate{RZ(\frac{5\pi}{2})} & \qw \\
					\lstick{} & \gate{RZ(\frac{\pi}{2})}  & \gate{RY(-\frac{\pi}{2})}  &  \gate{RZ(\frac{\pi}{2})} & \qw & \targ & \gate{RY(-\frac{\pi}{4})} & \targ & \gate{RZ(-1.23)} & \gate{RY(-\pi)} & \gate{RZ(-1.23)} & \qw & \qw & \qw & \qw & \qw\\
				} 
				
				\\
				
			\end{tabular}
			
		} 
		\\\hline
		
	\end{tabular}
	\caption{\textbf{Set 4}}
\end{table*}
\clearpage
In the above tables, we give detailed descriptions of all 16 2-qubit Pauli operators and their corresponding substitute operators. We also give concrete constructions of rotation circuits to make these substitute operators diagonal for direct measurements.

\section{Welcomed hardware structure}
\begin{figure*}[htbp]
\centering
\includegraphics[width=0.7\textwidth]{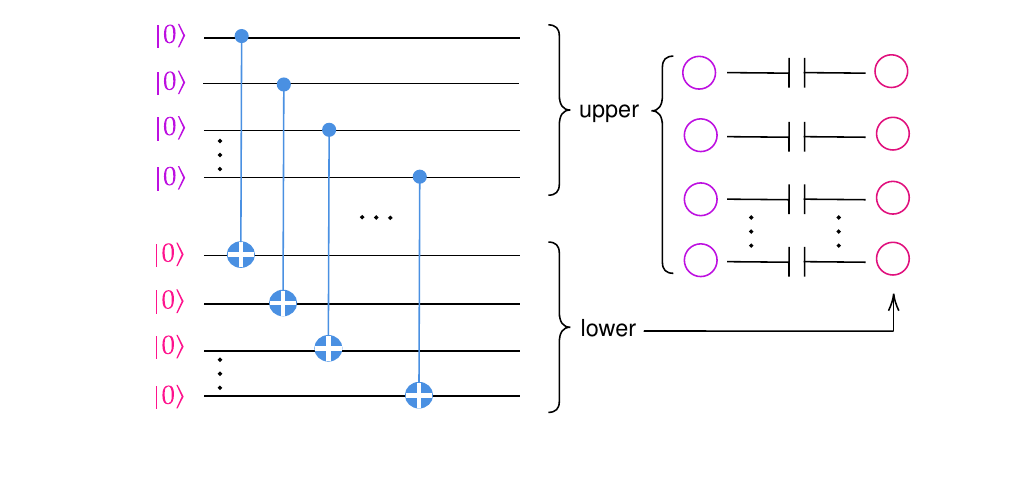}
\caption{Suitable hardware structure. \label{shs}}
\end{figure*}

Since the circuit ansatz we use contains the qubits in the upper system and the qubits in the lower system, and we may always need CNOT gates connecting the upper and the lower system, 1D quantum circuits will make these CNOT gates non-local. Thus, a more suitable architecture would be those with a ladder shape such that these CNOT gates are turned locally as shown in Fig. \ref{shs}.

\end{document}